\documentclass[review]{elsarticle}

\usepackage{hyperref}
\renewcommand{\baselinestretch}{1.8}
\usepackage{amssymb}
\newcommand\Tau{\mathcal{T}}
\usepackage{amsmath}
\usepackage{amssymb}
\usepackage{enumitem}
\usepackage{dsfont}
\usepackage{lscape}
\usepackage{rotating}
\usepackage{booktabs}
\usepackage[table]{xcolor}
\usepackage{threeparttable}

\usepackage{numcompress}\biboptions{authoryear}

\begin{document}

\begin{frontmatter}

\title{A Comparison of Testing Methods in Scalar-on-Function Regression}

\author{Merve Yasemin~Tekbudak\corref{cor1}}
\ead{mytekbud@ncsu.edu}
\author{Marcela Alfaro~C\'{o}rdoba\corref{}}
\ead{malfaro@ncsu.edu}
\author{Arnab~Maity\corref{}}
\ead{amaity@ncsu.edu}
\author{and Ana-Maria~Staicu\corref{}}
\ead{astaicu@ncsu.edu}

\address{Department of Statistics, North Carolina State University}

\cortext[cor1]{Corresponding author. \textit{Address:} North Carolina State University, 
               Department of Statistics, 5109 SAS Hall, 2301 Stinson Dr., Raleigh, NC 27695-8203}

% Address: University of Groningen, University Medical Center Groningen, Department
%of Epidemiology, Hanzeplein 1, 9700 RB, Groningen, The Netherlands. Tel: +31 503612273


%\tnotetext[mytitlenote]{Fully documented templates are available in the elsarticle package on 
%\href{http://www.ctan.org/tex-archive/macros/latex/contrib/elsarticle}{CTAN}.}


\begin{abstract}
A scalar-response functional model describes the association between a scalar response and a 
set of functional covariates. An important problem in the functional data literature is to 
test the nullity or linearity of the effect of the functional covariate in the context of 
scalar-on-function regression. This article provides an overview of the existing methods for 
testing both the null hypotheses that there is no relationship and that there is a linear 
relationship between the functional covariate and scalar response, and a comprehensive numerical 
comparison of their performance. The methods are compared for a variety of realistic scenarios: 
when the functional covariate is observed at dense or sparse grids and measurements include noise 
or not. Finally, the methods are illustrated on the Tecator data set.\\
\end{abstract}

\begin{keyword}
Functional regression \sep functional linear model \sep nonparametric regression 
\sep mixed-effects model \sep hypothesis testing
\end{keyword}

\end{frontmatter}

%\linenumbers

\section{Introduction}\label{introduction}

\noindent The scalar-on-function regression model refers to the situation where the response 
variable is a scalar, and the predictor variable is functional. Such models are generalizations 
of the usual regression models with a vector-valued covariate, both linear and nonlinear, to 
the case with functional covariates. The functional version of standard linear regression is 
the so-called \textit{functional linear model (FLM)}(see, e.g., \cite{RamsayDalzell1991}); 
various extensions to nonparametric functional regression models have also been developed (see, 
e.g., \cite{FerratyVieu2006}). In this article, we are concerned with hypothesis testing procedures 
in such scalar-on-function regression models. As in standard regression models, one important 
problem is to test whether there is any association between the functional covariate and the 
response, that is, the test for nullity. Also, for nonparametric or nonlinear functional 
regression models, another equally important question is to test linearity of the relationship 
between the functional covariate and the scalar response; this is primarily because of the 
interpretability and ease of fit of the FLM. There is a plethora of literature that develops 
statistical methods for testing nullity and linearity in both linear and nonlinear 
scalar-on-function regression models, respectively. Despite the various available methods, there is no 
clear guideline as to which method provides the best performance in different situations. In 
this article, our goal is to provide an overview of the available testing methods, perform an 
extensive numerical study to compare their size and power performance in various data generation 
models and provide a guideline as to which method yields to the best performance. We will illustrate 
the discussed methods via the Tecator data set.

Much of the literature on testing nullity has been developed under the assumption that there 
is a linear relationship between the functional covariate and the scalar response, that is, 
the functional linear model (FLM). First introduced by \cite{RamsayDalzell1991}, the FLM is 
one of the most commonly used functional regression models due to its interpretability and 
simplicity. It has received extensive attention in recent literature; see 
\cite{RamsaySilverman1997,RamsaySilverman2005}, \cite{Cardoetal1999}, \cite{MullerStadtmuller2005}, 
\cite{FerratyVieu2006}, \cite{CaiHall2006}, \cite{HallHorowitz2007}, \cite{Crambesetal2009}, 
and \cite{Goldsmithetal2011}. The FLM quantifies the effect of the functional covariate as an 
integral of the functional covariate weighted by an unknown coefficient function. The test for 
nullity, in this case, involves testing whether the coefficient function is zero or not. 

%%%Testing for the hypothesis that the functional covariate has no effect has been primarily 
%%%considered under the assumption of linear dependence, FLM. 

\cite{Cardotetal2003} proposed 
two testing methods that are based on the norm of the cross-covariance operator of the 
functional covariate and the scalar response for testing nullity in the FLM. They provided 
asymptotic normality and consistency results of the two proposed test statistics. Later, 
\cite{Cardotetal2004} considered an alternative approach based on a direct approximation of 
the distribution of the cross-covariance operator. Furthermore, they proposed a pseudo-likelihood 
test statistic for the situation when there are multiple functional predictors. Assuming 
an FLM, \cite{Swihartetal2014} proposed likelihood ratio-based test statistics, representing 
the model by using a mixed-effects modeling framework and rewriting the null hypothesis with 
zero-variance components. The major advantage of the mixed-effects model is that software is 
readily available for estimation and hypothesis testing. Recently, 
\cite{Kongetal2016} considered traditional testing methods---Wald, score, likelihood ratio, 
and \textit{F}, to test for no effect in the FLM. They derived the theoretical properties of 
each testing method and compared their performances for both densely and sparsely sampled 
functions.

The main disadvantage of the FLM is the assumption of linearity of the relationship between 
the functional covariate and the scalar response. Such a linear relationship may not be 
practical in many situations, and as a result, there is a substantial amount of literature 
on the development of nonlinear/nonparametric functional regression models. \cite{YaoMuller2010} 
considered a quadratic regression model as an extension of the FLM by including quadratic 
effects of the functional covariate. \cite{Garciaetal2014} considered testing in a nonparametric 
functional regression model, where the effect of the functional covariate was modeled via 
an unknown functional. \cite{McLeanetal2014} developed the so-called 
\textit{functional generalized additive model (FGAM)}, where the effect of the functional 
covariate is modeled using an integral of a bivariate smooth function involving the functional 
covariate at a specific time point and the time point itself. For testing nullity, 
\cite{Garciaetal2014} introduced the \textit{projected Cram\'{e}r-von Mises (PcVM)} test---a 
testing method which is derived by using random projection, and whose null distribution is 
approximated by bootstrap. \cite{McLeanetal2015} introduced a \textit{restricted likelihood ratio 
test (RLRT)} statistic for testing no effect under the assumption that the response and the 
predictor are related through an FGAM. The key idea is to use the mixed model formulation of 
the smooth effects and represent the null hypothesis as the test for a subset of variance 
components. 

Other than testing for nullity, another important problem is to test for linearity of the 
regression function. Motivated by the idea of a polynomial functional relationship (e.g., 
quadratic functional regression by \cite{YaoMuller2010}), \cite{HorvathReeder2013} developed 
a testing method by using functional principal component scores to test the null effect of 
the quadratic term and studied its asymptotic properties. The testing methods of 
\cite{Garciaetal2014} and \cite{McLeanetal2015} can also be used to investigate the problem 
of testing for the linear effect. 

In this article, our goal is to numerically compare the performance of all the existing methods 
for testing nullity as well as linearity of a functional covariate when the response is scalar 
in a variety of scenarios related to how the functional covariate is observed. The results are 
illustrated for varying sample sizes and situations of increasing complexity regarding the 
functional covariate. Additionally, we apply the methods to a commonly used data set, the Tecator 
data, to formally assess the relationship between the meat's spectrum of absorbances and the fat 
content, using 215 finely chopped pure meat samples.  

The article makes two key contributions to the literature. First, we study each of these methods 
under a wide variety of scenarios in which the functional predictor is observed either on a 
dense or sparse grid of points for each subject, with or without measurement error. Much of 
the previous work relies only on the assumption of densely observed functional predictors. 
Second, we provide a comprehensive comparison study of the existing approaches for testing 
nullity and linearity of scalar-on-function regression.

The remainder of the article is organized as follows. In Section 2, we introduce the data structure 
and model framework for scalar-on-function regression. In Section 3, we review each of the methods 
under comparison. Section 4 discusses the advantages and drawbacks of each method in greater detail. 
Simulation studies and the real data application follow in Section 5 and 6, respectively.

% % % % % % % % % % % % % % % % % % % % % % % % % % % % % % % % % %
% % % % % % % % % % % % % % % % % % % % % % % % % % % % % % % % % %

\section{Model Framework}\label{modelframework}

\noindent Suppose that for subject $i \in \{1,\ldots,n\}$ we observe data of the form 
$\{Y_i, (X_{ij},t_{ij}) : j=1,\ldots,J\}$, where $Y_i$ is a scalar response variable and 
$X_{ij}=X_i(t_{ij})$ are discrete realizations of a real-valued, square-integrable smooth 
curve $X_i (\cdot)$ at observation points $t_{ij}$. For simplicity of exposition, we assume 
that the full predictor trajectory $X_i(\cdot)$ is observed; however, the methods are 
investigated for the case when the true predictor is observed on a finite grid of points 
and corrupted with measurement error. Without loss of generality, we assume that the 
functional covariate is a zero-mean process. A scalar-response functional model can be defined as
\begin{equation}\label{eq1}
Y_i=  \alpha + m\{X_i(\cdot)\} + \varepsilon_i,
\end{equation}
\noindent where $m(\cdot)$ is an unknown functional and $\varepsilon_i$ are independent and 
identically distributed random errors with mean zero and variance $\sigma^2$. According to 
\cite{FerratyVieu2006}, $m(\cdot)$ can be classified as parametric and nonparametric, depending 
on the specific mean model at hand. An example of a functional parametric mean model is the 
functional linear model (FLM) where $m\{X_i(\cdot)\}=\int X_i(t)\beta(t)dt$ for some unknown 
continuous function $\beta(\cdot)$. In contrast, a functional nonparametric mean model assumes 
that the object $m(\cdot)$ is a continuous real-valued operator defined on a Hilbert space 
$\mathbb{H}$. In this article, we are interested in testing two important hypotheses about the 
mean structure: 
(i) $H_{01}: m\{X(\cdot)\} = \int X(t)\beta(t)dt$, the relationship between the covariate 
$X(\cdot)$ and the response $Y$ is linear, and 
(ii) $H_{02}: m\{X(\cdot)\} =  0 \; \text{for any} \; t \in \mathcal{T}$, there is no relationship 
between $X(\cdot)$ and $Y$.

The main focus of this article is to numerically compare the performance of the existing methods for 
testing $H_{01}$ and $H_{02}$ in a variety of realistic scenarios. For testing $H_{01}$, we study 
the nonparametric testing method of \cite{Garciaetal2014} (which we call GGF using first letters 
of the authors' names), the semi-parametric method of \cite{McLeanetal2015} (which we call MHR), 
and the parametric method of \cite{HorvathReeder2013} (which we call HR). For testing $H_{02}$, 
we study the GGF and MHR methods, and also the parametric method of \cite{Kongetal2016} 
(referred by KSM), which assumes a linear relationship between the response and the predictor. 
We assess the performance of the methods in the cases when the functional covariate is observed 
on a dense, moderately sparse, or sparse grid, with and without measurement error, using different 
sample sizes. This article offers a comprehensive comparison study of available approaches in the 
literature for testing nullity and linearity in scalar-on-function regression.

% % % % % % % % % % % % % % % % % % % % % % % % % % % % % % % %
% % % % % % % % % % % % % % % % % % % % % % % % % % % % % % % %

\section{Hypothesis Testing}\label{testing}

\noindent In this section, we review each of the methods under study. All the methods rely on the 
idea of using basis expansion to approximate the functional linear model by a simple mixed-effects 
model. The various methods use different test statistics and corresponding null distributions, and 
they have been developed to assess the null hypothesis in specific settings. First, we consider the 
problem of testing linearity. The GGF method considers this problem in the class of nonlinear models, 
which is the most general case considered in the literature. In contrast, MHR and HR consider this 
problem in a more restrictive class of models: MHR assumes a functional generalized additive model 
(FGAM), and HR assumes a functional quadratic model. For testing nullity, GGF assumes a general 
non-null relationship, MHR assumes an FGAM relationship, and KSM assumes a linear dependence. These 
assumptions are reflected in the form of the alternative hypothesis. The HR and KSM methods are 
parametric methods that require stronger assumptions in order to develop their corresponding null 
distributions.

\subsection{Testing for the Linear Effect of the Functional Covariate} \label{linearity}

\subsubsection{GGF Method for Testing Linearity $H_{01}$} \label{linearity_GGF}

\noindent The GGF method \citep{Garciaetal2014} for testing linearity is essentially a generalization 
of a goodness-of-fit test in regression models for scalar responses and vector covariates to the case 
when the covariate is functional. The interest is to test the null hypothesis $H_{01}$, which indicates 
that $m(\cdot)$ belongs to the family $\mathcal{M}= \{\langle \cdot, \beta \rangle: \beta \in \mathbb{H}=L^2[0,1] \}$ 
versus a general alternative of the form $H_{A1}: m \notin \{ \langle \cdot, \beta \rangle: \beta \in \mathbb{H}\}$ 
with positive probability. In other words, the alternative hypothesis can also be written as
\begin{equation}\label{eq2}
H_{A1}: E(Y)= m\{X(\cdot)\},
\end{equation} 
where $m(\cdot)$ is an unknown functional, while the null hypothesis is that 
$m\{X(\cdot)\}= \langle X(\cdot),\beta \rangle$. 

The key idea is to characterize the linear relationship in an equivalent way that is based on random 
projection. Specifically, \cite{Garciaetal2014} show that $ m\{X(\cdot)\}= \langle X(\cdot), \beta \rangle$ 
for $\beta$ an element in $L^2[0,1]$ is equivalent to 
\begin{equation}\label{eq3}
E[(Y-\langle X(\cdot),\beta \rangle) \mathds{1}_{\{\langle X(\cdot),\gamma \rangle \leq u\}}] = 0,
\end{equation}
almost everywhere for any $u\in \mathbb{R}$ and for all $\gamma \in \mathbb{S}_{\mathbb{H}}^p, \forall p \geq 1$, 
where $\mathbb{S}^p_{\mathbb{H}} = \{f= \sum_{j=1}^p x_j\Psi_j \in \mathbb{H}:||f||_{\mathbb{H}} = 1 \}$ and 
$f:[0,1] \rightarrow \mathbb{R}$ such that their norm $||f||_{\mathbb{H}}=(\int_{0}^{1} |f(t)|^2 dt)^{1/2}$, 
and $\{\Psi_1(\cdot), \Psi_2(\cdot), \ldots \Psi_p(\cdot)\}$ are orthogonal bases in $L^2[0,1]$. For more 
information, see Lemma 3.1 in \cite{Garciaetal2014}. The latter formulation essentially implies that the 
mean of the departure from a linear relationship---that is, concentrated on arbitrarily small neighborhoods, 
is zero. Thus, one approach to testing for linearity is to quantify the mean on the left-hand side of (\ref{eq3}) 
and assess how different it is from zero. 

The GGF method proposes to do this by first estimating $\beta(\cdot)$ by the best linear estimator, 
$\hat{\beta}(\cdot)$, using known basis functions to expand both the functional covariate and the 
coefficients, and then rewriting the functional linear model as a standard linear model, as proposed 
by \cite{Cardoetal1999}. The residuals under the null hypothesis are 
$\hat{\varepsilon_i}=Y_i-\hat{Y}_i = Y_i - \int X_i(t) \hat{\beta}(t) dt \; \text{for} \; i=1,\ldots,n$. 
Once the residuals are estimated, a projected Cram\'{e}r-von Mises (PcVM) test statistic with a plug-in 
estimator is used. Specifically, for fixed $u$ and $\gamma \in \mathbb{S}_{\mathbb{H}}$, a method of 
moment estimator of (\ref{eq3}) is $n^{-1/2} \hat{R}_n(u,\gamma)$, where
\begin{equation}\label{eq4}
\hat{R}_n(u,\gamma) = n^{-1/2} \sum_{i=1}^n \hat{\varepsilon_i} \mathds{1}_{ \int X_i(t)\gamma(t)dt \leq u}.
\end{equation} 
\noindent The PcVM test statistic is adapted to the projected space $\Pi=\mathbb{R}\times \mathbb{S}_{L^2[0,1]}$ 
and defined as
\begin{equation}\label{eq5}
\text{PCvM}_{n,p} = \int_{\Pi} \hat{R}_n(u,\gamma)^2 F_{n,\gamma}(du) \omega(d\gamma), 
\end{equation}
\noindent where $F_{n,\gamma}$ is the empirical cumulative distribution function of the data 
$\{\langle X_i(\cdot), \gamma \rangle\}_{i=1}^n$ and $\omega$ is a measure on $\mathbb{S}_{\mathbb{H}}$.

This expression is certainly complicated, and its derivation has numerous cumbersome steps. However, 
\cite{Garciaetal2014} show that in practice $\mbox{PCvM}_{n,p} $ is approximated by 
$n^{-2}\hat{\varepsilon}^T \boldsymbol{A}\hat{\varepsilon}$, where $\boldsymbol{A}=(\sum_{r=1}^n A_{ijr})_{ij}$ 
is an $n\times n$ matrix of the average over $i$ and $j$ of the three-dimensional array $A_{ijr}$. 
The array represents the product surface area of a spherical wedge of angle $A_{ijr}^{(0)}$ times the 
determinant of the matrix $R^{-1}$ (from the Cholesky decomposition of the basis functions). For details 
concerning the matrix $\boldsymbol{A}$ and the derivation of this approximation, we refer the reader to 
\cite{Garciaetal2014}. The null distribution of the test statistic is nonstandard and is approximated by 
a wild bootstrap on the residuals.

\subsubsection{MHR Method for Testing Linearity $H_{01}$}\label{linearity_MHR}

\noindent The MHR method \citep{McLeanetal2015} considers testing in the class of models called functional 
generalized additive model (FGAM), for which the response and covariate relate according to the following 
relationship:
\begin{equation}\label{eq6}
Y=\alpha + \int_\Tau F\{X(t),t\}dt + \varepsilon;
\end{equation}
\noindent we use the generic notation $\{Y,X(\cdot),\varepsilon\}$ respectively for the response, functional 
covariate, and Gaussian random error with zero mean and variance $\sigma^2$, and $F(\cdot,\cdot)$ represents 
an unknown bivariate function. It can be clearly seen that FGAM reduces to the FLM when $F(x,t)=x\beta(t)$; 
thus FLM is a special class of FGAM. Testing the hypothesis of interest in this class is equivalent to 
representing the alternative hypothesis as $H_{A1}: E(Y) = \alpha + \int_\Tau F\{X(t),t\}dt$. The key idea 
behind the test is to use the connection between the tensor product splines and mixed-effects modeling 
\citep{Woodetal2013} and to formulate the FGAM as a mixed model representation with two main parts: a 
component represented by unpenalized, fixed effects and a component represented by random effects.

Using low-rank spline bases, denoted as $\{B_j^X(x) : j=1,\ldots,K_x\}$ and $\{B^T_k(t) : k=1,\ldots,K_t\}$, 
the bivariate surface can be expanded as
\begin{equation}\label{eq8}
F(x,t) = \sum_{j=1}^{K_X} \sum_{k=1}^{K_t} B_j^X (x) B_k^T (t) \theta_{jk},
\end{equation}
\noindent where the $\theta_{jk}$ are unknown tensor product B-spline coefficients. Let $\mathbb{B}_x$ 
denote the $nJ \times K_x$ matrix of the $x$-axis B-splines that are evaluated at 
$\boldsymbol{x}=\mbox{vec}(\boldsymbol{X})$, where $\boldsymbol{X}=\{X_i(t_{im})\}$ is the $n \times J$ 
matrix whose rows include the observed functional predictor values for each subject. Similarly, let 
$\mathbb{B}_t$ denote the $nJ \times K_t$ matrix of the $t$-axis B-splines that are evaluated at 
$\boldsymbol{t}=\mbox{vec}(\boldsymbol{T})$, where $\boldsymbol{T}=\{t_{im}\}$ is the $n \times J$ matrix 
in which each row includes the observed time points for the functional predictor for each subject. The 
matrices $\mathbb{X}_x$, $\mathbb{Z}_x$, $\mathbb{X}_t$, and $\mathbb{Z}_t$ are derived from the 
eigendecompositions of marginal penalty matrices $\mathbb{P}_x$ and $\mathbb{P}_t$. After some mathematical 
manipulations, we can define the fixed-effects design matrix $\mathbb{X}=[\boldsymbol{1}:\boldsymbol{x}:
\boldsymbol{x} \otimes \boldsymbol{t}]$, and the random-effects design matrices $\mathbb{Z}_1=\boldsymbol{x}
\Box\mathbb{Z}_t$, $\mathbb{Z}_2=\mathbb{Z}_x\Box\mathbb{X}_t$, and $\mathbb{Z}_3=\mathbb{Z}_x\Box\mathbb{Z}_t$, 
where $\otimes$ denotes the Kronecker product and $\Box$ represents the box product (also known as the row-wise 
Kronecker product). Then, FGAM can be expressed in the form of a mixed-effects model with three pairwise 
independent vectors of random effects, each with a diagonal covariance matrix independent of the other effects:
\begin{equation}\label{eq9}
\textbf{Y} \approx \mathbb{LX}\boldsymbol{\beta} + \sum_{j=1}^3 \mathbb{LZ}_j\boldsymbol{b}_j+\boldsymbol{\varepsilon},
\end{equation}
where $\mathbb{L}$ is an $n \times nJ$ matrix of quadrature weights; 
$\boldsymbol{b}_j \sim N(\boldsymbol{0},\sigma^2_j\mathbb{I}_{q_j})$ with the dimensions $q_1=K_t-2$, $q_2=2(K_x-2)$, 
and $q_3=(K_x-2)(K_t-2)$; and $\boldsymbol{\varepsilon} \sim N(\boldsymbol{0}, \sigma^2_e \mathbb{I}_N)$. The matrix 
$\mathbb{X}$ forms a basis for functions of the form $\beta_0+\beta_1x + \beta_3xt$ without penalty, $\mathbb{Z}_1$ 
forms a basis for functions of the form $xf_2(t)$ and penalty $\int (\partial_{tt} f_2)^2$, $\mathbb{Z}_2$ forms a 
basis for functions of the form $g_1(x) + tg_2(x)$ and penalty $\int (\partial_{xx} g_1)^2 + \int (\partial_{xx} g_2)^2$, 
and $\mathbb{Z}_3$ forms a basis for functions of the form $h(x,t)$ without the previous terms and with penalty 
$\int (\partial_{xxtt} h)^2$. In addition, it can be shown that the FLM is nested within the FGAM in an explicit 
way, which allows the use of restricted likelihood ratio tests for zero-variance components to test the null 
hypothesis that the functional linear model holds, $H_{01}: \sigma_2^2=\sigma_3^2=0$. The testing is done via 
the restricted likelihood ratio test (RLRT) under the assumption that $\sigma_2 = \sigma_3$:
\[
\text{RLRT} = 2 \sup_{H_1} \ell_R(\boldsymbol{y}) - 2 \sup_{H_0} \ell_R(\boldsymbol{y}),
\]
where $\ell_R$ denotes the restricted log-likelihood function of the observed data vector $\boldsymbol{y}$ for 
model (\ref{eq9}). \cite{CrainiceanuRuppert2004} derive the finite-sample null distribution of the RLRT statistic 
and show that the distribution is different from the mixture of $\chi^2$ distributions.

\subsubsection{HR Method for Testing Linearity $H_{01}$}\label{linearity_HR}

\noindent The HR method \citep{HorvathReeder2013} considers the same problem and proposes a method based on 
projecting the predictor process onto a space of finite dimension by using the functional principal component 
analysis (FPCA). This approach assumes a functional quadratic regression model
\begin{equation}\label{eq10}
Y= \alpha + \int X(t)\beta(t)dt + \iint X(t) X(s)\gamma(s,t) dtds  + \varepsilon,
\end{equation}
\noindent where $\beta(t)$ and $\gamma(s,t)$ are unknown smooth univariate and bivariate functions, 
respectively. Notice that when $\gamma(s,t)=0$, model (\ref{eq10}) reduces to the FLM; equivalently the 
FLM is a subclass of model (\ref{eq10}). \cite{HorvathReeder2013} focus on testing the significance of 
the quadratic term in model (\ref{eq10}); that is, they focus on the null hypothesis $H_{01}: \gamma(s,t)=0$ 
versus $H_{A1}: \gamma(s,t) \neq 0$.

The regression coefficient functions are expanded using the eigenfunctions of the covariance function of 
the predictor $C(t,s)=E\{X_i(t)-\mu_x(t)\}\{X_i(s)-\mu_X(s)\}$ to represent them as $\beta(t)=\sum_{j=1}^p b_j v_j(t)$ 
and $\gamma(s,t) = \sum_{j=1}^p \sum_{k=1}^p a_{jk} v_k(s)v_j(t)$, where $v_j(t)$ denote the eigenfunctions 
of $C(t,s)$. By projecting the observations onto the space spanned by $\{v_j(t)\}_{j=1}^p$ and using the 
expansions given above, we can rewrite model (\ref{eq10}) as
\begin{equation}\label{eq11}
Y_i= \alpha + \sum_{j=1}^p b_j \langle X_i, v_j \rangle + \sum_{j=1}^p \sum_{k=1}^p 
\{2-\mathds{1}(j=k)\}a_{jk} \langle X_i, v_j \rangle \langle X_i, v_k \rangle + \varepsilon_i^*,
\end{equation} 
where $a_{jk}$ and $b_j$ are the coefficients, and 
\begin{equation*}
\begin{aligned}
\varepsilon_i^* & =\varepsilon_i+\sum_{j=p+1}^\infty b_j \langle X_i, v_j \rangle + \sum_{j=p+1}^\infty \sum_{k=j}^\infty 
\{2-\mathds{1}(j=k)\}a_{jk} \langle X_i, v_j \rangle \langle X_i, v_k \rangle \\ & + \sum_{j=1}^{p} \sum_{k=p+1}^\infty 
2a_{jk} \langle X_i, v_j \rangle \langle X_i, v_k \rangle.
\end{aligned}
\end{equation*}
Because the eigenfunctions and the mean process of the functional covariate are unknown, 
model (\ref{eq11}) is not adequate to make statistical inference. Substituting the estimates into 
(\ref{eq11}) results in
\begin{equation}\label{eq12}
Y_i= \alpha + \sum_{j=1}^p b_j \langle X_i- \bar{X}, \hat{c}_j\hat{v}_j \rangle + \sum_{j=1}^p \sum_{k=1}^p 
\{2-\mathds{1}(j=k)\}a_{jk} \langle X_i- \bar{X}, \hat{c}_j\hat{v}_j\rangle \langle X_i- \bar{X}, 
\hat{c}_k\hat{v}_k \rangle + \varepsilon_i^{**},
\end{equation}  
where $\bar{X}(t) = \frac{1}{n} \sum_{i=1}^n X_i(t)$, $\hat{v}_j(t)$ is the $j$th 
estimated eigenfunction of $\hat{C}(t,s)$, the $\hat{c_j}$ are random signs, 
and 
\begin{equation*}
\begin{aligned}
\varepsilon_i^{**} & =\varepsilon_i^*+\sum_{j=1}^{p} b_j \langle X_i, v_j-\hat{c}_j\hat{v}_j \rangle + 
\sum_{j=1}^{p} b_j \langle \bar{X}-\mu_X, \hat{c}_j\hat{v}_j \rangle \\& - \sum_{j=1}^{p} \sum_{k=j}^{p} 
\{2-\mathds{1}(j=k)\}a_{jk} (\langle X_i-\bar{X}, \hat{c}_j\hat{v}_j \rangle \langle X_i-\bar{X}, 
\hat{c}_k\hat{v}_k \rangle - \langle X_i, v_j \rangle \langle X_i, v_k \rangle).
\end{aligned}
\end{equation*}
The model can be rewritten as
\[
\textbf{Y} = \boldsymbol{\hat{Z}} \begin{pmatrix} \boldsymbol{\tilde{A}} \\ \boldsymbol{\tilde{B}} \\ 
\mu \end{pmatrix}+ \boldsymbol{\varepsilon^{**}} \quad \text{with} \quad \boldsymbol{\hat{Z}}= 
\begin{pmatrix}  \boldsymbol{\hat{D}}^T_1 & \boldsymbol{\hat{F}}^T_1 & 1 \\ 
\boldsymbol{\hat{D}}^T_2 & \boldsymbol{\hat{F}}^T_2 & 1 \\ 
\vdots & \vdots & \vdots \\ 
\boldsymbol{\hat{D}}^T_n & \boldsymbol{\hat{F}}^T_n & 1 \end{pmatrix}, \]                                                                         
where $\boldsymbol{Y} = (Y_1, Y_2,\ldots, Y_n)^T \in \mathbb{R}^n$; 
$\boldsymbol{\tilde{A}}=\mbox{vech}(\{\hat{c}_j \hat{c}_k a_{jk}\{2-\mathds{1}(j=k)\},1\leq j \leq k \leq p\}^T)
\in \mathbb{R}^{p(p+1)/2}$, where vech($\cdot$) denotes the half-vectorization (vectorization of the lower 
triangular portion of the matrix); $\boldsymbol{\tilde{B}}=(\hat{c}_1b_1,\ldots,\hat{c}_pb_p) \in \mathbb{R}^p$; 
and $\boldsymbol{\varepsilon^{**}}=(\varepsilon^{**}_1,\ldots,\varepsilon^{**}_n) \in \mathbb{R}^n$. 
$\boldsymbol{\hat{D}}_i^T$ is the half vectorization of the matrix constructed as a cross-product of each of the 
eigenfunctions $\hat{v}_j$ and the centered predictor $X_i$. $\boldsymbol{\hat{F}}_i^T$ is a vector constructed 
as $(\langle X_i-\bar{X},\hat{v}_1 \rangle,\ldots, \langle X_i-\bar{X},\hat{v}_p \rangle)$. $\boldsymbol{\tilde{A}}$, 
$\boldsymbol{\tilde{B}}$ and $\mu$ are estimated using the least squares estimator 
$(\boldsymbol{\hat{Z}}^T\boldsymbol{\hat{Z}})^{-1}\boldsymbol{\hat{Z} }^T\boldsymbol{Y}$.

\cite{HorvathReeder2013} construct their test by using summary quantities of $\boldsymbol{\hat{D}}^T$ and the 
sum of squared $\varepsilon_i^{**}$:
\[ 
U_n= \frac{n}{\tau^2} \boldsymbol{\hat{A}}^T(\boldsymbol{\hat{G}} - 
\boldsymbol{\hat{M}}\boldsymbol{\hat{M}}^T)\boldsymbol{\hat{A}},
\] 
where $\boldsymbol{\hat{G}}=\frac{1}{n} \sum_{i=1}^n\boldsymbol{\hat{D}}_i\boldsymbol{\hat{D}}_i^T$, 
$\boldsymbol{\hat{M}}=\frac{1}{n}\sum_{i=1}^n\boldsymbol{\hat{D}}_i$, and 
$\tau^2=\frac{1}{n}\sum_{i=1}^{n}\hat{\varepsilon}_i^2$. They show the null distribution of $U_N$ is a 
$\chi^2(r)$ with $r= p(p+1)/2$ degrees of freedom. $U_n$ measures the distance between $\boldsymbol{\hat{G}}$ 
and $\boldsymbol{\hat{M}}\boldsymbol{\hat{M}}^T$, scaled using the sample size, the estimated coefficients 
$\boldsymbol{\hat{A}}$ and the residuals $\tau^2$. The difference between $\boldsymbol{\hat{G}}$ and 
$\boldsymbol{\hat{M}}\boldsymbol{\hat{M}}^T$ corresponds to the interaction between different elements of 
$X(t)$, which represents the quadratic term. If the difference is too big, then there is evidence of a 
quadratic relationship. 

\subsection{Testing for the Null Effect of the Functional Covariate}\label{nullity}

\subsubsection{GGF Method for Testing Nullity $H_{02}$}\label{nullity_GGF}

\noindent Testing for the null effect of the functional covariate can be viewed as a special case of testing 
for the linear effect. \cite{Garciaetal2014} focus on testing for a specific functional linear model, 
$m\{X(\cdot)\}=\langle X,\beta_0 \rangle$, for a specified smooth function $\beta_0 \in \mathbb{H}$. When 
$\beta_0(t)=0$ for $t \in [0,1]$, an equivalent to the null hypothesis is $H_{02}: m\{X(\cdot)\}=0$ 
(versus $H_{A2}: m\{X(\cdot)\} \neq  0$).

By making minor modifications according to the choice of the null hypothesis, the GGF method uses the same 
procedure that is described in Section \ref{linearity_GGF} to compute the test statistic with the residuals 
under the null hypothesis, $\hat{\varepsilon}_i=\hat{Y}_i, \; i=1,\ldots,n$. The null distribution of the 
test statistic is again approximated by using a wild bootstrap sampling procedure on the residuals.

\cite{Garciaetal2014} compare the finite sample properties of the PcVM test statistic with two other 
competing methods proposed by \cite{Delsoletal2011} and \cite{Gonzalezetal2012}. Based on the numerical 
comparison, the PcVM test statistic is found to be the most powerful among these methods. Thus we focus on 
the PcVM (denoted by GGF) in this article.

\subsubsection{MHR Method for Testing Nullity $H_{02}$}\label{nullity_MHR}	

\noindent \cite{McLeanetal2015} also consider testing whether the functional covariate has any effect on 
the scalar response (that is, $H_{02}: \beta(t)=0$ for $t \in [0,1]$), where the alternative model is 
specified as the FLM, $H_{A2}: E(Y)= \alpha + \int_\Tau X(t)\beta(t)dt$. The MHR method tests for no effect 
by rewriting model (\ref{eq9}) without the random effects $\boldsymbol{b}_2$ and $\boldsymbol{b}_3$. Thus 
the null hypothesis is equivalent to testing $H_{02}: \beta_2=\beta_3=0, \; \sigma_1=0$ against the alternative 
hypothesis $H_{A2}: \beta_2 \neq 0 \; \mbox{or} \; \beta_3 \neq 0 \; \mbox{or} \; \sigma_1 > 0$. The 
likelihood ratio test (LRT) is more appropriate for this case, because the RLRT cannot be used for testing 
the fixed effects $\beta_2$ and $\beta_3$. The LRT statistic is:
\[
\text{LRT} = 2 \sup_{H_0 \cup H_1} \ell(\boldsymbol{y}) - 2 \sup_{H_0} \ell(\boldsymbol{y}),
\]
where $\ell$ denotes the log-likelihood function of the observed data vector $\boldsymbol{y}$ for the 
corresponding mixed-effects model. The exact null distribution for the LRT statistic is not a standard 
$\chi^2$ distribution, because the null value of the variance component is on the boundary of the parameter 
space. \cite{CrainiceanuRuppert2004} derive the finite-sample null distribution of the LRT statistic in detail.

\subsubsection{KSM Method for Testing Nullity $H_{02}$}\label{nullity_KSM}

\noindent The KSM method \citep{Kongetal2016} is an extension of classical testing methods in linear regression 
to functional linear regression with a scalar response and a functional covariate. \cite{Kongetal2016} 
are interested in testing for the null hypothesis given in Section \ref{nullity_MHR} against the alternative 
hypothesis $H_{A2}: \beta(t)\neq0$, $t \in [0,1]$, which yields the alternative model of the form, 
$H_{A2}: E(Y)= \alpha + \int_\Tau X(t)\beta(t)dt$.

This method uses a spectral decomposition of the covariance function to re-express the functional linear 
model as a standard linear model, where the effect of the functional covariate can be approximated by a 
finite linear combination of the functional principal component scores
\begin{equation}\label{eq13}
Y_i= \alpha + \sum_{j=1}^{s_n} \xi_{ij}\beta_j + \varepsilon_i,
\end{equation}
\noindent where $s_n$ is the number of principal components, $\{\xi_{ij}:i=1,\ldots,n\}$ are the functional 
principal component scores uncorrelated over $j$ with mean zero and variance decreasing with $j$, and 
$\beta_j$ denote the unknown basis coefficients in the expansion $\beta(t)=\sum_{j=1}^{s_n} \beta_j \phi_j(t)$. 
The functions $\{\phi_j(t)\}^{s_n}_{j=1}$ denote the eigenfunctions obtained from the spectral decomposition 
of the covariance operator of the functional predictor.

Testing the null hypothesis $H_{02}$ in Section \ref{modelframework} is equivalent to testing 
$H_{02}: \beta_1=\beta_2=\ldots=\beta_{s_n} = 0$ against the alternative hypothesis $H_{A2}: \beta_j \neq 0$ for 
at least one $j, \; i\leq j \leq  s_n$. The $F$ test is defined as 
\begin{equation}\label{eq14}
T_F = \frac{Y^T(P_1-P_B)Y/s_n}{Y^T(I_{n \times n}-P_B)Y/(n-s_n-1)} ,
\end{equation}
\noindent where $P_B$ and $P_1$ are the projection matrices under the alternative and the null models, 
respectively. Note that we need to fit both the alternative and null models in order to calculate the 
test statistic. \cite{Kongetal2016} show that the null distribution of $T_F$ behaves like $\chi^2_{s_n}$, 
which enables us to compute $p$-values by using $\chi^2$ quantiles.

\cite{Kongetal2016} theoretically and numerically investigate the finite sample performance of four 
tests---Wald, score, likelihood ratio, and $F$. Their study in finite sampling shows that the $F$ test 
provides reasonable Type I error rates and power values compared the other valid testing methods, and thus 
indicates that it is a robust testing method. On the basis of their results, we use only the $F$ test for 
testing the null effect of the functional covariate.

\section{General Discussion}\label{discussion}

\noindent The MHR method, proposed by \cite{McLeanetal2015}, considers an RLRT for testing the null 
hypothesis that the FLM is the true model versus the FGAM alternative. The main idea behind the MHR method 
is to represent the FGAM as a standard linear mixed model by taking advantage of the link between the 
mixed-effects model and penalized splines \citep{Woodetal2013}. This representation allows us to reduce 
the dimensionality of the testing problem and formulate the null hypothesis of an unknown function as 
a set of zero-variance components. The MHR method is computationally efficient because the finite sample 
null distribution of the RLRT statistic can be obtained very quickly by using a fast simulation algorithm.
The MHR method assumes that the functional predictor is observed at a dense and regular grid of points, 
without measurement error. This method can be modified in a straightforward way when there is more than 
one functional predictor in addition to the response from any exponential family distribution 
\citep{McLeanetal2014}.

The GGF method, introduced by \cite{Garciaetal2014}, considers a projected Cram\'{e}r-von Mises test 
statistic. The asymptotic null distribution of the test statistic is approximated by a wild bootstrap 
on the residuals. One advantage of the method is that it can be easily extended to any other 
scalar-on-function regression model, because the test statistic and its null distribution are obtained 
depending on the residuals under the null model. Unlike other methods in the study, the GGF method specifies 
a more general alternative model and provides a greater flexibility; as expected, this generality leads to 
a loss of power relative to competitors when simpler alternatives are true. Another drawback of the method 
is the fact that bootstrapping the null distribution of the test statistic is computationally intensive. 
We discuss these drawbacks in the context of our simulation study. The GGF method also makes the assumption 
that the functional predictor has a dense sampling design and is observed without measurement error. 

The HR method \citep{HorvathReeder2013} relaxes the restrictive assumption of a linear relationship 
between a scalar response and a functional predictor under the alternative by considering a functional 
quadratic regression model. The HR method is developed to test for linearity in a class of parametric 
scalar-on-function regression models. \cite{HorvathReeder2013} showed that HR provides good Type I error 
rates and power results when the sample size is greater than $n=200$ and the functional predictor is densely 
observed without measurement error. However, the question of whether the HR method still performs well when 
the sample size is small and the functional predictor is observed on irregular and/or sparse grids was not 
addressed by the authors. As we will see in our simulation study, the Type I error rate of the HR method is 
considerably inflated for small and moderate sample sizes, as well as for a sparsely observed functional 
predictor.

The KSM method \citep{Kongetal2016} extends the classical $F$ test from multiple linear regression to 
functional linear regression. This method uses the eigenbasis functions that are derived from the FPCA to 
reduce dimensionality and re-writes the FLM as a standard linear model. In contrast to the aforementioned 
methods, KSM is applicable to sparsely observed functional predictors that are corrupted with measurement 
error. \cite{Kongetal2016} indicate that the KSM method is a robust testing method that maintains the 
correct nominal level in various scenarios including small sample sizes and noisy and sparsely measured 
predictor trajectories. The power performance of the method relies mainly on the choice of the number of 
functional principal components (FPC). Choosing a large number of FPCs may cause a decrease in power 
\citep{Kongetal2016}. This problem has been considered recently by \cite{Suetal2017}.

% % % % % % % % % % % % % % % % % % % % % % % % % % % % % % % % % %
% % % % % % % % % % % % % % % % % % % % % % % % % % % % % % % % % %

\section{Numerical Investigation of Testing Methods}\label{simulation}

\noindent We conduct a simulation study to compare the finite sample performance of each testing method. 
In an effort to respect the simulation settings used by the original tests' proponents, we carry out two 
sets of simulations: one for testing the linear effect of the functional covariate and the other for 
testing the null effect. Each data set is generated under dense, moderately sparse, and sparse designs, 
and the number of units per subject is defined respecting their data generation settings. To investigate 
how the methods perform in moderately sparse and sparse designs, we randomly sample $m_i$ observation points 
per curve without replacement from the discrete uniform distributions $\mbox{Unif}(15,20)$ and 
$\mbox{Unif}(5,10)$ for both M1 and Y1 settings, $\mbox{Unif}(100,120)$ and $\mbox{Unif}(25,30)$ for both G1 
and G2 settings, and $\mbox{Unif}(50,60)$ and $\mbox{Unif}(15,20)$ for the H1 setting. After generating the 
designs, we carried out functional principal component analysis with 99\% of the total variance explained to 
impute functional data that were sparsely observed.

Assuming that the settings of each method will highlight the characteristics of its respective test, we use 
all the testing methods with each of the data sets that are generated. We assess the size and power of the 
tests for sample sizes that vary from $n=50$ to $n=500$; the results are based on 5,000 simulations for size 
assessment and 1,000 simulations for power assessment. 

\subsection{Simulation Designs for Testing No Covariate Effect}\label{simul_nullity}

\noindent For the no-effect null hypothesis, the model used to generate the data under the alternative 
is different in each scenario, but they have in common the use of $\delta$ to control the departure from 
the model without the covariate effect. For all settings, $\delta = 0$ corresponds to the null hypothesis 
of no effect and $\delta \textgreater 0$ corresponds to the alternative hypothesis of a non-null effect. 
The scenarios {\textbf{(G0, M0)}} are as follows:

\begin{itemize}[leftmargin=*]
	\item {\textbf{Setting 1 (G0)}}. The functional process for the functional 
	covariate $X$ in this case is a Brownian motion with functional mean 
	$\mu(t)= 0 \; \text{for all} \; t \in [0,1]$ 
	and $Cov(X(s),X(t)) = \frac{\sigma^2}{2\theta}e^{\{-\theta(s+t)\}}(e^{\{2\theta \min(s,t)\}}-1)$, 
	with $\theta = 1/3$ and $\sigma=1$. We use $201$ equidistant points in the interval $[0,1]$. 
	The model that generates the data is
	\begin{equation}\label{eq15}
	Y_i= \delta  \int X_i(t)\beta(t)dt + \varepsilon_i, 
	\end{equation}
	where $\delta \in \{0.02,\ldots,0.9\}$, $\beta(t) = \sin(2\pi t)-\cos(2\pi t)$, 
	and $\varepsilon_i \sim N(0,\sigma^2_{\varepsilon}=0.01)$ \citep{Garciaetal2014}.
	
	\item {\textbf{Setting 2 (M0)}}. In this setting the functional covariate is generated as  
	$X(t)=\sum_{j=1}^4 \xi_j\phi_j(t)$, with $\xi_j \sim N(0,8j^{-2})$ and 
	$\{\phi_1(t),\ldots, \phi_4(t)\} = \{ \sin(\pi t), \\\cos(\pi t), \sin(2\pi t), \cos(2 \pi t) \}$. 
	We use $30$ equidistant points in the interval $[0,1]$. The model that generates the data 
	uses a bivariate function linear in $x$,
	\begin{equation}\label{eq16}
	Y_i = \alpha + \delta \int F\{X_i(t),t\}  dt + \varepsilon_i,
	\end{equation}
	where $F(x,t) = 2x \sin(\pi t)$, $\delta \in \{0.005,\ldots,0.04\}$, $\alpha=1$, 
	and $\varepsilon_i \sim N(0,\sigma^2_{\varepsilon}=1)$ \citep{McLeanetal2015}.
	
\end{itemize}

\subsection{Simulation Designs for Testing a Linear Covariate Effect}\label{simul_linearity}

\noindent In this section, we consider five simulation scenarios: four of them are defined by the articles 
under study, and the last one is inspired from \cite{YaoMuller2010} and used as a baseline. As in Section 
\ref{simul_nullity}, the index $\delta$ is used to control the departure from the null hypothesis. 
Specifically, $\delta = 0$ corresponds to the null hypothesis of linear effect and $\delta \textgreater 0$ 
corresponds to the alternative hypothesis of nonlinear effect. The scenarios {\textbf{(G1, G2, M1, H1, and Y1)}} 
are as follows:

\begin{itemize}[leftmargin=*]
	\item {\textbf{Setting 1 (G1)}}. The functional process for the functional 
	covariate $X$ in this case is a Brownian motion with functional mean 
	$\mu(t)= 0 \; \text{for all} \; t \in [0,1]$, 
	and $Cov(X(s),X(t)) = \frac{\sigma^2}{2\theta}e^{\{-\theta(s+t)\}}(e^{\{2\theta \min(s,t)\}}-1)$, 
	with $\theta = 1/3$ and $\sigma=1$. We use $201$ equidistant points in the interval $[0,1]$. 
	The model that generates the data is
	\begin{equation}\label{eq17}
	Y_i=\int X_i(t)\beta(t)dt + \delta  \int X_i(t)X_i(t)dt+ \varepsilon_i.
	\end{equation} 
	We consider $\delta \in \{0.01,\ldots,0.2\}$, $\varepsilon_i \sim N(0,\sigma^2_{\varepsilon}=0.01)$, 
	and evaluate the model with $\beta(t) = \sin(2\pi t)-\cos(2\pi t)$ \citep{Garciaetal2014}.
	
	\item {\textbf{Setting 2 (G2)}}. This setting is like \textbf{G1}, except that $\beta$ 
	is defined as $\beta(t) = t-(t-0.75)^2$ \citep{Garciaetal2014}.
	
	\item {\textbf{Setting 3 (M1)}}. In this setting, the functional covariate $X$ is given by 
	$X(t)=\sum_{j=1}^4 \xi_j\phi_j(t)$, with $\xi_j \sim N(0,8j^{-2})$ 
	and $\{\phi_1(t),\ldots, \phi_4(t)\} = \{ \sin(\pi t), \\\cos(\pi t), \sin(2\pi t), \cos(2 \pi t) \}$. 
	We use $30$ equidistant points in the interval $[0,1]$. The model that generates 
	the data uses a convex combination of a bivariate function linear in $x$ and one 
	nonlinear in $x$, in the following form:
	\begin{equation}\label{eq18}
	Y_i = \int \left[ (1-\delta) F_1\{X_i(t),t\} + \delta F_2\{X_i(t),t\} \right] dt + \varepsilon_i,
	\end{equation}
	where $F_1(x,t) = 2x \sin(\pi t) \quad \text{and} \quad F_2(x,t) = 10 \cos(-0.125x+0.25t-5)$. 
	The departure in this case has a factor $\delta$ that can control how much the generated data 
	deviates from the linear function. We consider several values for 
	$\delta \in \{0.05,\ldots,0.4\}$ and assume $\varepsilon_i \sim N(0,\sigma^2_{\varepsilon}=1)$ 
	\citep{McLeanetal2015}.
	
	\item {\textbf{Setting 4 (H1)}}. The functional covariate $X$ is given by an independent 
	standard Brownian motion. We use $100$ equidistant points in the interval $[0,1]$. The model 
	that generates the data is
	\begin{equation}\label{eq19}
	Y_i= \alpha + \int X_i(t)\beta(t)dt + \delta \int \int X_i(t) X_i(s)dtds  + \varepsilon_i,
    \end{equation} 
    where $\beta(t) =1$ in all cases. We consider  $\delta \in \{0.1,\ldots,1.8\}$, $\alpha=4$, 
    and assume $\varepsilon_i \sim N(0,\sigma^2_{\varepsilon}=1)$ \citep{HorvathReeder2013}.
	
	\item {\textbf{Setting 5 (Y1)}}. This last scenario is used as a baseline comparison, 
	because there is no testing method associated with it. The functional 
	process for the functional covariate $X$ is generated as 
	$X(t) = \mu(t) + \sum_{j=1}^2 \xi_j \phi_j(t) + \epsilon_i(t)$, 
	where $\mu(t) = t + \sin(t) $, where $\phi_1(t) = -\cos(\pi t / 10) / \sqrt{5}$, 
	$\phi_2(t) = \sin(\pi t / 10) / \sqrt{5}$, $\lambda_1 = 4$, and $\lambda_2 = 1$, 
	and $\epsilon_i(t) \sim N(0,0.5^2)$ is a measurement error for $X$. We use $101$ 
	equidistant points in the interval $[0,10]$. The model that generates the data is 
	\begin{equation}\label{eq20}
	Y_i = \sum_{j=1}^2 \xi_{ij}\beta+\delta \{\sum_{j=1}^2 \xi_{ij}^2+\xi_{i1}\xi_{i2}\}+\varepsilon_i,
	\end{equation} 
	where $\beta =1$ and $\xi_{ij} \sim N (0,\lambda_j)$. 
	We consider  $\delta \in \{0.005,\ldots,0.14\}$ and 
	assume $\varepsilon_i \sim N(0,\sigma^2_{\varepsilon}=0.1)$ \citep{YaoMuller2010}.
	
\end{itemize}

\subsection{Computational Implementation}\label{computation}

\noindent The GGF method was implemented through the \textit{flm.test} function in the R package 
\textit{fda.usc} version 1.2.3. The software fits the FLM and estimates the coefficient function 
by using B-spline basis functions without penalization. The number of basis functions can be 
predetermined by the user or be chosen via the generalized cross-validation criterion 
\citep{RamsaySilverman2005}. However, it is worth mentioning that for the M1 setting, the 
\textit{flm.test} function encounters singularity errors and fails when more than four basis functions 
are used. We therefore used $p=4$ basis functions to approximate the functional covariate. The number 
of bootstrap replicates was $B=5,000$.

The HR testing method requires the number of functional principal components (FPCs) to be decided 
initially. \cite{HorvathReeder2013} reported simulation results for several components. In our simulation 
study, we fixed the number of FPCs at 3.

We implemented the MHR method by using the \textit{pseudo.rlr.test} function of the R package 
\textit{lmeVarComp} version 1.0 and considered 10,000 runs for approximating the null distribution of 
the test statistic. 

\subsection{Results}

\noindent We evaluate the size and power performance of the described testing methods under a wide 
variety of scenarios. The Type I error rates and power are estimated as the proportion of rejecting 
the null hypothesis in the 5,000 and 1,000 simulated samples, respectively. % -------- only moderate ------ 

For testing linearity, Table S1 in the Supplementary Material (Appendix A.1) shows the performance of 
the testing methods for dense sampling design by comparing their empirical Type I error rates for nominal 
levels of 1\%, 5\%, and 10\% and also for varying sample sizes. The results indicate that all three 
methods behave satisfactorily in terms of empirical levels under all settings, when the sample size 
is large ($n=500$). The HR method slightly overestimates the highest nominal level (10\%), especially 
under the G1 and G2 settings. The GGF and MHR methods perform similarly. Power curves for dense sampling 
design are included in the Supplementary Material, Appendix A.1. Figure S2 shows that when $n=500$, the 
MHR method appears to outperform GGF and HR under all data generation settings. For the G1 and G2 settings, 
there is a small difference in power between MHR and GGF. However, the difference between the two methods 
becomes more distinguishable under the H1, M1, and Y1 settings. The GGF and HR methods perform similarly 
in terms of power under the H1 and Y1 settings. For the other settings, GGF is more powerful than HR. We 
also investigate how the methods behave when the sample size changes. As the sample size decreases to 
$n=100$, both GGF and MHR still provide good Type I error rates. The empirical levels of MHR are fairly 
close to the nominal levels regardless of the data generation setting. However, HR performs very poorly 
compared to the other two methods. The HR method tends to overestimate all nominal levels for the moderate 
sample size. When $n=50$, all three methods in general overestimate the nominal levels. The empirical 
levels are only slightly higher than the nominal ones for the MHR method, but not for the GGF and HR 
methods. Particularly, the performance of HR deteriorates considerably as the sample size becomes smaller. 
Because the empirical levels for the small sample size are significantly inflated, power comparison for 
$n=50$ would not be appropriate. In Figure S1 (Appendix A.1), we observe that MHR is more powerful than 
GGF for the moderate sample size under all settings.
% %The empirical power increases at a faster rate for equal sample sizes than unequal sample sizes,

Table \ref{tab_lin_mod} summarizes the rejection rates for testing linearity under moderately sparse design. 
For a large sample size ($n=500$), GGF and MHR maintain the correct nominal levels. The HR method still tends 
to overestimate the nominal ones, but provides close results to the desired levels. Figure \ref{plot_lin_mod_500} 
displays the simulated power curves for testing the null hypothesis of the linear covariate effect under the 
moderate sampling design with the large sample size. The power performance of the methods is a little affected 
by the change of the sampling design. The methods exhibit minor power loss compared to the results obtained for 
densely sampled design. The MHR method shows a general advantage over the GGF and HR methods as $\delta$ increases. 
It performs slightly better than the GGF method under the G1 and G2 settings. For the other settings, the 
difference between the two methods is more significant. The HR and GGF methods appear to perform very similarly 
under the H1 and Y1 settings. However, the power of HR is consistently lower than that of GGF for the other data 
generation settings. For a small sample size ($n=50$), the GGF method tends to underestimate lower nominal levels 
(1\% and 5\%), while the MHR method produces significantly higher empirical rejection rates than the nominal 
levels of 5\% and 10\%. As in the dense case, there is an especially pronounced difference between the empirical 
levels for the HR method and the nominal levels. Both GGF and MHR result in more stable Type I errors for 
$n=100$. We notice that the empirical levels decrease significantly as the sample size increases, but the 
HR method still produces inflated Type I error rates. Similar to the dense case with the moderate sample 
size, MHR performs better than GGF in terms of power under all settings. The power of the methods increases 
at a slower rate as the sample size decreases.

\begin{center}
	\begin{minipage}{2in} 
		\framebox[2\width]{Table 1 about here.}
	\end{minipage} 
\end{center}

Tables S2 in the Supplementary Material (Appendix A.3) reports the probability of rejecting the null hypothesis 
of linear relationship for sparse sampling design. The Type I error rates are similar to those of the moderately 
sparse design except for the M1 data generation setting. For the M1 setting, we notice that all three methods 
have very inflated rejection probabilities. Moreover, the rejection rates for the GGF and MHR methods are not 
decreasing as the sample size increases. The problem here might be that there are very few observations per curve, 
so the estimation performance of the FPCA is affected by the sparsity level of the data. Hence, the methods fail 
to estimate the Type I error rates accurately. The Supplementary Material includes additional simulation results 
(Table S3), which indicate that adding few more observations per curve---that is, making the data less sparse, 
improves the performance of GGF and MHR considerably. As for power comparison, the ordering of the methods does 
not change except that the HR method produces sightly better results than GGF under the Y1 setting for the large 
sample size. In general, all three methods lose power as the functional data becomes more sparse, as expected.

For testing nullity, Table \ref{tab_null_mod} shows the Type I error rates of the GGF, MHR, and KSM methods. 
Our results indicate that the rejection probabilities do not appear to change much as the grid of points for 
the functional covariate becomes more sparse. The rejection rates are mostly within two standard errors of 
the correct levels for all the designs and various sample sizes, which means that all three methods result 
in reasonable Type I errors. For all sampling designs, GGF provides more conservative results for the G0 
setting than those for the M0 setting. Furthermore, when $n=500$, GGF provides more conservative results 
than those for MHR and KSM under the G0 data generation setting. The methods still provide good rejection 
probabilities as the sample size decreases. Only MHR seems to provide relatively conservative results for 
$n=50$ under the G0 data generation setting. The methods have comparable power for all sample sizes. 
According to Figure S6 (Appendix B.1), for dense sampling design with a large sample size, the power functions 
for MHR and KSM are very close to each other such that they overlap. The GGF method is falling dramatically 
behind these two methods in terms of the power performance. When $n=100$, the KSM outperforms MHR under the G0 
setting. Figures \ref{plot_null_mod_500} and S7 (Appendix B.2)show the power performance of the methods for 
moderately sampled data with large and moderate sample sizes, respectively. Similar to the previous results, 
the MHR and KSM methods have good power properties and that they outperform the GGF method substantially both 
under the G0 and M0 data generation settings when $n=500$.  For a moderate sample size ($n=100$), in particular, 
KSM is more powerful than MHR under the G0 setting. We notice that the power curves for sparse design are very 
similar to those obtained for moderately sparse design. 

\begin{center}
	\begin{minipage}{2in} 
		\framebox[2\width]{Table 2 about here.}
	\end{minipage} 
\end{center}

%The power curves in Figure S7 indicate that, the power difference between MHR and KSM becomes negligible as 
%the sample size increases, as expected. 

We also compare the computational costs of the four methods for 10 simulation runs of the M0 and M1 data 
generation settings under dense design with the sample size $n=100$. The simulations were run on a 2.3 GHz 
DELL Quad Processor AMD Opteron with 512 Gb of RAM. The KSM method simulated the data in approximately 3 
seconds and the MHR method did so in 4 seconds, which indicates that MHR runs almost as fast as KSM. The GGF 
method took roughly 12 seconds. The HR method was by far the slowest method, with a computation time of 159 
seconds. 

To sum up, the HR method falls well behind the MHR and GGF methods, because it provides inaccurate size and 
power results and has computational complexity. Despite the fact that the GGF method produces results rather 
close to those for the MHR method for some cases, it still has the disadvantage of being computationally more 
expensive. Our extensive simulation studies indicate that, for testing linearity, the MHR method outperforms 
its competitors with regard to approximately close empirical levels, high power rates, and computational 
efficiency. For testing nullity, both MHR and KSM perform similarly in terms size and power performance for 
the large sample size. For a moderate sample size, there is no uniform best method; however, based on the 
results we recommend the KSM method.

\renewcommand{\arraystretch}{0.75}
\begin{sidewaystable}[!t]\centering 
	\renewcommand{\baselinestretch}{1.3}
	\caption{Testing linearity: Comparison of the estimated Type I error rates of the GGF, MHR, and HR methods 
		in the context of moderately sparse functional data. The data generation settings are G1, G2, H1, M1, 
		and Y1. The number of Monte Carlo experiments is 5,000, and the sample sizes are 50, 100, and 500. 
		Standard errors are shown in parentheses.}
	\label{tab_lin_mod}
	\renewcommand{\baselinestretch}{1.8}
	\scalebox{0.8}{
	\begin{tabular}{ccccc|ccc|ccc} 
		
		\toprule 
		\toprule
		& & \multicolumn{3}{c}{\textbf{GGF}} & \multicolumn{3}{c}{\textbf{MHR}} & \multicolumn{3}{c}{\textbf{HR}} \\ 
		\cmidrule(r){3-5} \cmidrule(r){6-8} \cmidrule{9-11}
		& $\boldsymbol{\alpha}$ & $\textbf{n=50}$ & $\textbf{n=100}$ & $\textbf{n=500}$  
		& $\textbf{n=50}$ & $\textbf{n=100}$ & $\textbf{n=500}$  & $\textbf{n=50}$ & $\textbf{n=100}$ & $\textbf{n=500}$ \\ 
		\midrule 
		
		           & $\textbf{0.01}$ & 0.005\textcolor{blue}{(0.001)} & 0.008\textcolor{blue}{(0.001)} & 0.010\textcolor{blue}{(0.001)} 
		                             & 0.011\textcolor{blue}{(0.001)} & 0.011\textcolor{blue}{(0.001)} & 0.009\textcolor{blue}{(0.001)} 
		                             & 0.087\textcolor{blue}{(0.004)} & 0.036\textcolor{blue}{(0.003)} & 0.012\textcolor{blue}{(0.002)}\\ 
		\textbf{G1}& $\textbf{0.05}$ & 0.049\textcolor{blue}{(0.003)} & 0.050\textcolor{blue}{(0.003)} & 0.051\textcolor{blue}{(0.003)} 
		                             & 0.058\textcolor{blue}{(0.003)} & 0.048\textcolor{blue}{(0.003)} & 0.048\textcolor{blue}{(0.003)} 
		                             & 0.199\textcolor{blue}{(0.006)} & 0.105\textcolor{blue}{(0.004)} & 0.060\textcolor{blue}{(0.003)}\\ 
		           & $\textbf{0.10}$ & 0.118\textcolor{blue}{(0.005)} & 0.108\textcolor{blue}{(0.004)} & 0.100\textcolor{blue}{(0.004)} 
		                             & 0.115\textcolor{blue}{(0.005)} & 0.100\textcolor{blue}{(0.004)} & 0.098\textcolor{blue}{(0.004)} 
		                             & 0.288\textcolor{blue}{(0.006)} & 0.179\textcolor{blue}{(0.005)} & 0.123\textcolor{blue}{(0.005)}\\
		\midrule 
		
		           & $\textbf{0.01}$ & 0.004\textcolor{blue}{(0.001)} & 0.008\textcolor{blue}{(0.001)} & 0.011\textcolor{blue}{(0.001)} 
		                             & 0.013\textcolor{blue}{(0.002)} & 0.012\textcolor{blue}{(0.002)} & 0.011\textcolor{blue}{(0.001)} 
		                             & 0.093\textcolor{blue}{(0.004)} & 0.035\textcolor{blue}{(0.003)} & 0.013\textcolor{blue}{(0.002)}\\ 
		\textbf{G2}& $\textbf{0.05}$ & 0.049\textcolor{blue}{(0.003)} & 0.048\textcolor{blue}{(0.003)} & 0.049\textcolor{blue}{(0.003)} 
		                             & 0.054\textcolor{blue}{(0.003)} & 0.047\textcolor{blue}{(0.003)} & 0.051\textcolor{blue}{(0.003)} 
		                             & 0.209\textcolor{blue}{(0.006)} & 0.105\textcolor{blue}{(0.004)} & 0.059\textcolor{blue}{(0.003)}\\ 
		           & $\textbf{0.10}$ & 0.119\textcolor{blue}{(0.005)} & 0.108\textcolor{blue}{(0.004)} & 0.103\textcolor{blue}{(0.004)} 
		                             & 0.109\textcolor{blue}{(0.004)} & 0.103\textcolor{blue}{(0.004)} & 0.104\textcolor{blue}{(0.004)} 
		                             & 0.299\textcolor{blue}{(0.006)} & 0.173\textcolor{blue}{(0.005)} & 0.119\textcolor{blue}{(0.005)}\\
		\midrule 
		
				   & $\textbf{0.01}$ & 0.006\textcolor{blue}{(0.001)} & 0.008\textcolor{blue}{(0.001)} & 0.012\textcolor{blue}{(0.002)} 
				                     & 0.015\textcolor{blue}{(0.002)} & 0.014\textcolor{blue}{(0.002)} & 0.010\textcolor{blue}{(0.001)} 
				                     & 0.093\textcolor{blue}{(0.004)} & 0.035\textcolor{blue}{(0.003)} & 0.012\textcolor{blue}{(0.002)}\\ 
		\textbf{H1}& $\textbf{0.05}$ & 0.046\textcolor{blue}{(0.003)} & 0.052\textcolor{blue}{(0.003)} & 0.056\textcolor{blue}{(0.003)} 
				                     & 0.063\textcolor{blue}{(0.003)} & 0.053\textcolor{blue}{(0.003)} & 0.051\textcolor{blue}{(0.003)} 
				                     & 0.206\textcolor{blue}{(0.006)} & 0.115\textcolor{blue}{(0.005)} & 0.060\textcolor{blue}{(0.003)}\\ 
				   & $\textbf{0.10}$ & 0.113\textcolor{blue}{(0.004)} & 0.109\textcolor{blue}{(0.004)} & 0.099\textcolor{blue}{(0.004)} 
				                     & 0.113\textcolor{blue}{(0.004)} & 0.105\textcolor{blue}{(0.004)} & 0.100\textcolor{blue}{(0.004)} 
				                     & 0.295\textcolor{blue}{(0.006)} & 0.184\textcolor{blue}{(0.005)} & 0.112\textcolor{blue}{(0.004)}\\
		\midrule
		
				   & $\textbf{0.01}$ & 0.002\textcolor{blue}{(0.001)} & 0.005\textcolor{blue}{(0.001)} & 0.011\textcolor{blue}{(0.001)} 
				                     & 0.014\textcolor{blue}{(0.002)} & 0.012\textcolor{blue}{(0.002)} & 0.011\textcolor{blue}{(0.001)} 
				                     & 0.095\textcolor{blue}{(0.004)} & 0.039\textcolor{blue}{(0.003)} & 0.015\textcolor{blue}{(0.002)}\\ 
		\textbf{M1}& $\textbf{0.05}$ & 0.036\textcolor{blue}{(0.003)} & 0.039\textcolor{blue}{(0.003)} & 0.050\textcolor{blue}{(0.003)} 
				                     & 0.064\textcolor{blue}{(0.003)} & 0.052\textcolor{blue}{(0.003)} & 0.051\textcolor{blue}{(0.003)} 
				                     & 0.210\textcolor{blue}{(0.006)} & 0.115\textcolor{blue}{(0.005)} & 0.059\textcolor{blue}{(0.003)}\\ 
				   & $\textbf{0.10}$ & 0.102\textcolor{blue}{(0.004)} & 0.094\textcolor{blue}{(0.004)} & 0.097\textcolor{blue}{(0.004)} 
				                     & 0.121\textcolor{blue}{(0.005)} & 0.107\textcolor{blue}{(0.004)} & 0.103\textcolor{blue}{(0.004)} 
				                     & 0.300\textcolor{blue}{(0.006)} & 0.187\textcolor{blue}{(0.005)} & 0.115\textcolor{blue}{(0.005)}\\
		\midrule 
		
		           & $\textbf{0.01}$ & 0.005\textcolor{blue}{(0.001)} & 0.014\textcolor{blue}{(0.002)} & 0.013\textcolor{blue}{(0.002)} 
		                             & 0.016\textcolor{blue}{(0.002)} & 0.011\textcolor{blue}{(0.001)} & 0.010\textcolor{blue}{(0.001)} 
		                             & 0.098\textcolor{blue}{(0.004)} & 0.038\textcolor{blue}{(0.003)} & 0.016\textcolor{blue}{(0.002)}\\ 
		\textbf{Y1}& $\textbf{0.05}$ & 0.058\textcolor{blue}{(0.003)} & 0.055\textcolor{blue}{(0.003)} & 0.067\textcolor{blue}{(0.004)} 
		                             & 0.062\textcolor{blue}{(0.003)} & 0.049\textcolor{blue}{(0.003)} & 0.050\textcolor{blue}{(0.003)} 
		                             & 0.222\textcolor{blue}{(0.006)} & 0.118\textcolor{blue}{(0.005)} & 0.061\textcolor{blue}{(0.003)}\\ 
		           & $\textbf{0.10}$ & 0.125\textcolor{blue}{(0.005)} & 0.116\textcolor{blue}{(0.005)} & 0.119\textcolor{blue}{(0.004)} 
		                             & 0.126\textcolor{blue}{(0.005)} & 0.100\textcolor{blue}{(0.004)} & 0.100\textcolor{blue}{(0.004)} 
		                             & 0.320\textcolor{blue}{(0.007)} & 0.193\textcolor{blue}{(0.006)} & 0.116\textcolor{blue}{(0.005)}\\
		\midrule 
		\bottomrule 
		
	\end{tabular}
}
\end{sidewaystable}

\renewcommand{\arraystretch}{0.75}
\begin{sidewaystable}[!t]\centering 
	\renewcommand{\baselinestretch}{1.3}
	\caption{Testing nullity: Comparison of the estimated Type I error rates of the GGF, MHR, and KSM methods 
		in the context of dense, moderately sparse, and sparse functional data. The data generation settings 
		are G0 and M0. The number of Monte Carlo experiments is 5,000, and the sample sizes are 50, 100, and 500. 
		Standard errors are shown in parentheses.}
	\label{tab_null_mod}
	\renewcommand{\baselinestretch}{1.8}
	\scalebox{0.8}{
	\begin{tabular}{ccccc|ccc|ccc} 
		
		\toprule 
		\toprule
		& & \multicolumn{3}{c}{\textbf{GGF}} & \multicolumn{3}{c}{\textbf{MHR}} & \multicolumn{3}{c}{\textbf{KSM}} \\ 
		\cmidrule(r){3-5} \cmidrule(r){6-8} \cmidrule{9-11}
		& $\boldsymbol{\alpha}$ & $\textbf{n=50}$ & $\textbf{n=100}$ & $\textbf{n=500}$  
		& $\textbf{n=50}$ & $\textbf{n=100}$ & $\textbf{n=500}$  & $\textbf{n=50}$ & $\textbf{n=100}$ & $\textbf{n=500}$ \\  
		\midrule \addlinespace
		\rowcolor{gray!20}\multicolumn{11}{c}{\textbf{Dense}}\\
		
		           & $\textbf{0.01}$ & 0.008\textcolor{blue}{(0.001)} & 0.010\textcolor{blue}{(0.001)} & 0.010\textcolor{blue}{(0.001)} 
		                             & 0.008\textcolor{blue}{(0.001)} & 0.009\textcolor{blue}{(0.001)} & 0.010\textcolor{blue}{(0.001)} 
		                             & 0.009\textcolor{blue}{(0.001)} & 0.013\textcolor{blue}{(0.002)} & 0.008\textcolor{blue}{(0.001)}\\ 
		\textbf{G0}& $\textbf{0.05}$ & 0.048\textcolor{blue}{(0.003)} & 0.049\textcolor{blue}{(0.003)} & 0.046\textcolor{blue}{(0.003)} 
		                             & 0.043\textcolor{blue}{(0.003)} & 0.049\textcolor{blue}{(0.003)} & 0.050\textcolor{blue}{(0.003)} 
	 	                             & 0.051\textcolor{blue}{(0.003)} & 0.055\textcolor{blue}{(0.003)} & 0.049\textcolor{blue}{(0.003)}\\ 
		           & $\textbf{0.10}$ & 0.101\textcolor{blue}{(0.004)} & 0.097\textcolor{blue}{(0.004)} & 0.095\textcolor{blue}{(0.004)} 
		                             & 0.084\textcolor{blue}{(0.004)} & 0.097\textcolor{blue}{(0.004)} & 0.108\textcolor{blue}{(0.004)} 
		                             & 0.100\textcolor{blue}{(0.004)} & 0.105\textcolor{blue}{(0.004)} & 0.098\textcolor{blue}{(0.004)}\\
		\midrule
		
		           & $\textbf{0.01}$ & 0.008\textcolor{blue}{(0.001)} & 0.010\textcolor{blue}{(0.001)} & 0.012\textcolor{blue}{(0.002)} 
		                             & 0.011\textcolor{blue}{(0.001)} & 0.010\textcolor{blue}{(0.001)} & 0.010\textcolor{blue}{(0.001)} 
		                             & 0.010\textcolor{blue}{(0.001)} & 0.009\textcolor{blue}{(0.001)} & 0.009\textcolor{blue}{(0.001)}\\ 
		\textbf{M0}& $\textbf{0.05}$ & 0.047\textcolor{blue}{(0.003)} & 0.052\textcolor{blue}{(0.003)} & 0.052\textcolor{blue}{(0.003)} 
		                             & 0.050\textcolor{blue}{(0.003)} & 0.048\textcolor{blue}{(0.003)} & 0.055\textcolor{blue}{(0.003)}
		                             & 0.049\textcolor{blue}{(0.003)} & 0.048\textcolor{blue}{(0.003)} & 0.054\textcolor{blue}{(0.003)}\\ 
		           & $\textbf{0.10}$ & 0.103\textcolor{blue}{(0.004)} & 0.103\textcolor{blue}{(0.004)} & 0.104\textcolor{blue}{(0.004)} 
		                             & 0.099\textcolor{blue}{(0.004)} & 0.097\textcolor{blue}{(0.004)} & 0.106\textcolor{blue}{(0.004)} 
		                             & 0.101\textcolor{blue}{(0.004)} & 0.099\textcolor{blue}{(0.004)} & 0.107\textcolor{blue}{(0.004)}\\
		\midrule \addlinespace
		\rowcolor{gray!20}\multicolumn{11}{c}{\textbf{Moderate}}\\ 
		
		           & $\textbf{0.01}$ & 0.009\textcolor{blue}{(0.001)} & 0.010\textcolor{blue}{(0.001)} & 0.010\textcolor{blue}{(0.001)} 
		                             & 0.008\textcolor{blue}{(0.001)} & 0.009\textcolor{blue}{(0.001)} & 0.011\textcolor{blue}{(0.001)} 
		                             & 0.010\textcolor{blue}{(0.001)} & 0.011\textcolor{blue}{(0.001)} & 0.008\textcolor{blue}{(0.001)}\\ 
		\textbf{G0}& $\textbf{0.05}$ & 0.048\textcolor{blue}{(0.003)} & 0.048\textcolor{blue}{(0.003)} & 0.046\textcolor{blue}{(0.003)} 
		                             & 0.043\textcolor{blue}{(0.003)} & 0.050\textcolor{blue}{(0.003)} & 0.052\textcolor{blue}{(0.003)} 
		                             & 0.050\textcolor{blue}{(0.003)} & 0.055\textcolor{blue}{(0.003)} & 0.051\textcolor{blue}{(0.003)}\\ 
		           & $\textbf{0.10}$ & 0.102\textcolor{blue}{(0.004)} & 0.096\textcolor{blue}{(0.004)} & 0.095\textcolor{blue}{(0.004)} 
		                             & 0.086\textcolor{blue}{(0.004)} & 0.098\textcolor{blue}{(0.004)} & 0.104\textcolor{blue}{(0.004)}
		                             & 0.103\textcolor{blue}{(0.004)} & 0.104\textcolor{blue}{(0.004)} & 0.098\textcolor{blue}{(0.004)}\\
		
		\midrule			           
		           & $\textbf{0.01}$ & 0.009\textcolor{blue}{(0.001)} & 0.010\textcolor{blue}{(0.001)} & 0.011\textcolor{blue}{(0.001)} 
		                             & 0.010\textcolor{blue}{(0.001)} & 0.009\textcolor{blue}{(0.001)} & 0.009\textcolor{blue}{(0.001)} 
		                             & 0.012\textcolor{blue}{(0.002)} & 0.008\textcolor{blue}{(0.001)} & 0.008\textcolor{blue}{(0.001)}\\ 
		\textbf{M0}& $\textbf{0.05}$ & 0.050\textcolor{blue}{(0.003)} & 0.051\textcolor{blue}{(0.003)} & 0.052\textcolor{blue}{(0.003)} 
		                             & 0.049\textcolor{blue}{(0.003)} & 0.049\textcolor{blue}{(0.003)} & 0.054\textcolor{blue}{(0.003)} 
		                             & 0.048\textcolor{blue}{(0.003)} & 0.046\textcolor{blue}{(0.003)} & 0.055\textcolor{blue}{(0.003)}\\ 
		           & $\textbf{0.10}$ & 0.104\textcolor{blue}{(0.004)} & 0.104\textcolor{blue}{(0.004)} & 0.104\textcolor{blue}{(0.004)} 
		                             & 0.098\textcolor{blue}{(0.104)} & 0.098\textcolor{blue}{(0.004)} & 0.108\textcolor{blue}{(0.004)} 
		                             & 0.099\textcolor{blue}{(0.004)} & 0.093\textcolor{blue}{(0.004)} & 0.108\textcolor{blue}{(0.004)}\\
		\midrule \addlinespace
		\rowcolor{gray!20}\multicolumn{11}{c}{\textbf{Sparse}}\\ 
		
		           & $\textbf{0.01}$ & 0.009\textcolor{blue}{(0.001)} & 0.009\textcolor{blue}{(0.001)} & 0.010\textcolor{blue}{(0.001)} 
		                             & 0.008\textcolor{blue}{(0.001)} & 0.009\textcolor{blue}{(0.001)} & 0.011\textcolor{blue}{(0.001)}
		                             & 0.010\textcolor{blue}{(0.001)} & 0.012\textcolor{blue}{(0.002)} & 0.011\textcolor{blue}{(0.001)}\\ 
		\textbf{G0}& $\textbf{0.05}$ & 0.048\textcolor{blue}{(0.003)} & 0.047\textcolor{blue}{(0.003)} & 0.046\textcolor{blue}{(0.003)} 
		                             & 0.044\textcolor{blue}{(0.003)} & 0.050\textcolor{blue}{(0.003)} & 0.050\textcolor{blue}{(0.003)} 
		                             & 0.057\textcolor{blue}{(0.003)} & 0.046\textcolor{blue}{(0.003)} & 0.046\textcolor{blue}{(0.003)}\\ 
		           & $\textbf{0.10}$ & 0.101\textcolor{blue}{(0.004)} & 0.095\textcolor{blue}{(0.004)} & 0.094\textcolor{blue}{(0.004)} 
		                             & 0.092\textcolor{blue}{(0.004)} & 0.097\textcolor{blue}{(0.004)} & 0.106\textcolor{blue}{(0.004)}
		                             & 0.102\textcolor{blue}{(0.004)} & 0.099\textcolor{blue}{(0.004)} & 0.099\textcolor{blue}{(0.004)}\\
		
		\midrule			           	
		           & $\textbf{0.01}$ & 0.009\textcolor{blue}{(0.001)} & 0.011\textcolor{blue}{(0.001)} & 0.011\textcolor{blue}{(0.001)} 
		                             & 0.011\textcolor{blue}{(0.001)} & 0.010\textcolor{blue}{(0.001)} & 0.011\textcolor{blue}{(0.001)} 
		                             & 0.012\textcolor{blue}{(0.002)} & 0.009\textcolor{blue}{(0.001)} & 0.010\textcolor{blue}{(0.001)}\\ 
		\textbf{M0}& $\textbf{0.05}$ & 0.049\textcolor{blue}{(0.003)} & 0.050\textcolor{blue}{(0.003)} & 0.053\textcolor{blue}{(0.003)} 
		                             & 0.050\textcolor{blue}{(0.003)} & 0.049\textcolor{blue}{(0.003)} & 0.054\textcolor{blue}{(0.003)}
		                             & 0.049\textcolor{blue}{(0.003)} & 0.048\textcolor{blue}{(0.003)} & 0.052\textcolor{blue}{(0.003)}\\ 
		           & $\textbf{0.10}$ & 0.104\textcolor{blue}{(0.004)} & 0.103\textcolor{blue}{(0.004)} & 0.105\textcolor{blue}{(0.004)} 
		                             & 0.102\textcolor{blue}{(0.004)} & 0.096\textcolor{blue}{(0.004)} & 0.108\textcolor{blue}{(0.004)}
		                             & 0.100\textcolor{blue}{(0.004)} & 0.101\textcolor{blue}{(0.004)} & 0.105\textcolor{blue}{(0.004)}\\
		
		\bottomrule 
		
	\end{tabular}
}
\end{sidewaystable}

\begin{figure}[htp]
	\centering
	\includegraphics[width=.48\textwidth]{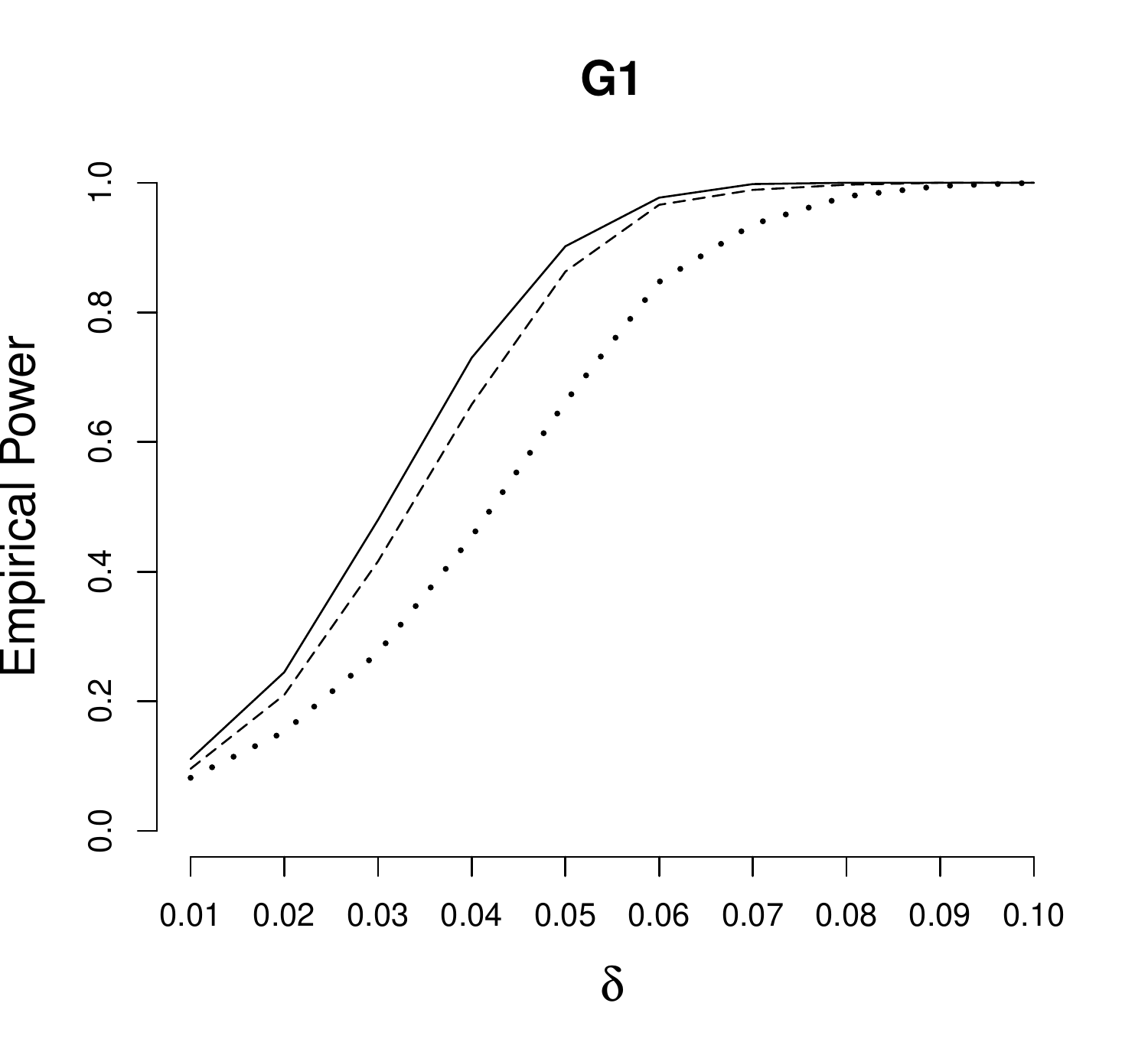} \quad
	\includegraphics[width=.48\textwidth]{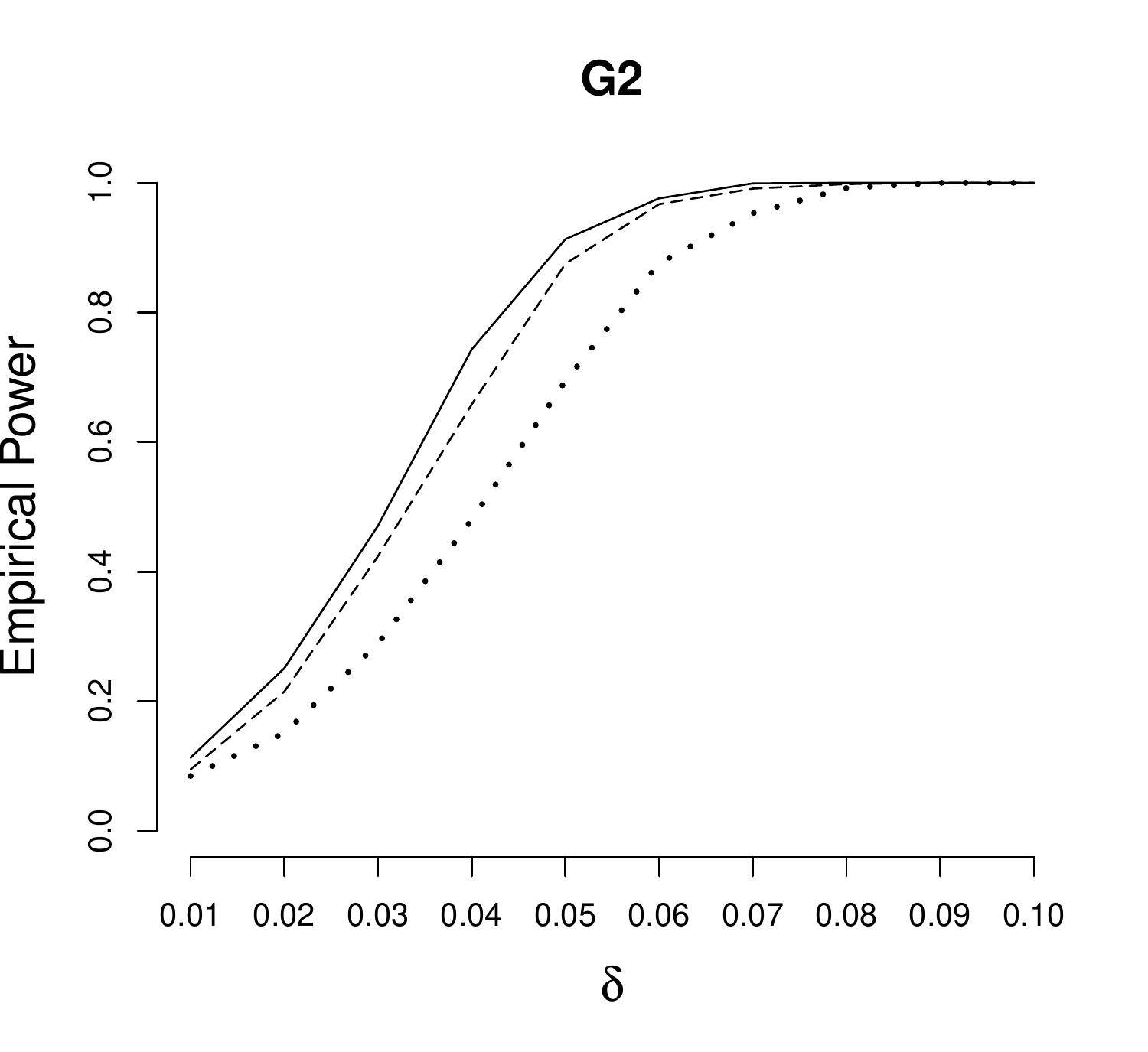}
	
	\medskip
	
	\includegraphics[width=.48\textwidth]{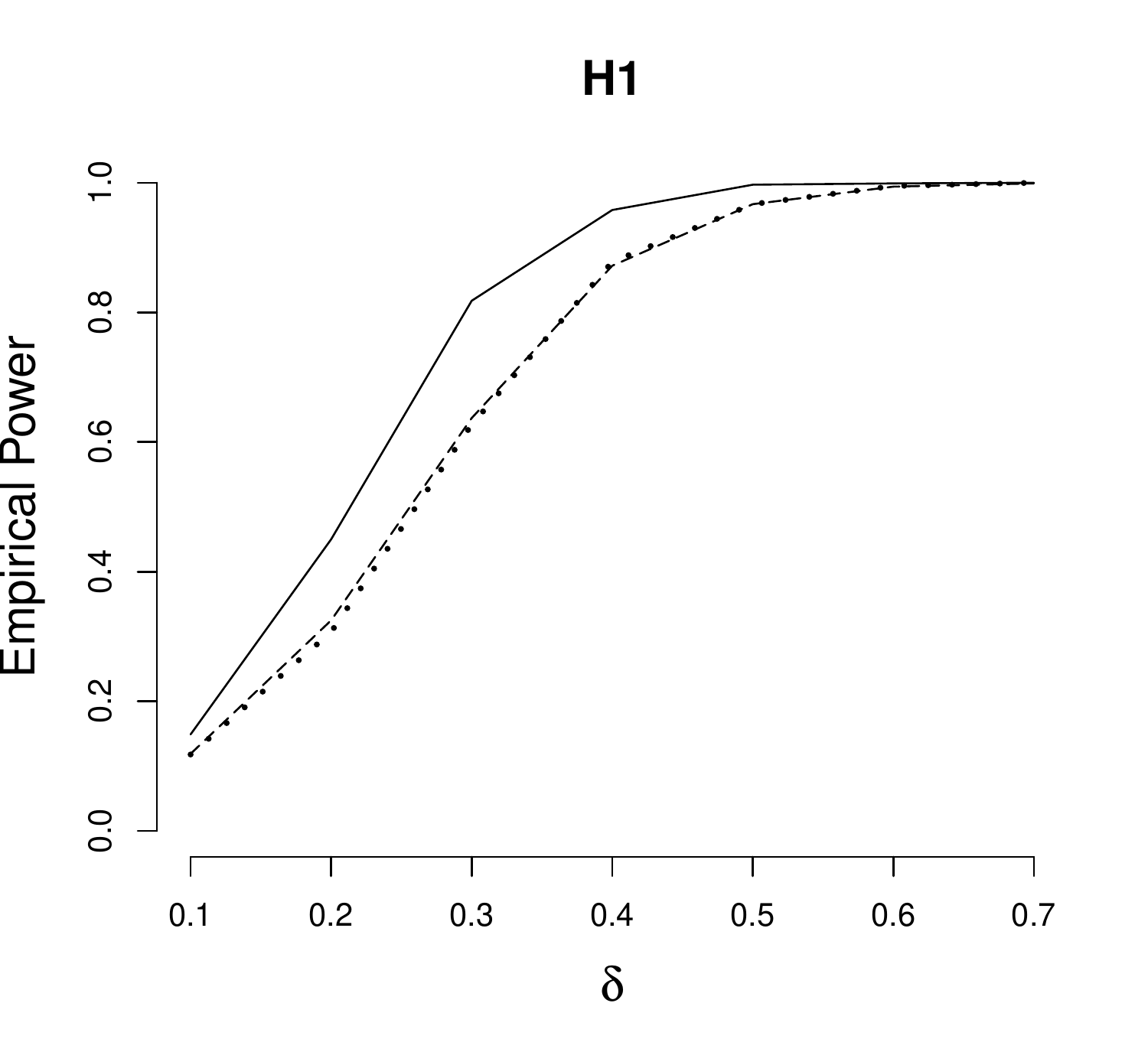}\quad
	\includegraphics[width=.48\textwidth]{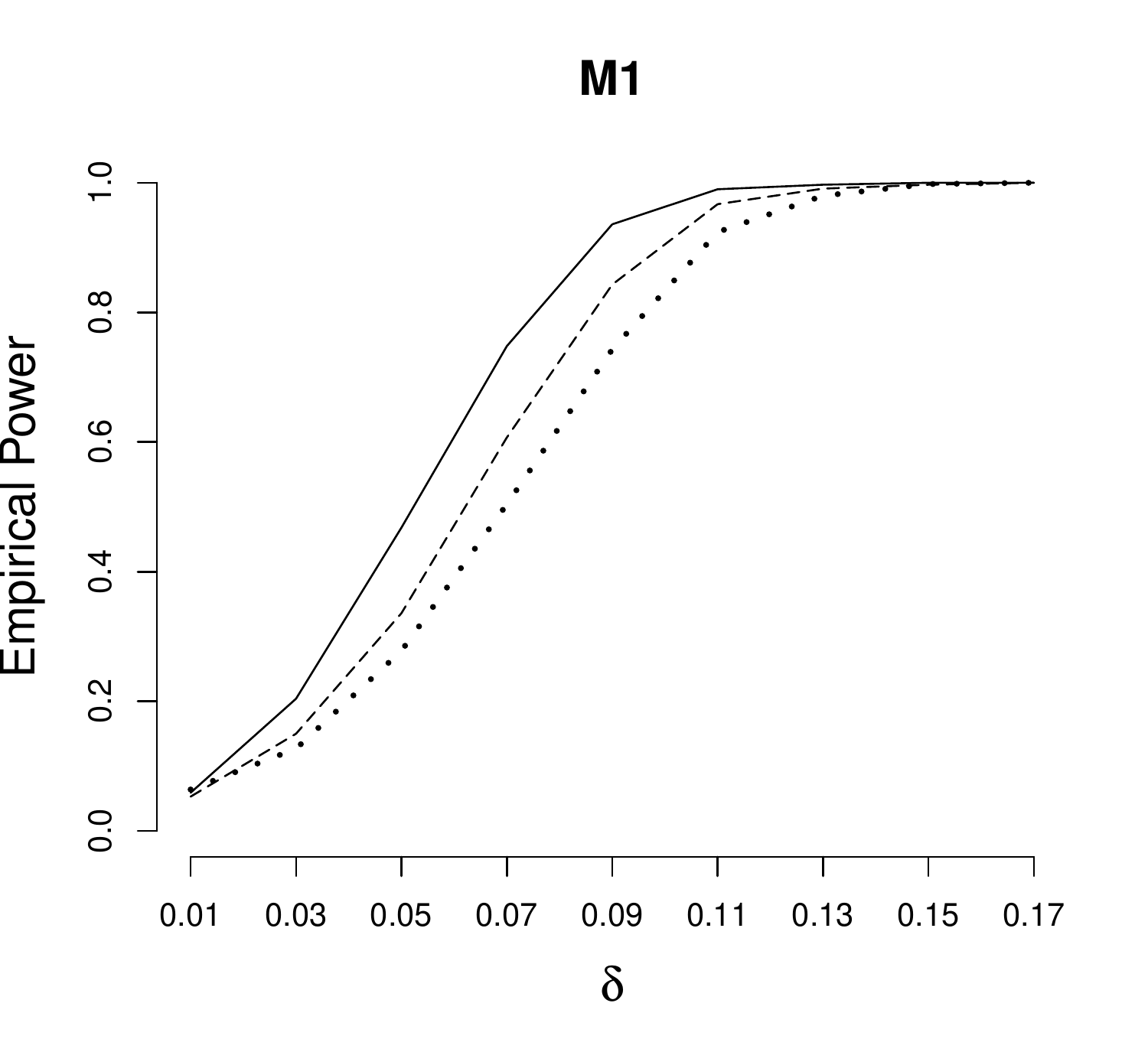}
	
	\medskip
	
	\includegraphics[width=.48\textwidth]{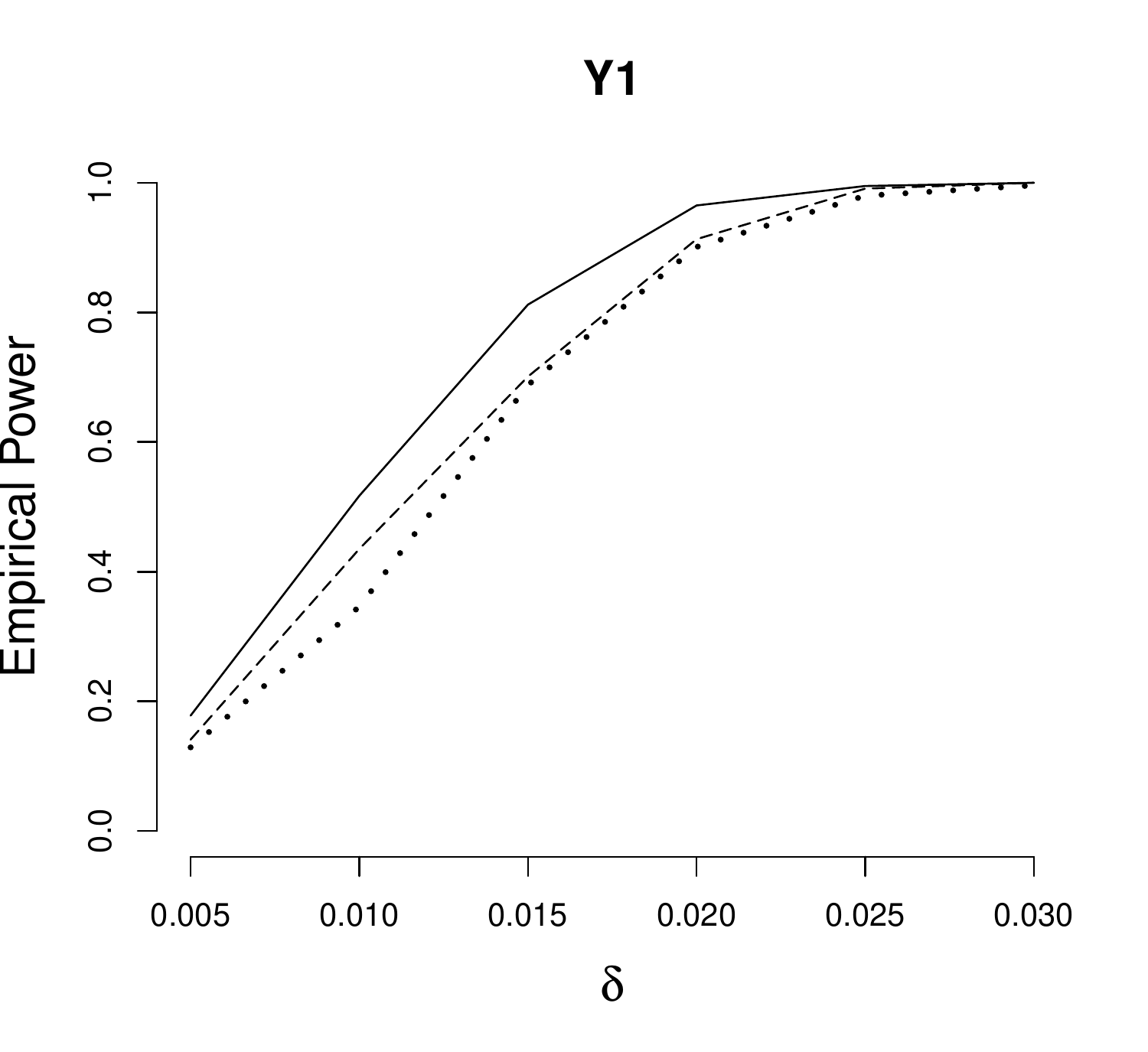}
	
	\renewcommand{\baselinestretch}{1.3}
	\caption{Empirical power of the competing GGF, MHR, and HR methods for 
		testing linear effect for the moderately sparse sampling design. \textit{Solid lines} 
		indicate results for the MHR method, \textit{dashed lines} indicate results 
		for the GGF method, and \textit{dotted lines} indicate results for the HR method. 
		The significance level is $\alpha=0.05$. The number of Monte Carlo experiments is 
		1,000, and the sample size is $n=500$.}
	\label{plot_lin_mod_500}
\end{figure}

\begin{figure}[htp]
	\centering
	\includegraphics[width=.48\textwidth]{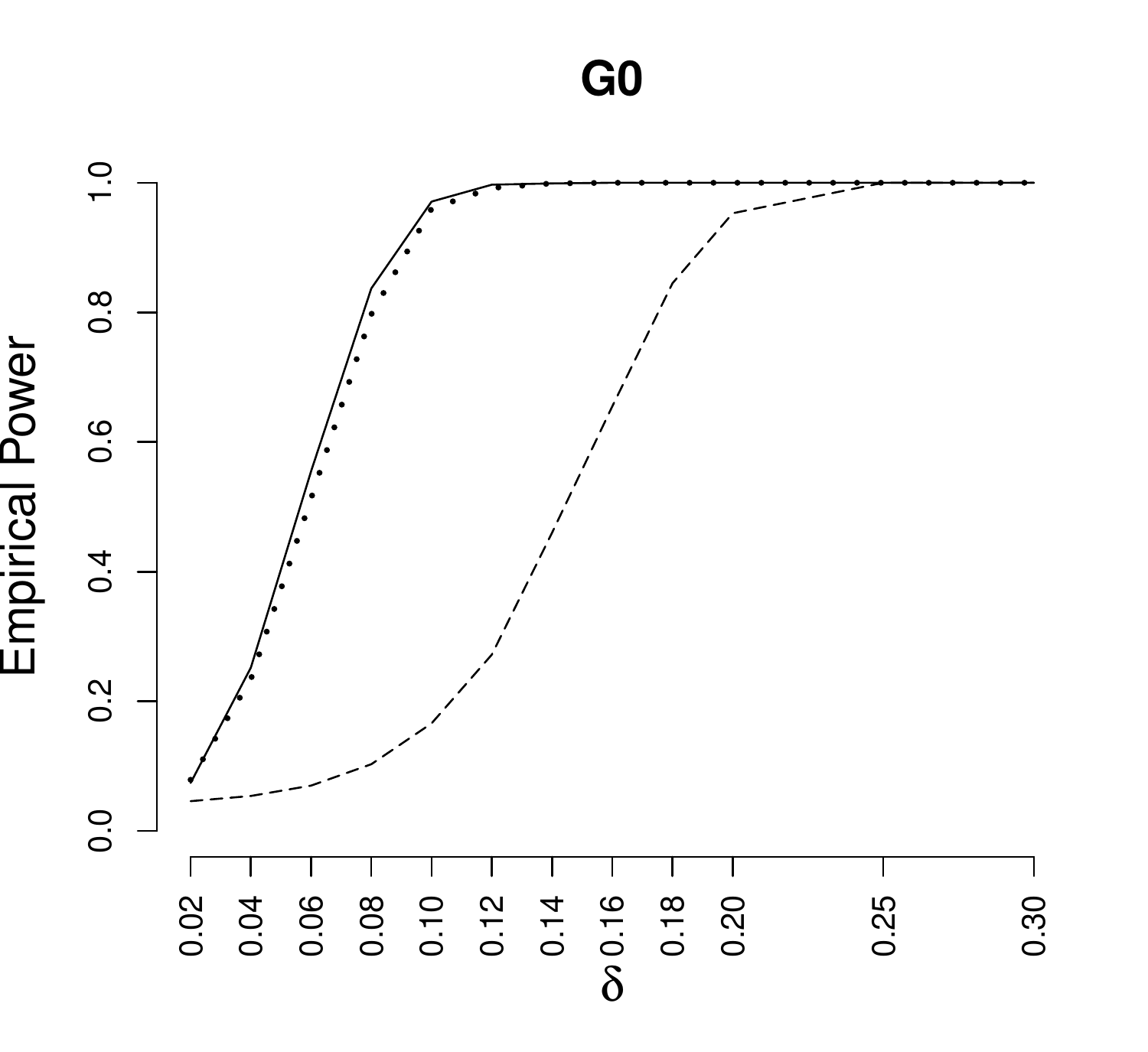}\quad
	\includegraphics[width=.49\textwidth]{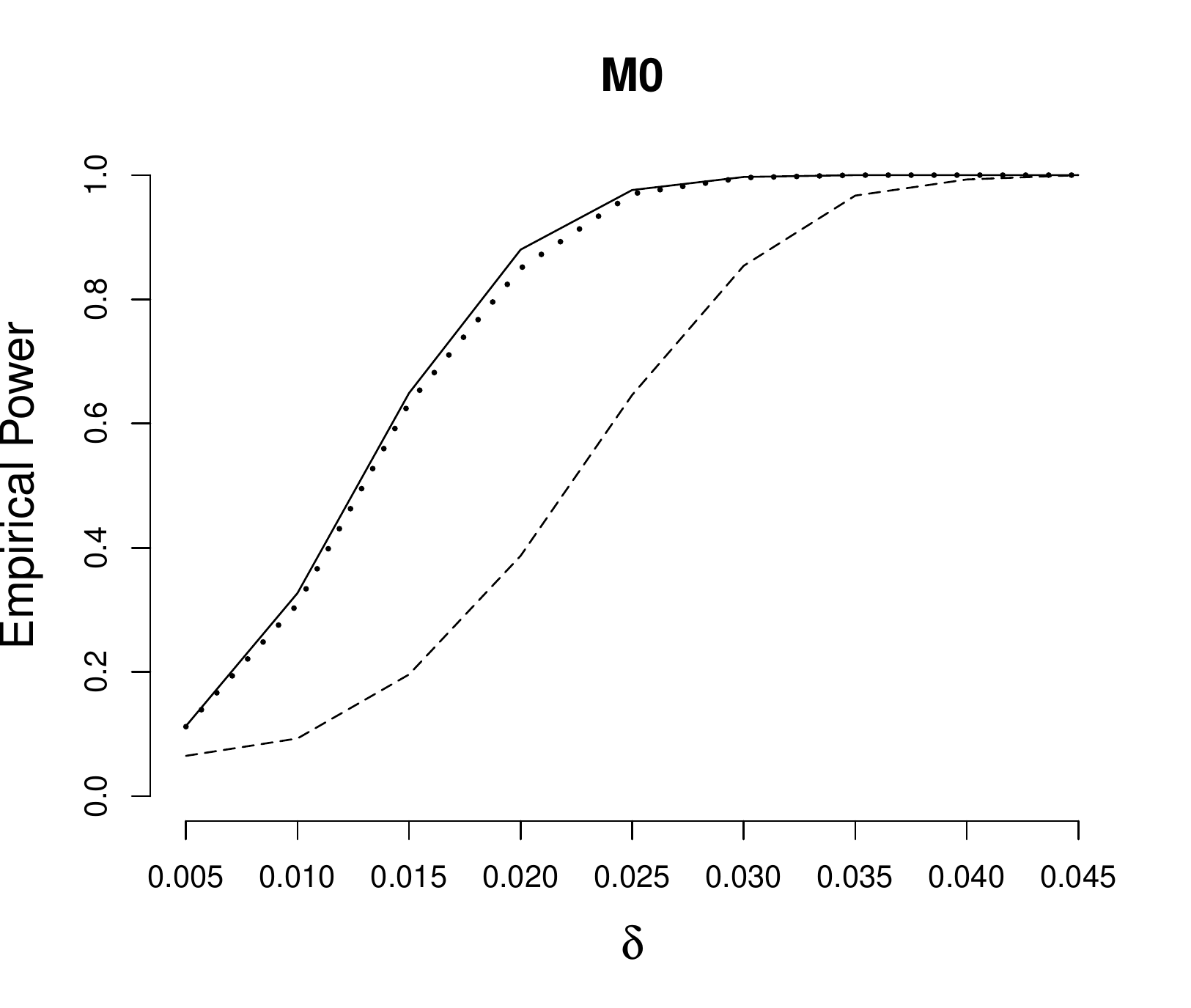}
	
	\renewcommand{\baselinestretch}{1.3}
	\caption{Empirical power of the competing GGF, MHR, and KSM methods for 
		testing no effect for the moderately sparse sampling design. \textit{Solid lines} 
		indicate results for the MHR method, \textit{dashed lines} indicate results 
		for the GGF method, and \textit{dotted lines} indicate results for the KSM method. 
		The significance level is $\alpha=0.05$. The number of Monte Carlo experiments is 
		1,000, and the sample size is $n=500$.}
	\label{plot_null_mod_500}
\end{figure}

\section{Data Analysis}\label{data_analysis}

\noindent We consider the application of these methods to a food quality control problem. The Tecator data 
set has been commonly used to predict the fat content of meat samples and is found at 
\hyperlink{data}{http://lib.stat.cmu.edu/datasets/tecator}. 

\begin{figure}
	\begin{center}
		\includegraphics[scale=0.5]{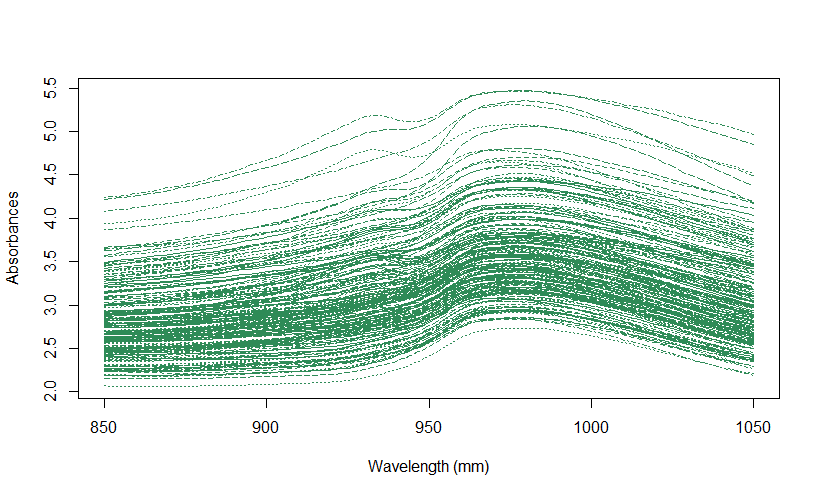}
		\caption{Absorbance trajectories concerning 215 samples of finely chopped pure meat.}
		\label{fig2}
	\end{center}
\end{figure}

This data set includes measurements of a 100-channel spectrum of absorbances, in addition to fat, protein, 
and moisture (water) content from $n=215$ finely chopped pure meat samples. For each sample of meat, a 
100-channel near-infrared (NIR) spectrum of absorbances is calculated as a log transform of the transmittance 
obtained by the analyzer and recorded. The absorbances for a meat sample can be deemed to be discrete 
realizations of random smooth curves, $X_i(\cdot)$. The absorbance trajectories versus wavelength are 
displayed in Figure~\ref{fig2}.

The data were first analyzed by \cite{BorggaardThodberg1992}, who trained neural network models to predict 
the fat content. \cite{YaoMuller2010} proposed a functional quadratic regression model to predict the fat 
content depending on the absorbance trajectories. For the same purpose, \cite{FebreroGonzalez2013} developed 
an algorithm for functional regression models whose response variable comes from an exponential family. 
Rather than focusing on prediction, \cite{HorvathReeder2013} and \cite{Garciaetal2014} used this data set to 
investigate whether a linear dependence existed between the functional covariate and the scalar response.  

The goal of this section is to test whether the association between the spectra of absorbances 
(functional predictor) and each of the measures of fat, protein, or water content of the meat samples 
(scalar responses) is null or not. In this regard, we employ the three methods---GGF, MHR and KSM---and 
we discuss whether there is evidence against a null association. In addition, we investigate whether 
the existing association is linear by employing the three methods---GGF, MHR and HR. A significance 
level of $\alpha=0.05$ is used. Because we use the same data set with three different methods for each 
response (fat, protein, and moisture), we apply a Bonferroni correction to account for multiple testing. 
The adjusted significance level is $\alpha= 0.05/3 = 0.0167$.

\begin{center}
	\begin{minipage}{2in} 
		\framebox[2\width]{Table 3 about here.}
	\end{minipage} 
\end{center}

The results for the nullity test are shown in Table~\ref{tab1}. These results are not very surprising, because 
previous analyses \citep{HorvathReeder2013,Garciaetal2014} have determined that an association exists between 
each of the responses and the functional covariate. Our analysis confirms this, because all $p$-values are 
less than $\alpha = 0.0167$. More interesting results are obtained by the linear tests (Table~\ref{tab1}), 
because we can draw different conclusions depending on the test we use. The $p$-values of the GGF method 
are greater than $\alpha$ for both fat and water content, which means that this test produces no evidence 
to reject the null hypothesis of a linear relationship between percentage of fat and the absorbance 
trajectories, or between the moisture and the same functional covariate. A different conclusion can be 
drawn if we use the MHR or the HR method, because their $p$-values are less than $\alpha$ for all responses. 
These results are expected as in our simulation study; the GGF method showed consistently less power than the 
MHR test for various data structures. 
%Furthermore, to better understand where to 
%find the evidence of $H_A$ being true, we can use the alternative hypothesis for the HR method: quadratic association between the scalar response and the functional covariate. This quadratic association is a subset of FGAM, which is the alternative hypothesis for MHR: a functional generalized additive model. To do so, we plot $\gamma(s,t)$, which we use as an estimate of fat percentage as a response. Figure~\ref{fig3} shows how the surface $\gamma(s,t)$ is very far from being a flat zero surface, thus providing evidence to affirm that the alternative hypothesis is true for both HR and MHR methods. Even if $\gamma(s,t) \neq 0$ is part of the alternative hypothesis in the GGF method, we interpret this result as a consequence of having such a large space contained in the alternative hypothesis.

\begin{table}
	\centering
	\renewcommand{\baselinestretch}{1.3}
	\caption{$p$-values for each of the methods for testing null effect and linear effect 
		of the spectra of absorbances on response variables fat, protein, and water content. 
	    The significance level is $\alpha = 0.0167$.}
    \renewcommand{\baselinestretch}{1.8}
	\scalebox{0.9}{
	\begin{tabular}{cccc | cccc}
		\hline
		&  \multicolumn{3}{c}{\textbf{Nullity}} && \multicolumn{3}{c}{\textbf{Linearity}}\\ 
		\hline
		& \textbf{Fat} & \textbf{Water} & \textbf{Protein} && \textbf{Fat} & \textbf{Water} & \textbf{Protein}\\ 
		\hline
		\textbf{GGF} & 0.000* & 0.000* & 0.000* & \textbf{GGF} & 0.029  & 0.017  & 0.009* \\ 
		\textbf{MHR} & 0.000* & 0.000* & 0.000* & \textbf{MHR} & 0.000* & 0.000* & 0.000* \\ 
		\textbf{KSM} & 0.000* & 0.000* & 0.000* & \textbf{HR}  & 0.000* & 0.000* & 0.000* \\ 
		\hline
	\end{tabular}
}
\footnotesize
\begin{tablenotes}   
	\item \textit{Note}. *Significant at the $p < 0.0167$ level.   
\end{tablenotes}
	\label{tab1}
\end{table}

%\section{Conclusion}

\section*{References}

\bibliography{MerveTekbudak_Manuscript}

	\begin{frontmatter}
		
		\title{Supplementary Material for ``A Comparison of Testing Methods in Scalar-on-Function Regression"}
%		\author{Merve Yasemin~Tekbudak\corref{cor1}}
%		\author{Marcela Alfaro~C\'{o}rdoba\corref{}}
%		\author{Arnab~Maity\corref{}}
%		\author{and Ana-Maria~Staicu\corref{}}
%		
%		\address{Department of Statistics, North Carolina State University}
%		
%		\cortext[cor1]{Corresponding author. Email:mytekbud@ncsu.edu}
	\end{frontmatter}
	
	The Supplement Material contains two sections. Section A presents additional simulation 
	results (empirical rejection rates and power curves) for testing linearity in the context 
	of dense, moderately sparse, and sparse functional data. Section B includes additional 
	power curves for testing nullity in the context of dense, moderately sparse, and sparse 
	functional data.
	
	\newpage
	\renewcommand{\thesection}{\Alph{section}}  
	\renewcommand{\thesubsection}{\thesection.\arabic{subsection}}
	
	\section{Simulation results for testing linearity}
	
	\subsection{\textbf{Type I error rates and power curves for dense sampling design}}
	
	\renewcommand{\thetable}{S\arabic{table}}
	\renewcommand{\thefigure}{S\arabic{figure}}
	\renewcommand{\arraystretch}{0.75}
	\begin{sidewaystable}[!t]\centering 
		\renewcommand{\baselinestretch}{1.3}
		\caption{Testing linearity: Comparison of the estimated Type I error rates of the GGF, MHR, and HR methods 
			in the context of dense functional data. The data generation settings are G1, G2, H1, M1, and Y1. The 
			number of Monte Carlo experiments is 5,000, and the sample sizes are 50, 100, and 500. Standard errors 
			are shown in parentheses.}
		\renewcommand{\baselinestretch}{1.8}
		\scalebox{0.8}{
			\begin{tabular}{ccccc|ccc|ccc} 
				
				\toprule 
				\toprule
				& & \multicolumn{3}{c}{\textbf{GGF}} & \multicolumn{3}{c}{\textbf{MHR}} & \multicolumn{3}{c}{\textbf{HR}} \\ 
				\cmidrule(r){3-5} \cmidrule(r){6-8} \cmidrule{9-11}
				& $\boldsymbol{\alpha}$ & $\textbf{n=50}$ & $\textbf{n=100}$ & $\textbf{n=500}$  
				& $\textbf{n=50}$ & $\textbf{n=100}$ & $\textbf{n=500}$  & $\textbf{n=50}$ & $\textbf{n=100}$ & $\textbf{n=500}$ \\ 
				\midrule 
				
				& $\textbf{0.01}$ & 0.011\textcolor{blue}{(0.001)} & 0.009\textcolor{blue}{(0.001)} & 0.009\textcolor{blue}{(0.001)} 
				& 0.011\textcolor{blue}{(0.001)} & 0.012\textcolor{blue}{(0.002)} & 0.009\textcolor{blue}{(0.001)} 
				& 0.090\textcolor{blue}{(0.004)} & 0.033\textcolor{blue}{(0.003)} & 0.011\textcolor{blue}{(0.001)}\\ 
				\textbf{G1}& $\textbf{0.05}$ & 0.073\textcolor{blue}{(0.004)} & 0.054\textcolor{blue}{(0.003)} & 0.052\textcolor{blue}{(0.003)} 
				& 0.061\textcolor{blue}{(0.003)} & 0.052\textcolor{blue}{(0.003)} & 0.048\textcolor{blue}{(0.003)} 
				& 0.200\textcolor{blue}{(0.006)} & 0.104\textcolor{blue}{(0.004)} & 0.060\textcolor{blue}{(0.003)}\\ 
				& $\textbf{0.10}$ & 0.157\textcolor{blue}{(0.005)} & 0.112\textcolor{blue}{(0.004)} & 0.099\textcolor{blue}{(0.004)} 
				& 0.116\textcolor{blue}{(0.005)} & 0.098\textcolor{blue}{(0.004)} & 0.099\textcolor{blue}{(0.004)} 
				& 0.288\textcolor{blue}{(0.006)} & 0.176\textcolor{blue}{(0.005)} & 0.121\textcolor{blue}{(0.005)}\\
				\midrule 
				
				& $\textbf{0.01}$ & 0.010\textcolor{blue}{(0.001)} & 0.009\textcolor{blue}{(0.001)} & 0.011\textcolor{blue}{(0.001)} 
				& 0.012\textcolor{blue}{(0.002)} & 0.011\textcolor{blue}{(0.001)} & 0.012\textcolor{blue}{(0.002)} 
				& 0.091\textcolor{blue}{(0.004)} & 0.033\textcolor{blue}{(0.003)} & 0.012\textcolor{blue}{(0.002)}\\ 
				\textbf{G2}& $\textbf{0.05}$ & 0.066\textcolor{blue}{(0.004)} & 0.056\textcolor{blue}{(0.003)} & 0.051\textcolor{blue}{(0.003)} 
				& 0.054\textcolor{blue}{(0.003)} & 0.048\textcolor{blue}{(0.003)} & 0.053\textcolor{blue}{(0.003)} 
				& 0.203\textcolor{blue}{(0.006)} & 0.105\textcolor{blue}{(0.004)} & 0.060\textcolor{blue}{(0.003)}\\ 
				& $\textbf{0.10}$ & 0.151\textcolor{blue}{(0.005)} & 0.115\textcolor{blue}{(0.005)} & 0.104\textcolor{blue}{(0.004)} 
				& 0.110\textcolor{blue}{(0.004)} & 0.101\textcolor{blue}{(0.004)} & 0.106\textcolor{blue}{(0.004)} 
				& 0.294\textcolor{blue}{(0.006)} & 0.170\textcolor{blue}{(0.005)} & 0.122\textcolor{blue}{(0.005)}\\
				\midrule 
				
				& $\textbf{0.01}$ & 0.013\textcolor{blue}{(0.002)} & 0.009\textcolor{blue}{(0.001)} & 0.012\textcolor{blue}{(0.002)} 
				& 0.016\textcolor{blue}{(0.002)} & 0.014\textcolor{blue}{(0.002)} & 0.010\textcolor{blue}{(0.001)} 
				& 0.097\textcolor{blue}{(0.004)} & 0.036\textcolor{blue}{(0.003)} & 0.012\textcolor{blue}{(0.002)}\\ 
				\textbf{H1}& $\textbf{0.05}$ & 0.076\textcolor{blue}{(0.004)} & 0.055\textcolor{blue}{(0.003)} & 0.055\textcolor{blue}{(0.003)} 
				& 0.062\textcolor{blue}{(0.003)} & 0.052\textcolor{blue}{(0.003)} & 0.052\textcolor{blue}{(0.003)} 
				& 0.208\textcolor{blue}{(0.006)} & 0.111\textcolor{blue}{(0.004)} & 0.059\textcolor{blue}{(0.003)}\\ 
				& $\textbf{0.10}$ & 0.154\textcolor{blue}{(0.005)} & 0.120\textcolor{blue}{(0.005)} & 0.105\textcolor{blue}{(0.004)} 
				& 0.114\textcolor{blue}{(0.004)} & 0.107\textcolor{blue}{(0.004)} & 0.102\textcolor{blue}{(0.004)} 
				& 0.297\textcolor{blue}{(0.006)} & 0.185\textcolor{blue}{(0.005)} & 0.112\textcolor{blue}{(0.004)}\\ 
				\midrule 
				
				& $\textbf{0.01}$ & 0.002\textcolor{blue}{(0.001)} & 0.005\textcolor{blue}{(0.001)} & 0.011\textcolor{blue}{(0.001)} 
				& 0.015\textcolor{blue}{(0.002)} & 0.011\textcolor{blue}{(0.001)} & 0.011\textcolor{blue}{(0.001)} 
				& 0.087\textcolor{blue}{(0.004)} & 0.038\textcolor{blue}{(0.003)} & 0.017\textcolor{blue}{(0.002)}\\ 
				\textbf{M1}& $\textbf{0.05}$ & 0.034\textcolor{blue}{(0.003)} & 0.038\textcolor{blue}{(0.003)} & 0.049\textcolor{blue}{(0.003)} 
				& 0.065\textcolor{blue}{(0.003)} & 0.052\textcolor{blue}{(0.003)} & 0.051\textcolor{blue}{(0.003)} 
				& 0.202\textcolor{blue}{(0.006)} & 0.114\textcolor{blue}{(0.004)} & 0.058\textcolor{blue}{(0.003)}\\ 
				& $\textbf{0.10}$ & 0.099\textcolor{blue}{(0.004)} & 0.092\textcolor{blue}{(0.004)} & 0.097\textcolor{blue}{(0.004)} 
				& 0.122\textcolor{blue}{(0.005)} & 0.107\textcolor{blue}{(0.004)} & 0.103\textcolor{blue}{(0.004)} 
				& 0.291\textcolor{blue}{(0.006)} & 0.180\textcolor{blue}{(0.005)} & 0.112\textcolor{blue}{(0.004)}\\
				\midrule 
				
				& $\textbf{0.01}$ & 0.012\textcolor{blue}{(0.002)} & 0.008\textcolor{blue}{(0.001)} & 0.008\textcolor{blue}{(0.001)} 
				& 0.013\textcolor{blue}{(0.002)} & 0.008\textcolor{blue}{(0.001)} & 0.010\textcolor{blue}{(0.001)} 
				& 0.084\textcolor{blue}{(0.004)} & 0.038\textcolor{blue}{(0.003)} & 0.013\textcolor{blue}{(0.002)}\\ 
				\textbf{Y1}& $\textbf{0.05}$ & 0.074\textcolor{blue}{(0.004)} & 0.053\textcolor{blue}{(0.003)} & 0.055\textcolor{blue}{(0.003)} 
				& 0.057\textcolor{blue}{(0.003)} & 0.046\textcolor{blue}{(0.003)} & 0.046\textcolor{blue}{(0.003)} 
				& 0.199\textcolor{blue}{(0.006)} & 0.116\textcolor{blue}{(0.005)} & 0.059\textcolor{blue}{(0.003)}\\ 
				& $\textbf{0.10}$ & 0.160\textcolor{blue}{(0.005)} & 0.124\textcolor{blue}{(0.005)} & 0.103\textcolor{blue}{(0.004)} 
				& 0.107\textcolor{blue}{(0.004)} & 0.097\textcolor{blue}{(0.004)} & 0.093\textcolor{blue}{(0.004)} 
				& 0.290\textcolor{blue}{(0.006)} & 0.183\textcolor{blue}{(0.005)} & 0.111\textcolor{blue}{(0.004)}\\
				\midrule 
				\bottomrule 
				
			\end{tabular}
		}
	\end{sidewaystable}

	% % % % % % % % % % % % % % % % % % % % % % % % % % % % % % % % %
	% % % % % % % % % % % % % % % % % % % % % % % % % % % % % % % % %
	\bigskip
	\bigskip
	\begin{center}
		\begin{minipage}{2in} 
			\framebox[2\width]{Table S1 about here.}
		\end{minipage} 
	\end{center}
	
	\begin{figure}[htp]
		\centering
		\includegraphics[width=.48\textwidth]{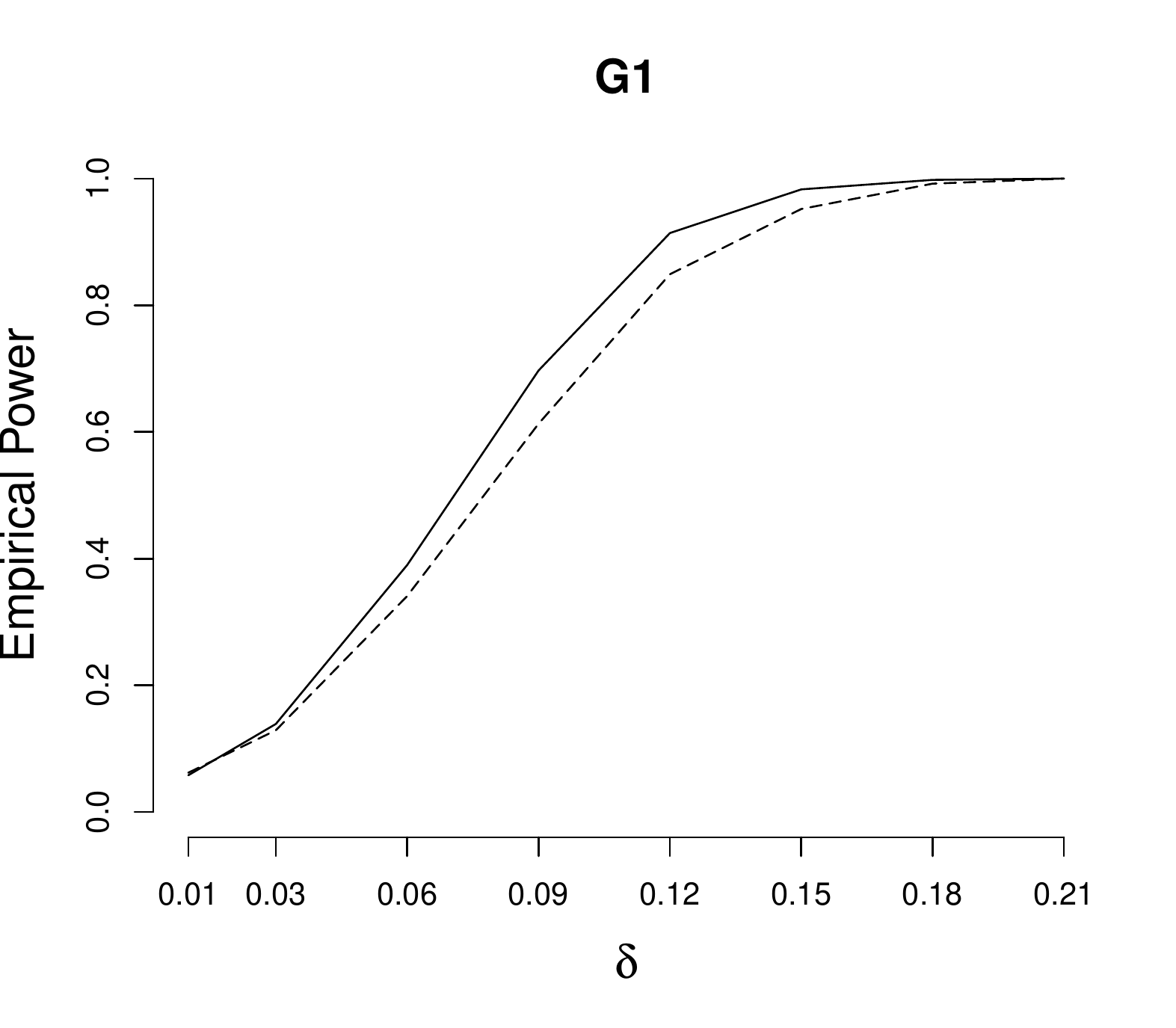}
		\includegraphics[width=.48\textwidth]{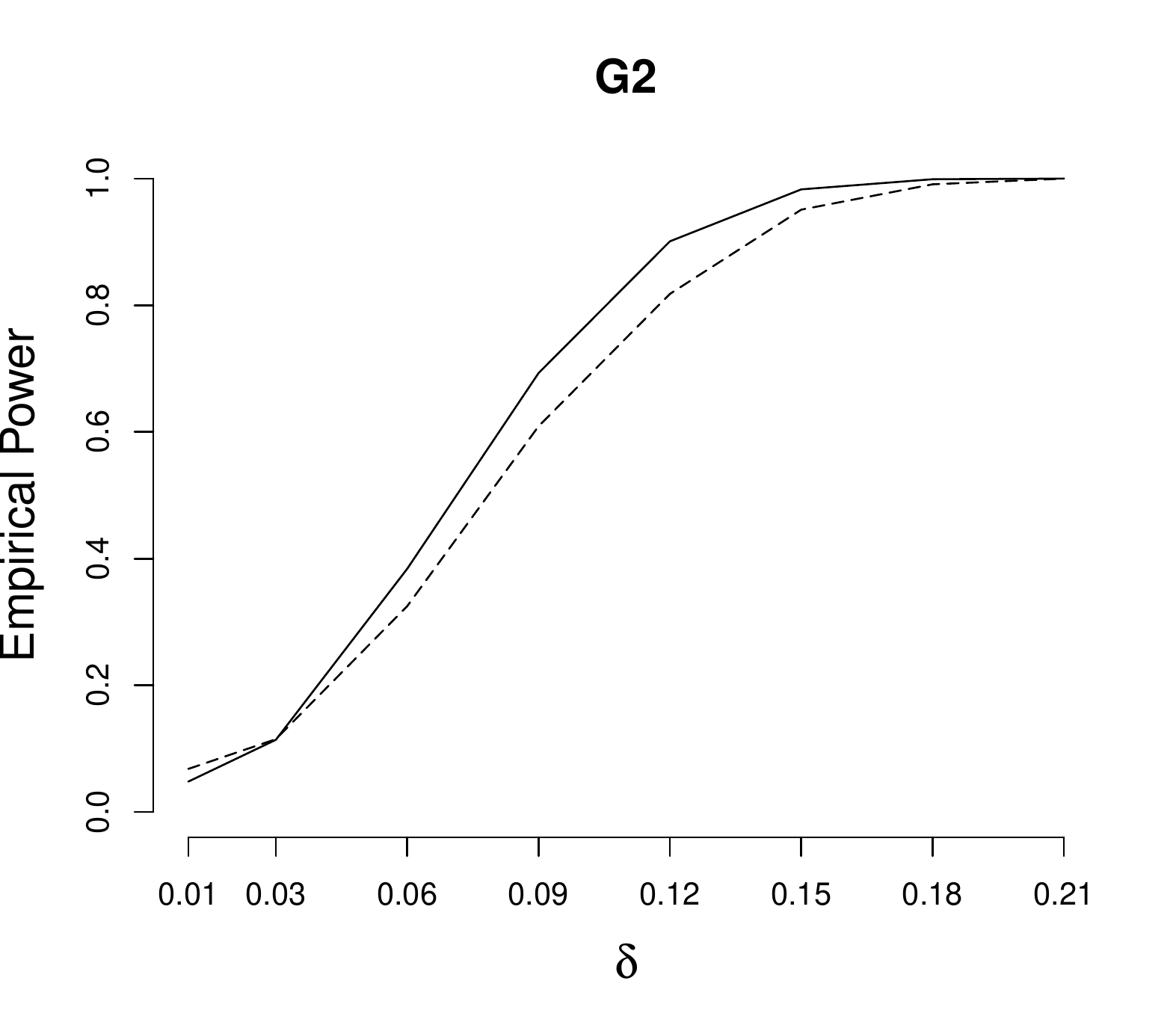}
		
		\medskip
		
		\includegraphics[width=.48\textwidth]{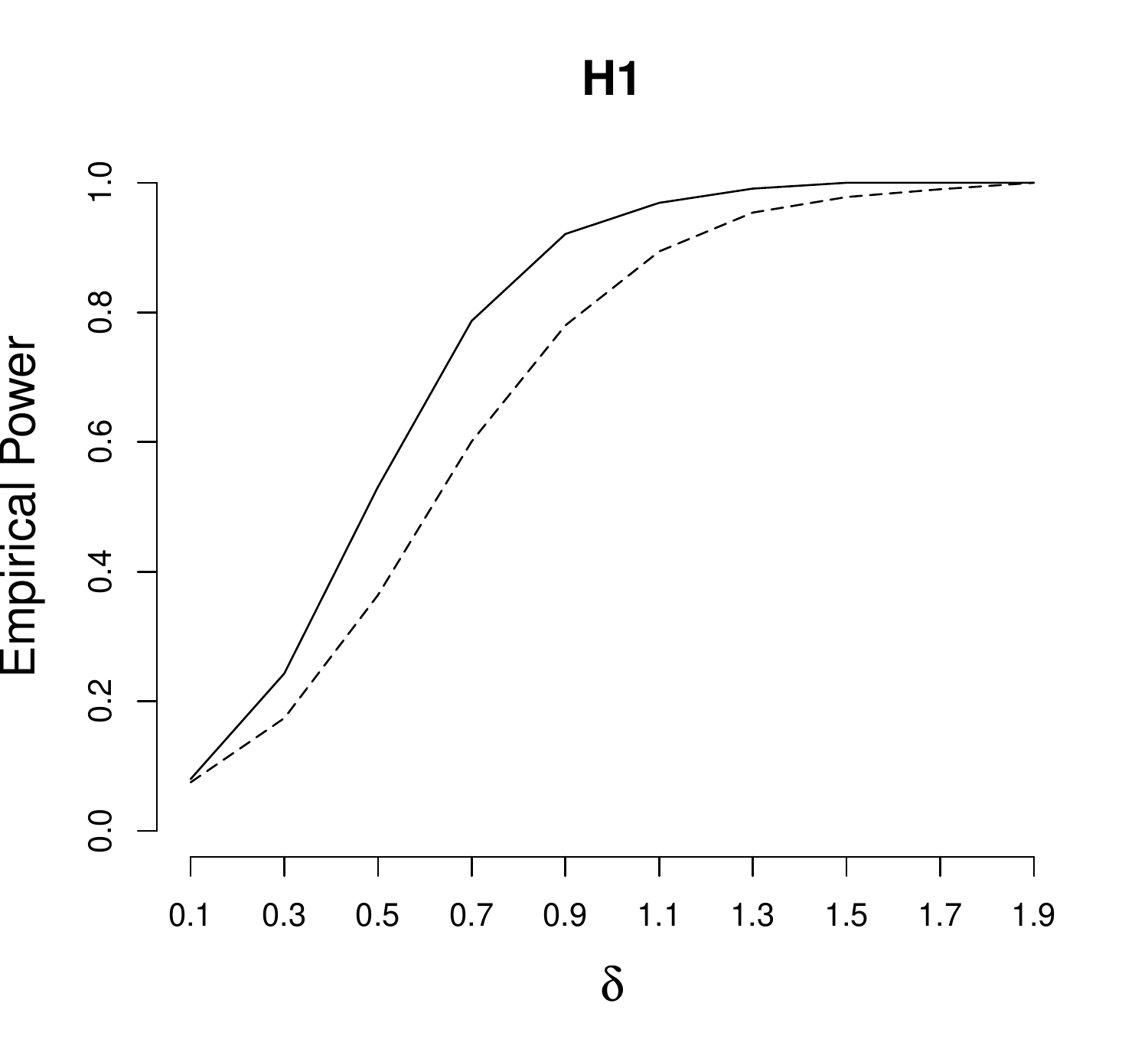}
		\includegraphics[width=.48\textwidth]{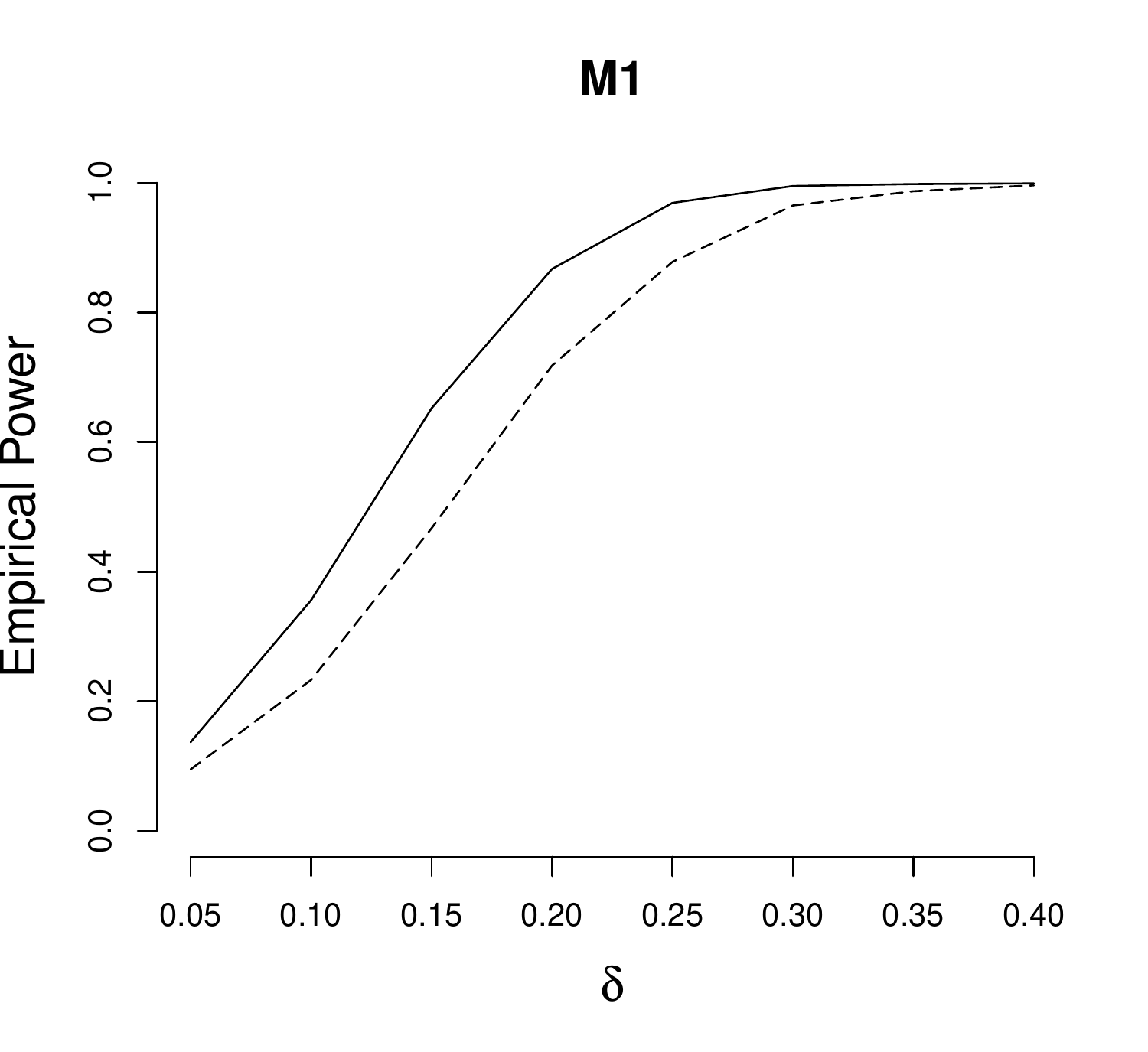}
		
		\medskip
		
		\includegraphics[width=.48\textwidth]{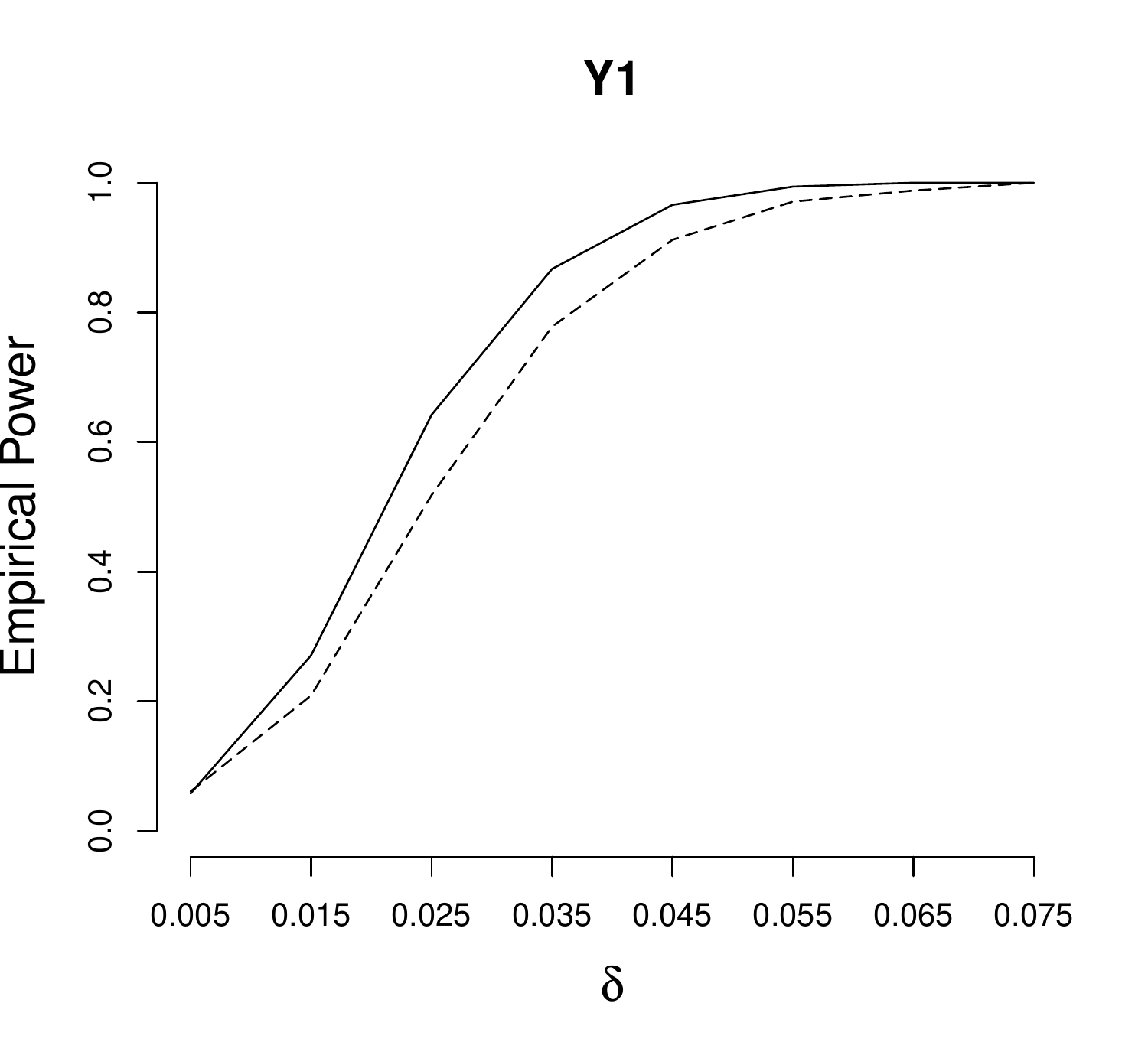}
		\renewcommand{\baselinestretch}{1.2}
		\caption{Empirical power of the competing GGF and MHR methods for testing linear effect 
			for the dense sampling design. \textit{Solid lines} indicate results for the MHR method 
			and \textit{dashed lines} indicate results for the GGF method. The significance level is 
			$\alpha=0.05$. The number of Monte Carlo experiments is 1,000, and the sample size is $n=100$.}
		\label{plot_lin_dns_100}
	\end{figure}
	% % % % % % % % % % % % % % % % % % % % % % % % % % % % % % % % % % % % % % % % % % % % % % % % % %
	% % % % % % % % % % % % % % % % % % % % % % % % % % % % % % % % % % % % % % % % % % % % % % % % % %
	\begin{figure}[htp]
		\centering
		\includegraphics[width=.48\textwidth]{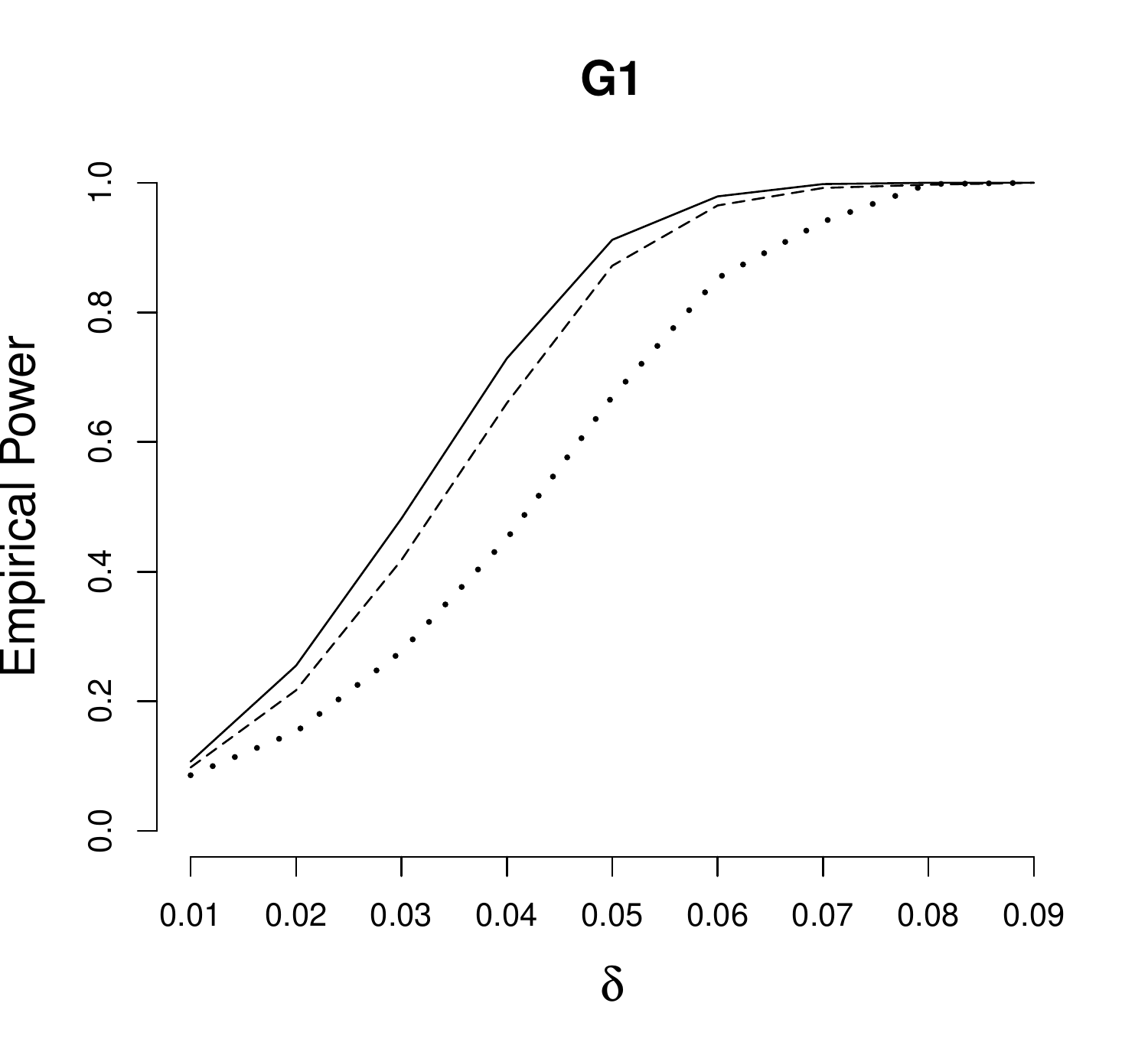}\quad
		\includegraphics[width=.48\textwidth]{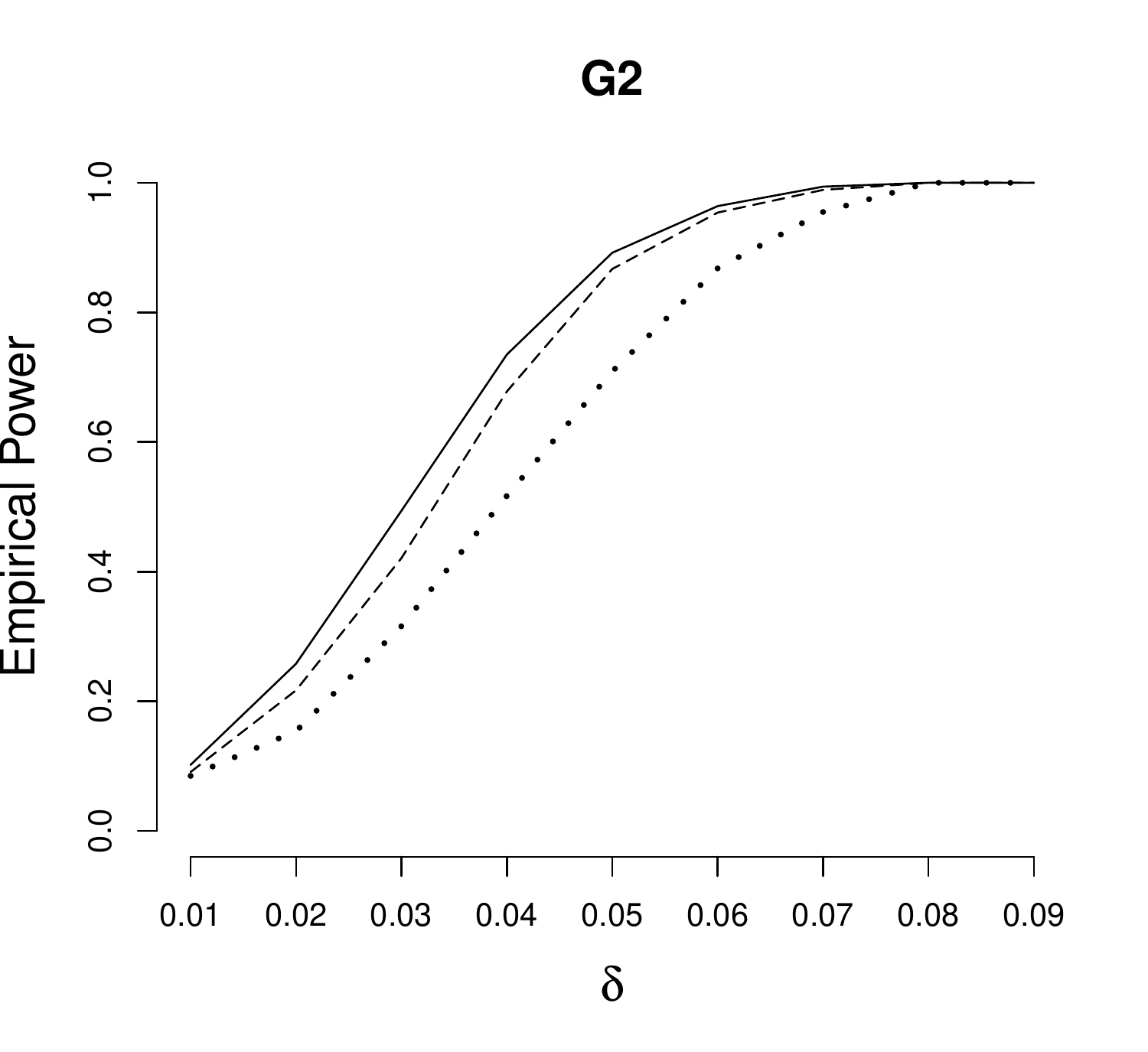}
		
		\medskip
		
		\includegraphics[width=.48\textwidth]{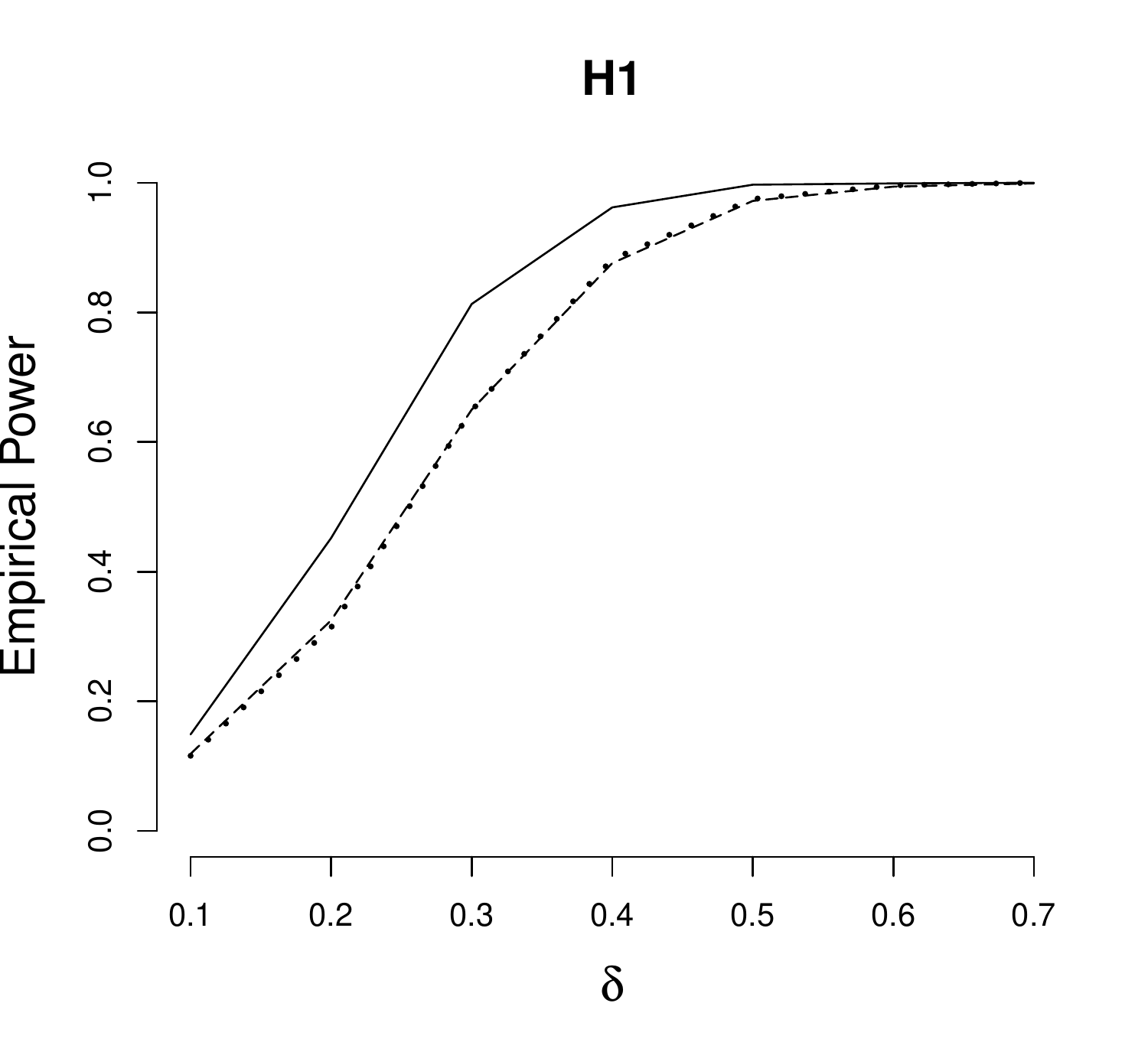}\quad
		\includegraphics[width=.48\textwidth]{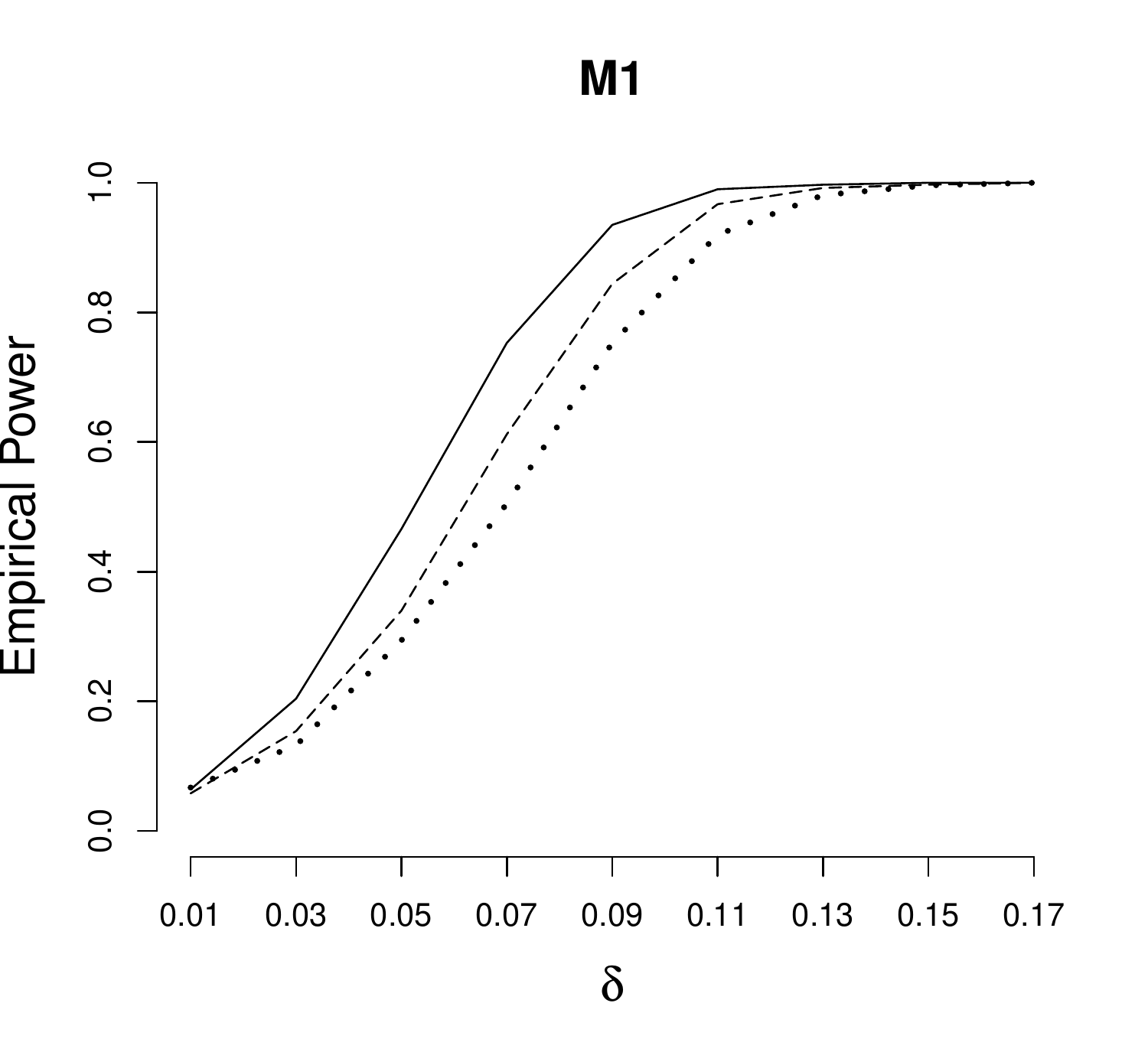}
		
		\medskip
		
		\includegraphics[width=.48\textwidth]{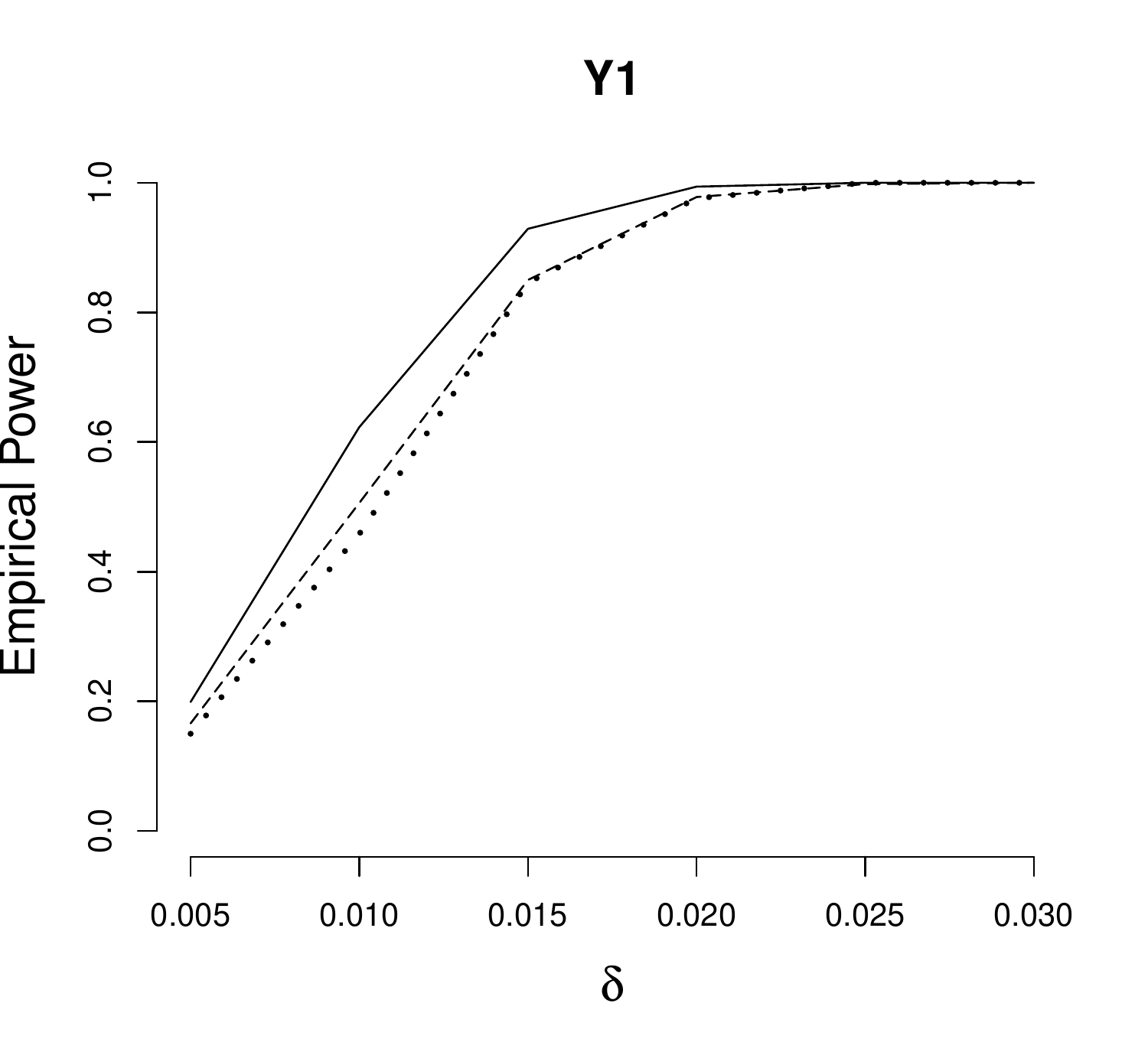}\quad
		\renewcommand{\baselinestretch}{1.2}
		\caption{Empirical power of the competing GGF, MHR, and HR methods for testing linear effect 
			for the dense sampling design. \textit{Solid lines} indicate results for the MHR method, 
			\textit{dashed lines} indicate results for the GGF method, and \textit{dotted lines} indicate 
			results for the HR method. The significance level is $\alpha=0.05$. The number of Monte Carlo 
			experiments is 1,000, and the sample size is $n=500$.}
		\label{plot_lin_dns_500}
	\end{figure}
	
	\newpage
	\subsection{\textbf{Power curves for moderate sampling design}}
	
	% % % % % % % % % % % % % % % % % % % % % % % % % % % % % % % % % % % % % % % % % % % % % % % % % %
	% % % % % % % % % % % % % % % % % % % % % % % % % % % % % % % % % % % % % % % % % % % % % % % % % %
	\begin{figure}[htp]
		\centering
		\includegraphics[width=.48\textwidth]{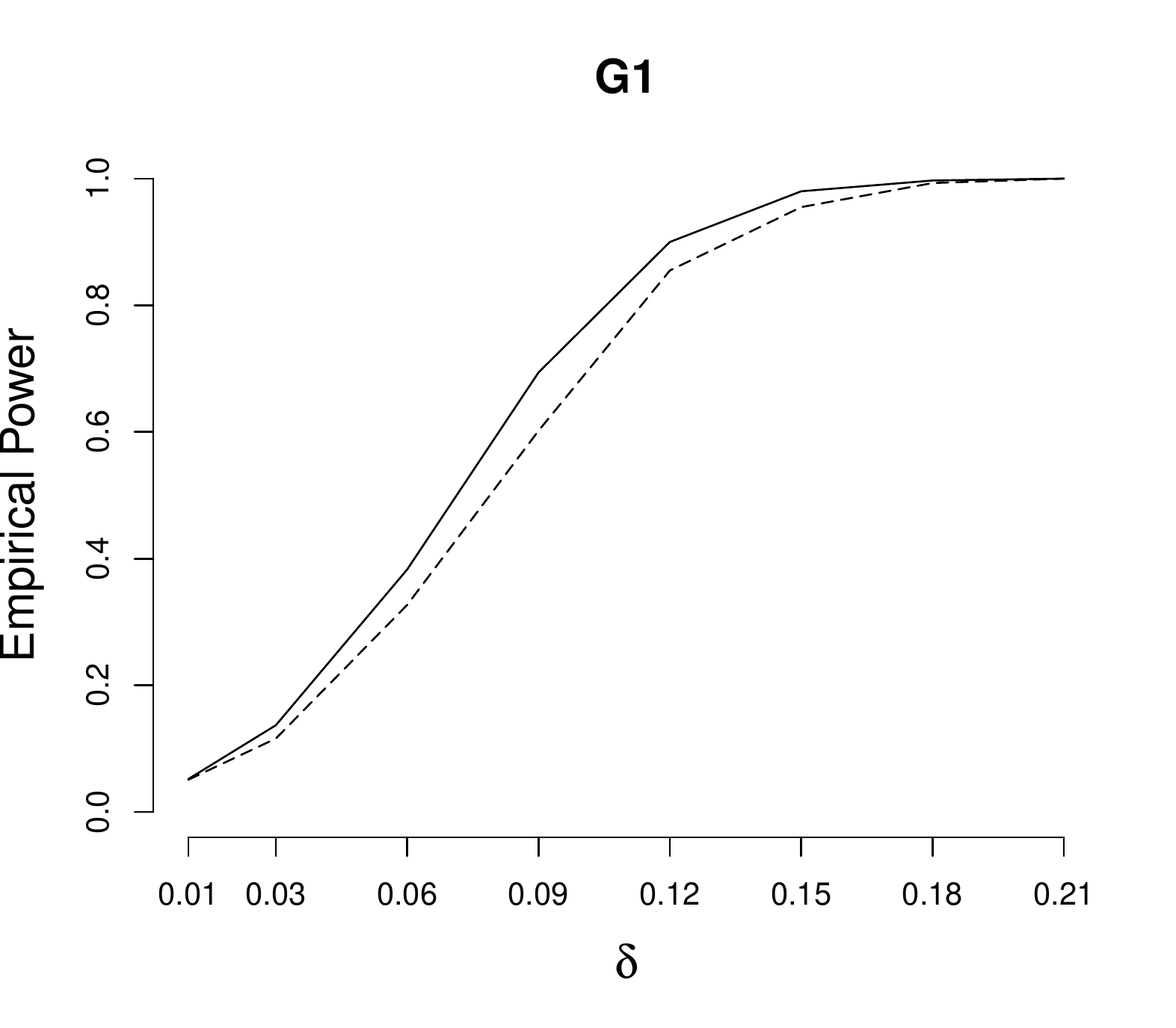}\quad
		\includegraphics[width=.48\textwidth]{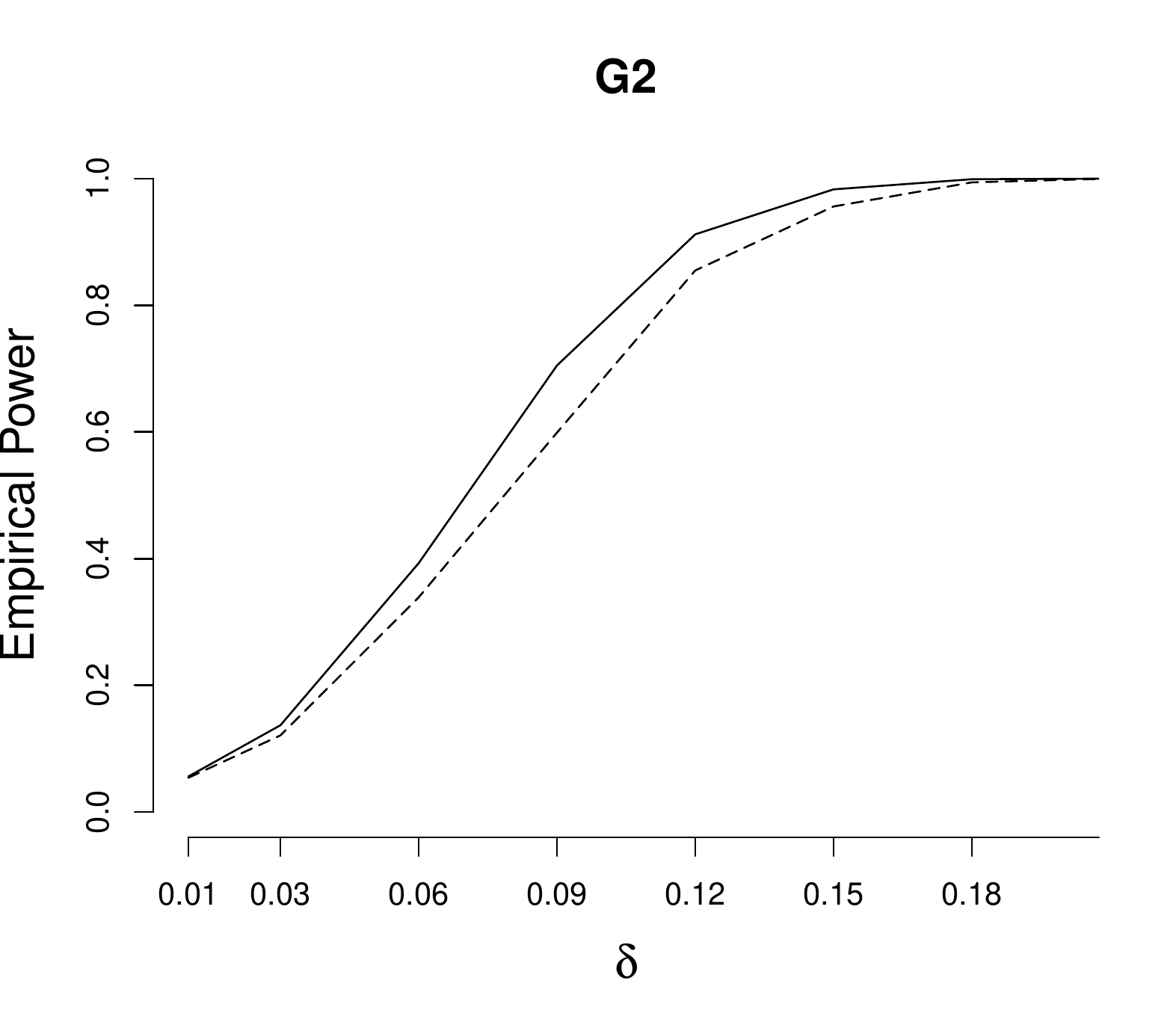}
		
		\medskip
		
		\includegraphics[width=.48\textwidth]{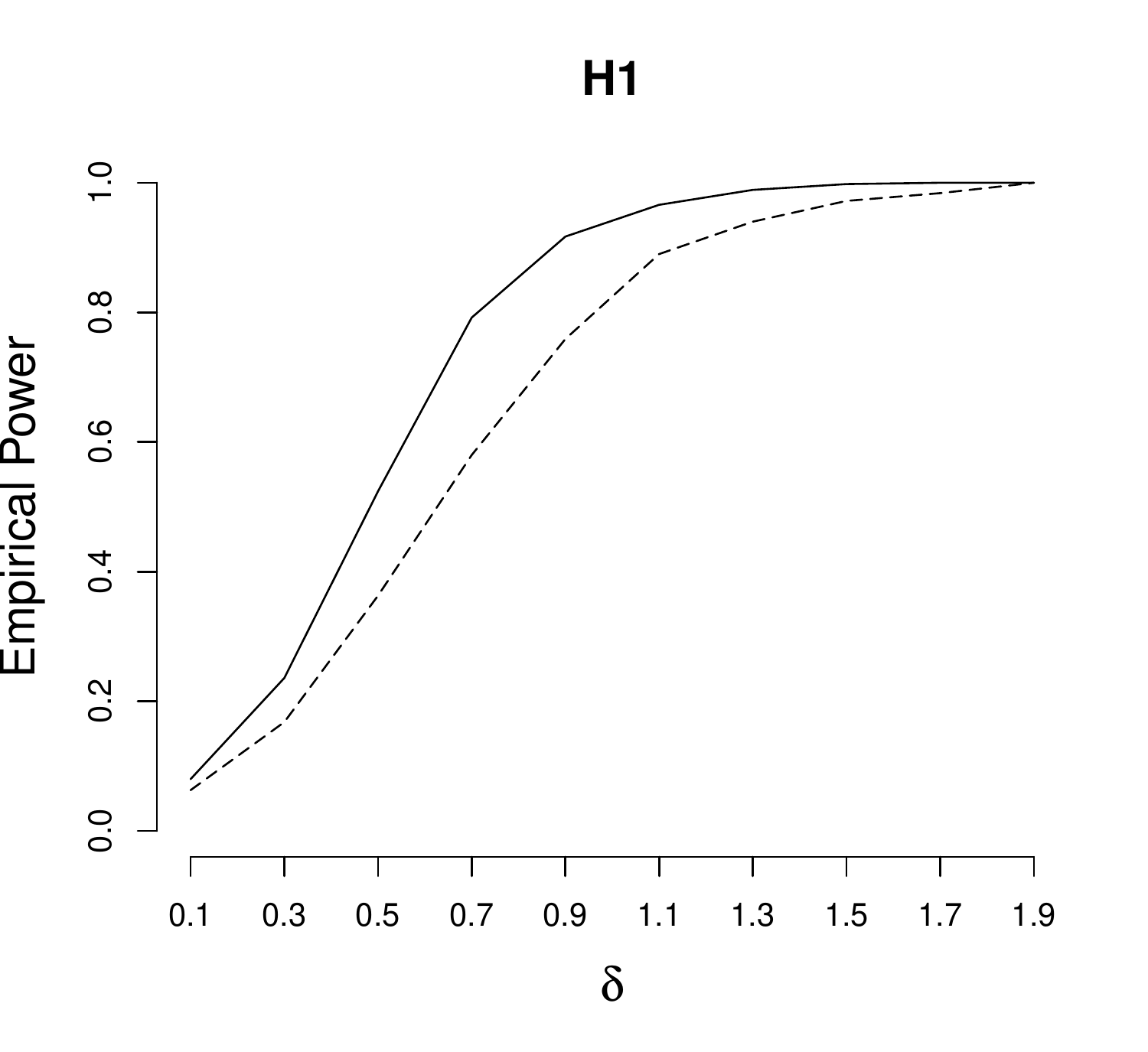}\quad
		\includegraphics[width=.48\textwidth]{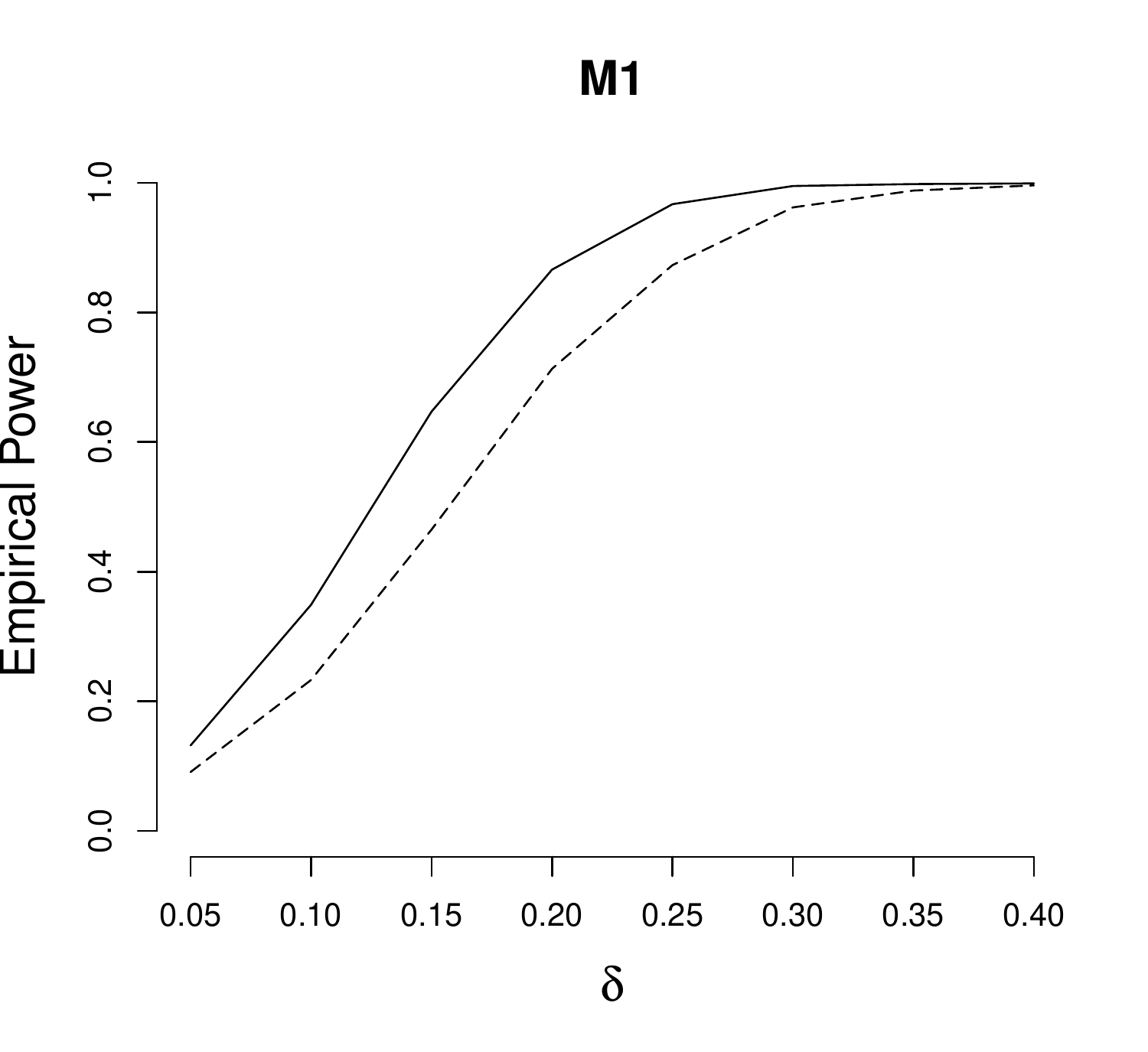}
		
		\medskip
		
		\includegraphics[width=.56\textwidth]{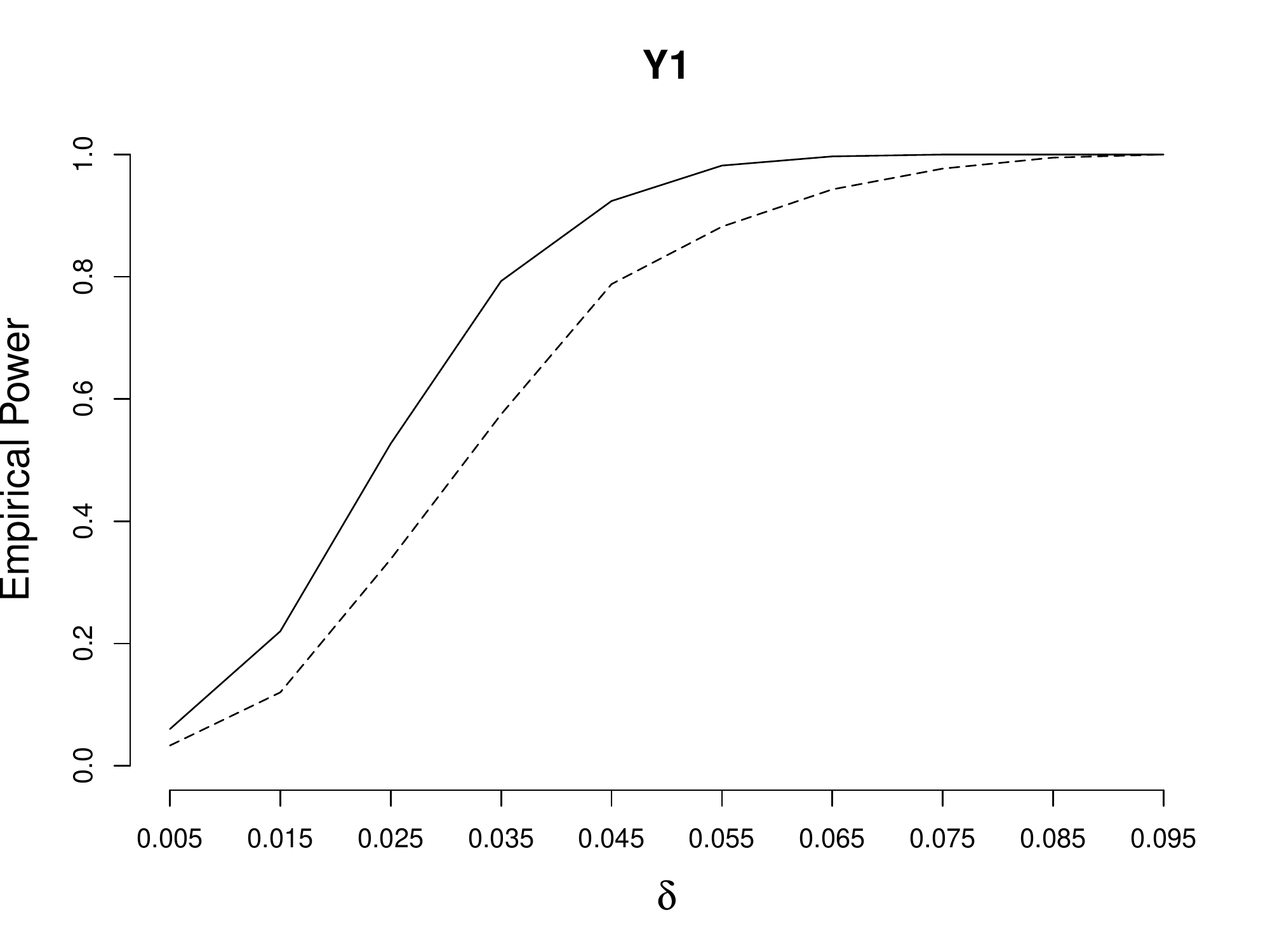}\quad
		\renewcommand{\baselinestretch}{1.2}
		\caption{Empirical power of the competing GGF and MHR methods for testing linear effect 
			for the moderately sparse sampling design. \textit{Solid lines} indicate results for 
			the MHR method and \textit{dashed lines} indicate results for the GGF method. 
			The significance level is $\alpha=0.05$. The number of Monte Carlo experiments is 
			1,000, and the sample size is $n=100$.}
		\label{plot_lin_mod_100}
	\end{figure}
	
	% % % % % % % % % % % % % % % % % % % % % % % % % % % % % % % % % % % % % % % % % % % % % % % % % %
	% % % % % % % % % % % % % % % % % % % % % % % % % % % % % % % % % % % % % % % % % % % % % % % % % %
	\newpage
	\subsection{\textbf{Type I error rates and power curves for sparse sampling design}}
	
	\renewcommand{\arraystretch}{0.75}
	\begin{sidewaystable}[!t]\centering 
		\renewcommand{\baselinestretch}{1.3}
		\caption{Testing linearity: Comparison of the estimated Type I error rates of the GGF, MHR, and HR methods 
			in the context of sparse functional data. The data generation settings are G1, G2, H1, M1, and Y1. The 
			number of Monte Carlo experiments is 5,000, and the sample sizes are 50, 100, and 500. Standard errors 
			are shown in parentheses.}
		\renewcommand{\baselinestretch}{1.8}
		\scalebox{0.8}{
			\begin{tabular}{ccccc|ccc|ccc} 
				
				\toprule 
				\toprule
				& & \multicolumn{3}{c}{\textbf{GGF}} & \multicolumn{3}{c}{\textbf{MHR}} & \multicolumn{3}{c}{\textbf{HR}} \\ 
				\cmidrule(r){3-5} \cmidrule(r){6-8} \cmidrule{9-11}
				& $\boldsymbol{\alpha}$ & $\textbf{n=50}$ & $\textbf{n=100}$ & $\textbf{n=500}$  
				& $\textbf{n=50}$ & $\textbf{n=100}$ & $\textbf{n=500}$  & $\textbf{n=50}$ & $\textbf{n=100}$ & $\textbf{n=500}$ \\ 
				\midrule 
				
				& $\textbf{0.01}$ & 0.006\textcolor{blue}{(0.001)} & 0.008\textcolor{blue}{(0.001)} & 0.009\textcolor{blue}{(0.001)} 
				& 0.015\textcolor{blue}{(0.002)} & 0.013\textcolor{blue}{(0.002)} & 0.011\textcolor{blue}{(0.001)} 
				& 0.105\textcolor{blue}{(0.004)} & 0.039\textcolor{blue}{(0.003)} & 0.012\textcolor{blue}{(0.002)}\\ 
				\textbf{G1}& $\textbf{0.05}$ & 0.045\textcolor{blue}{(0.003)} & 0.047\textcolor{blue}{(0.003)} & 0.051\textcolor{blue}{(0.003)} 
				& 0.059\textcolor{blue}{(0.003)} & 0.049\textcolor{blue}{(0.003)} & 0.056\textcolor{blue}{(0.003)} 
				& 0.224\textcolor{blue}{(0.006)} & 0.120\textcolor{blue}{(0.005)} & 0.060\textcolor{blue}{(0.003)}\\ 
				& $\textbf{0.10}$ & 0.110\textcolor{blue}{(0.004)} & 0.106\textcolor{blue}{(0.004)} & 0.102\textcolor{blue}{(0.004)} 
				& 0.116\textcolor{blue}{(0.005)} & 0.097\textcolor{blue}{(0.004)} & 0.104\textcolor{blue}{(0.004)} 
				& 0.315\textcolor{blue}{(0.007)} & 0.191\textcolor{blue}{(0.006)} & 0.117\textcolor{blue}{(0.005)}\\
				\midrule 
				
				& $\textbf{0.01}$ & 0.004\textcolor{blue}{(0.001)} & 0.007\textcolor{blue}{(0.001)} & 0.009\textcolor{blue}{(0.001)} 
				& 0.013\textcolor{blue}{(0.002)} & 0.012\textcolor{blue}{(0.002)} & 0.013\textcolor{blue}{(0.002)} 
				& 0.097\textcolor{blue}{(0.004)} & 0.036\textcolor{blue}{(0.003)} & 0.013\textcolor{blue}{(0.002)}\\ 
				\textbf{G2}& $\textbf{0.05}$ & 0.038\textcolor{blue}{(0.003)} & 0.053\textcolor{blue}{(0.003)} & 0.049\textcolor{blue}{(0.003)} 
				& 0.056\textcolor{blue}{(0.003)} & 0.054\textcolor{blue}{(0.003)} & 0.053\textcolor{blue}{(0.003)} 
				& 0.213\textcolor{blue}{(0.006)} & 0.117\textcolor{blue}{(0.005)} & 0.063\textcolor{blue}{(0.003)}\\ 
				& $\textbf{0.10}$ & 0.101\textcolor{blue}{(0.004)} & 0.108\textcolor{blue}{(0.004)} & 0.103\textcolor{blue}{(0.004)} 
				& 0.117\textcolor{blue}{(0.005)} & 0.104\textcolor{blue}{(0.004)} & 0.105\textcolor{blue}{(0.004)} 
				& 0.310\textcolor{blue}{(0.007)} & 0.190\textcolor{blue}{(0.006)} & 0.119\textcolor{blue}{(0.005)}\\
				\midrule 
				
				& $\textbf{0.01}$ & 0.006\textcolor{blue}{(0.001)} & 0.008\textcolor{blue}{(0.001)} & 0.012\textcolor{blue}{(0.002)} 
				& 0.015\textcolor{blue}{(0.002)} & 0.013\textcolor{blue}{(0.002)} & 0.012\textcolor{blue}{(0.002)} 
				& 0.094\textcolor{blue}{(0.004)} & 0.036\textcolor{blue}{(0.003)} & 0.013\textcolor{blue}{(0.002)}\\ 
				\textbf{H1}& $\textbf{0.05}$ & 0.050\textcolor{blue}{(0.003)} & 0.049\textcolor{blue}{(0.003)} & 0.052\textcolor{blue}{(0.003)} 
				& 0.061\textcolor{blue}{(0.003)} & 0.055\textcolor{blue}{(0.003)} & 0.051\textcolor{blue}{(0.003)} 
				& 0.218\textcolor{blue}{(0.006)} & 0.113\textcolor{blue}{(0.004)} & 0.057\textcolor{blue}{(0.003)}\\ 
				& $\textbf{0.10}$ & 0.117\textcolor{blue}{(0.005)} & 0.105\textcolor{blue}{(0.004)} & 0.100\textcolor{blue}{(0.004)} 
				& 0.119\textcolor{blue}{(0.005)} & 0.108\textcolor{blue}{(0.004)} & 0.100\textcolor{blue}{(0.004)} 
				& 0.312\textcolor{blue}{(0.007)} & 0.182\textcolor{blue}{(0.005)} & 0.115\textcolor{blue}{(0.005)}\\
				\midrule 
				
				& $\textbf{0.01}$ & 0.035\textcolor{blue}{(0.003)} & 0.069\textcolor{blue}{(0.004)} & 0.155\textcolor{blue}{(0.005)} 
				& 0.108\textcolor{blue}{(0.004)} & 0.169\textcolor{blue}{(0.005)} & 0.293\textcolor{blue}{(0.006)} 
				& 0.170\textcolor{blue}{(0.005)} & 0.110\textcolor{blue}{(0.004)} & 0.085\textcolor{blue}{(0.004)}\\ 
				\textbf{M1}& $\textbf{0.05}$ & 0.101\textcolor{blue}{(0.004)} & 0.135\textcolor{blue}{(0.005)} & 0.218\textcolor{blue}{(0.006)} 
				& 0.181\textcolor{blue}{(0.005)} & 0.230\textcolor{blue}{(0.006)} & 0.335\textcolor{blue}{(0.007)} 
				& 0.297\textcolor{blue}{(0.006)} & 0.203\textcolor{blue}{(0.006)} & 0.150\textcolor{blue}{(0.005)}\\ 
				& $\textbf{0.10}$ & 0.189\textcolor{blue}{(0.006)} & 0.208\textcolor{blue}{(0.006)} & 0.278\textcolor{blue}{(0.006)} 
				& 0.239\textcolor{blue}{(0.006)} & 0.285\textcolor{blue}{(0.006)} & 0.378\textcolor{blue}{(0.007)} 
				& 0.389\textcolor{blue}{(0.007)} & 0.280\textcolor{blue}{(0.006)} & 0.214\textcolor{blue}{(0.006)}\\
				\midrule 
				
				& $\textbf{0.01}$ & 0.006\textcolor{blue}{(0.002)} & 0.008\textcolor{blue}{(0.002)} & 0.008\textcolor{blue}{(0.002)} 
				& 0.013\textcolor{blue}{(0.002)} & 0.014\textcolor{blue}{(0.002)} & 0.008\textcolor{blue}{(0.001)} 
				& 0.105\textcolor{blue}{(0.004)} & 0.038\textcolor{blue}{(0.003)} & 0.010\textcolor{blue}{(0.001)}\\ 
				\textbf{Y1}& $\textbf{0.05}$ & 0.045\textcolor{blue}{(0.004)} & 0.052\textcolor{blue}{(0.004)} & 0.042\textcolor{blue}{(0.004)} 
				& 0.064\textcolor{blue}{(0.003)} & 0.054\textcolor{blue}{(0.003)} & 0.047\textcolor{blue}{(0.003)} 
				& 0.237\textcolor{blue}{(0.006)} & 0.118\textcolor{blue}{(0.005)} & 0.058\textcolor{blue}{(0.003)}\\ 
				& $\textbf{0.10}$ & 0.114\textcolor{blue}{(0.006)} & 0.102\textcolor{blue}{(0.006)} & 0.087\textcolor{blue}{(0.006)} 
				& 0.121\textcolor{blue}{(0.005)} & 0.103\textcolor{blue}{(0.004)} & 0.097\textcolor{blue}{(0.004)} 
				& 0.337\textcolor{blue}{(0.007)} & 0.195\textcolor{blue}{(0.006)} & 0.104\textcolor{blue}{(0.004)}\\
				\midrule 
				\bottomrule 
				
			\end{tabular}
		}
	\end{sidewaystable}
	
	\renewcommand{\arraystretch}{0.75}
	\begin{table}[!t]\centering 
		\renewcommand{\baselinestretch}{1.3}
		\caption{Testing linearity: Comparison of the estimated Type I error rates of the GGF and MHR methods 
			in the context of sparse functional data under the M1 setting. The observation points are sampled 
			per curve without replacement from the discrete uniform distribution $\mbox{Unif}\{9,\ldots,12\}$. 
			The number of Monte Carlo experiments is 5,000, and the sample sizes are 50, 100, and 500. Standard 
			errors are shown in parentheses.}
		\renewcommand{\baselinestretch}{1.8}
		\scalebox{0.8}{
			\begin{tabular}{cccc|ccc} 
				\toprule 
				\toprule
				&\multicolumn{3}{c}{\textbf{GGF}} & \multicolumn{3}{c}{\textbf{MHR}} \\ 
				\cmidrule(r){2-4} \cmidrule(r){5-7} 
				$\boldsymbol{\alpha}$ & $\textbf{n=50}$ & $\textbf{n=100}$ & $\textbf{n=500}$ & $\textbf{n=50}$ & $\textbf{n=100}$ & $\textbf{n=500}$ \\ 
				\midrule 
				
				$\textbf{0.01}$ & 0.003\textcolor{blue}{(0.001)} & 0.006\textcolor{blue}{(0.001)} & 0.011\textcolor{blue}{(0.001)}
				& 0.019\textcolor{blue}{(0.001)} & 0.018\textcolor{blue}{(0.002)} & 0.018\textcolor{blue}{(0.001)} \\ 
				$\textbf{0.05}$ & 0.037\textcolor{blue}{(0.003)} & 0.043\textcolor{blue}{(0.003)} & 0.051\textcolor{blue}{(0.003)} 
				& 0.072\textcolor{blue}{(0.003)} & 0.065\textcolor{blue}{(0.003)} & 0.059\textcolor{blue}{(0.003)} \\ 
				$\textbf{0.10}$ & 0.106\textcolor{blue}{(0.004)} & 0.098\textcolor{blue}{(0.004)} & 0.099\textcolor{blue}{(0.004)} 
				& 0.138\textcolor{blue}{(0.005)} & 0.116\textcolor{blue}{(0.004)} & 0.111\textcolor{blue}{(0.004)} \\
				
				\midrule 
				\bottomrule 
				
			\end{tabular}
		}
	\end{table}
	
	% % % % % % % % % % % % % % % % % % % % % % % % % % % % % % % % % % % % % % % % % % % % % % % % % %
	% % % % % % % % % % % % % % % % % % % % % % % % % % % % % % % % % % % % % % % % % % % % % % % % % %
	
	\bigskip
	\bigskip
	\begin{center}
		\begin{minipage}{2in} 
			\framebox[2\width]{Table S2 about here.}
		\end{minipage} 
	\end{center}
	
	\begin{figure}[htp]
		\centering
		\includegraphics[width=.48\textwidth]{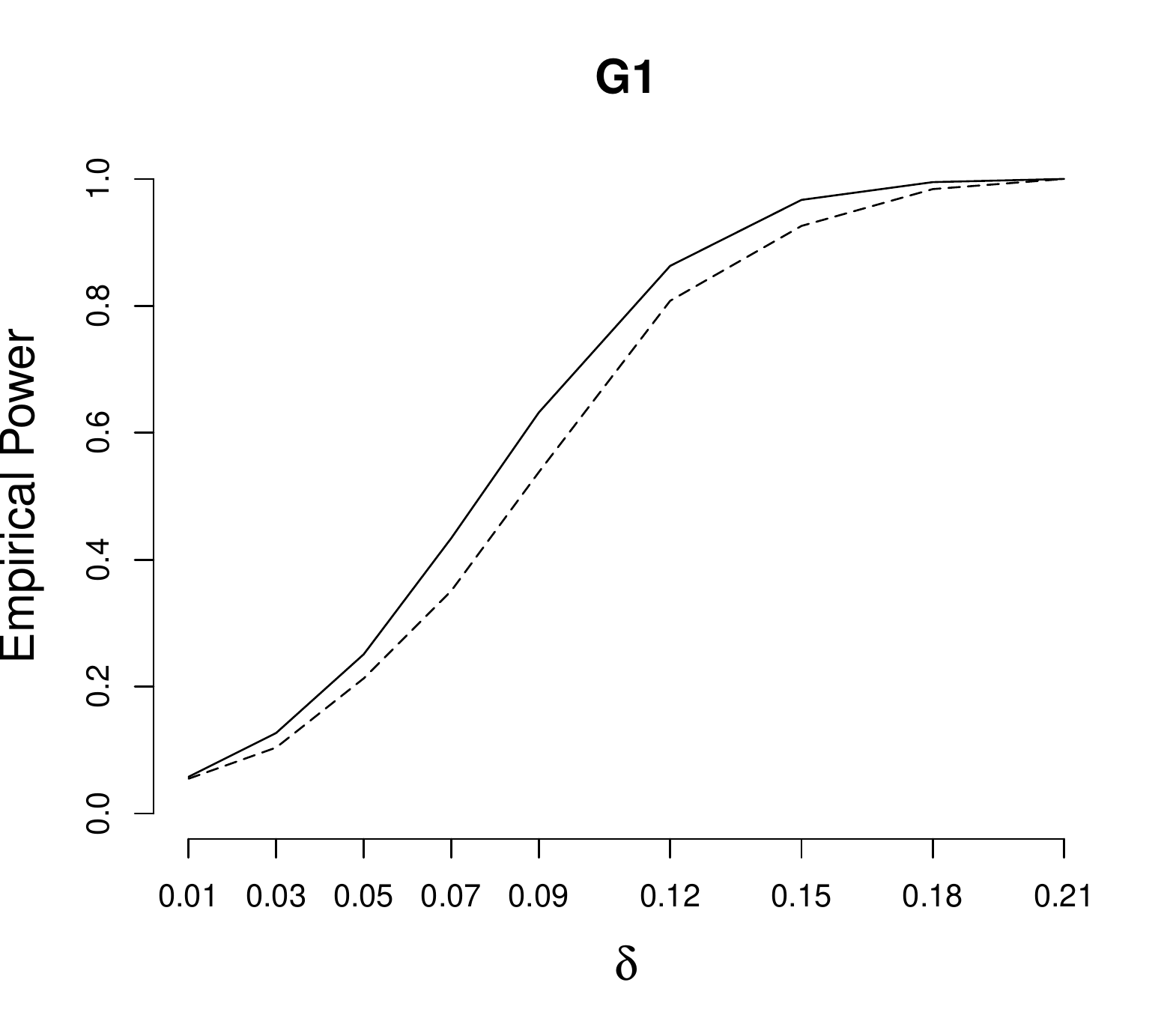}\quad
		\includegraphics[width=.48\textwidth]{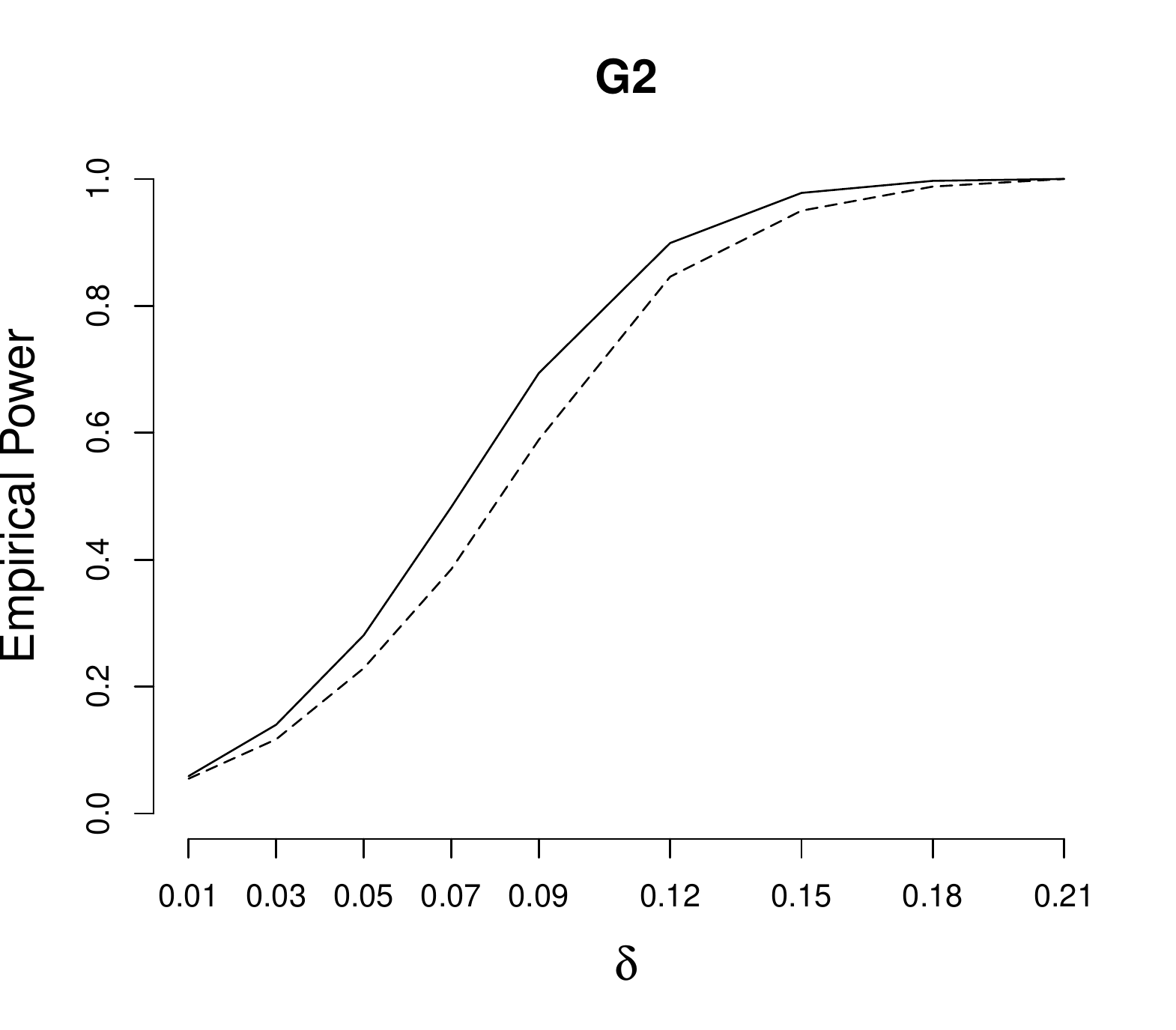}
		
		\medskip
		
		\includegraphics[width=.48\textwidth]{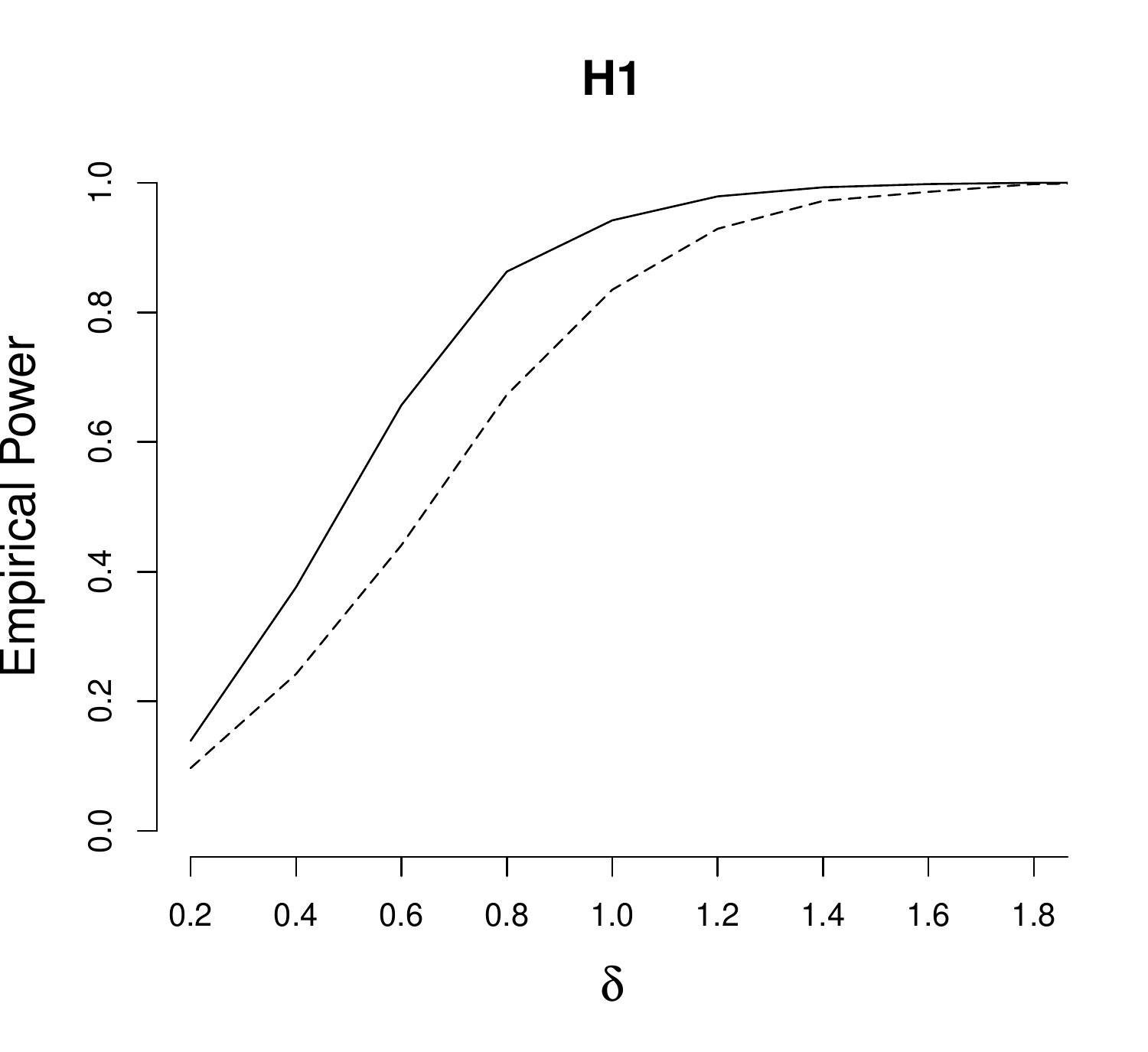}\quad
		\includegraphics[width=.48\textwidth]{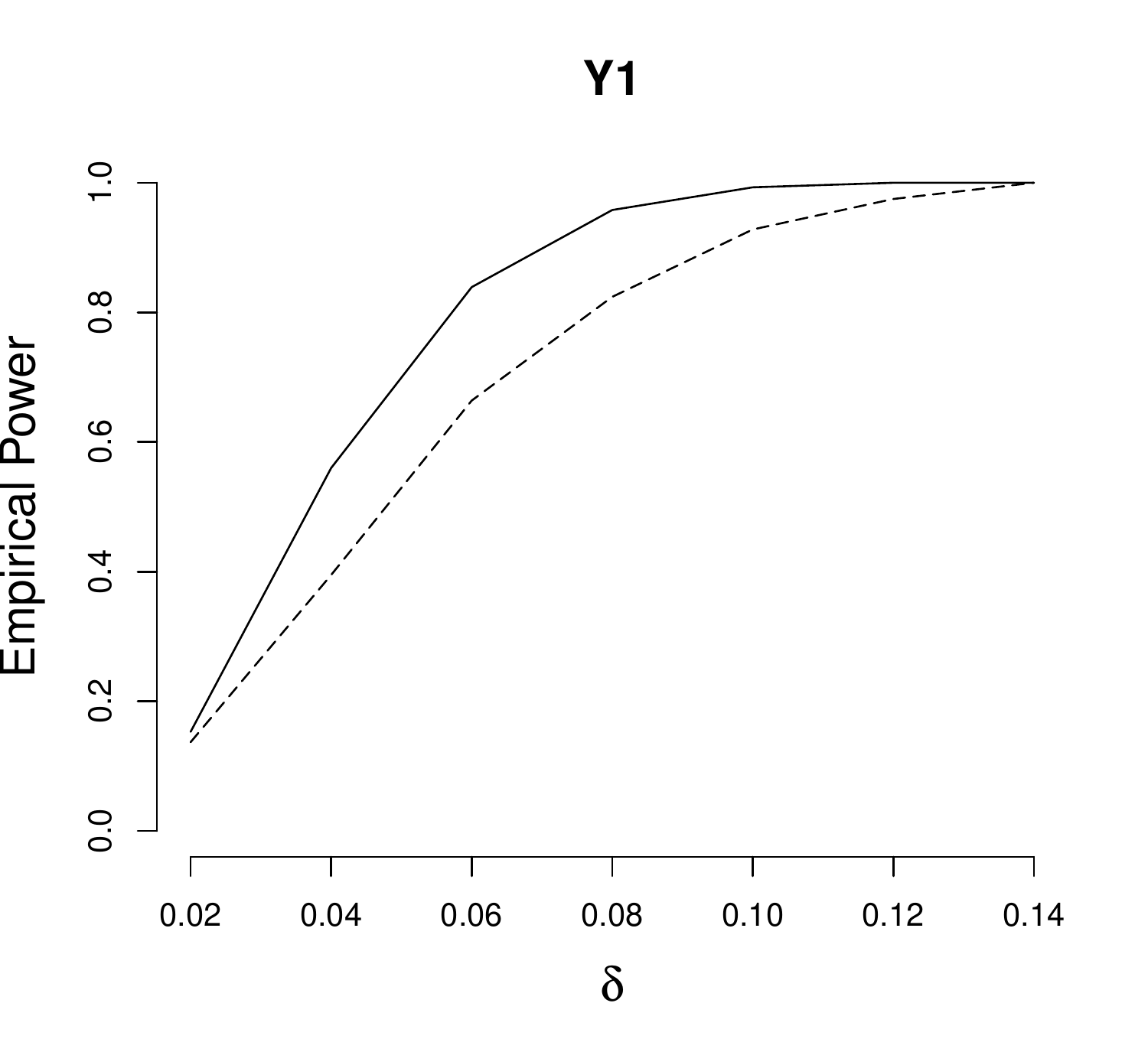}
		\renewcommand{\baselinestretch}{1.2}
		\caption{Empirical power of the competing GGF and MHR methods for testing linear effect 
			for the sparse sampling design. \textit{Solid lines} indicate results for the MHR method 
			and \textit{dashed lines} indicate results for the GGF method. The significance level is 
			$\alpha=0.05$. The number of Monte Carlo experiments is 1,000, and the sample size is $n=100$.}
		\label{plot_lin_spr_100}
	\end{figure}
	% % % % % % % % % % % % % % % % % % % % % % % % % % % % % % % % % % % % % % % % % % % % % % % % % %
	% % % % % % % % % % % % % % % % % % % % % % % % % % % % % % % % % % % % % % % % % % % % % % % % % %
	\begin{figure}[htp]
		\centering
		\includegraphics[width=.48\textwidth]{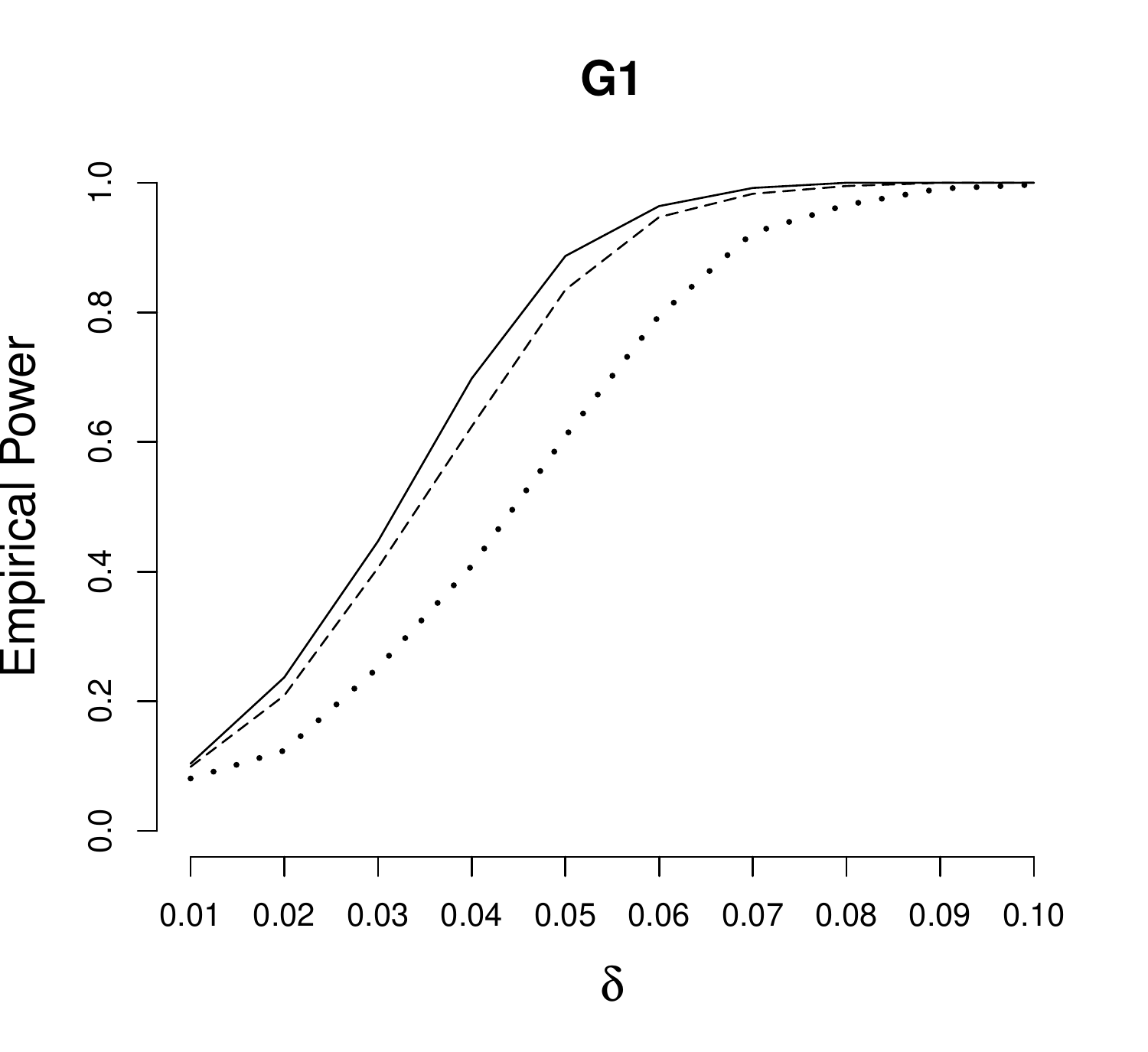}\quad
		\includegraphics[width=.48\textwidth]{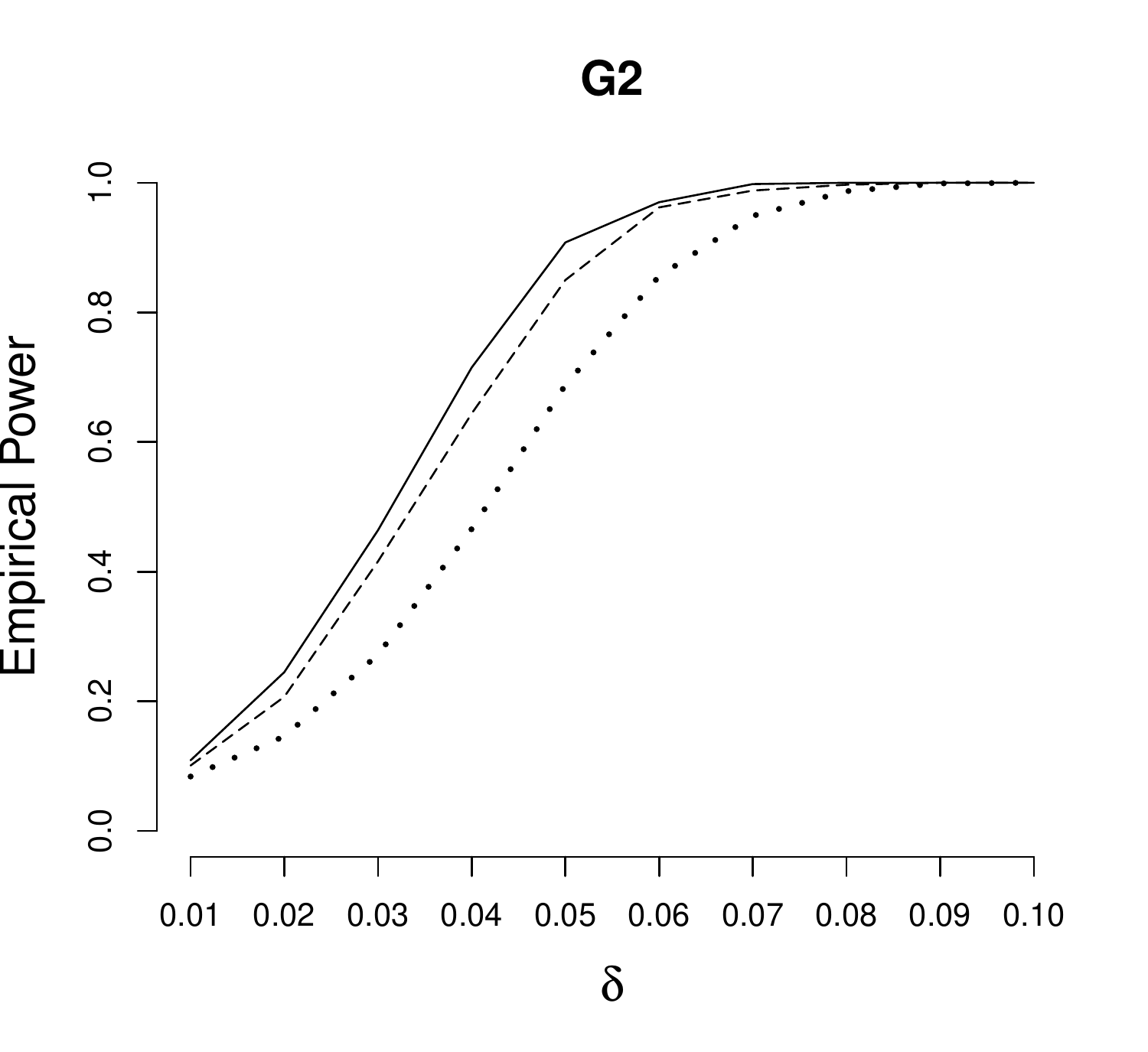}
		
		\medskip
		
		\includegraphics[width=.48\textwidth]{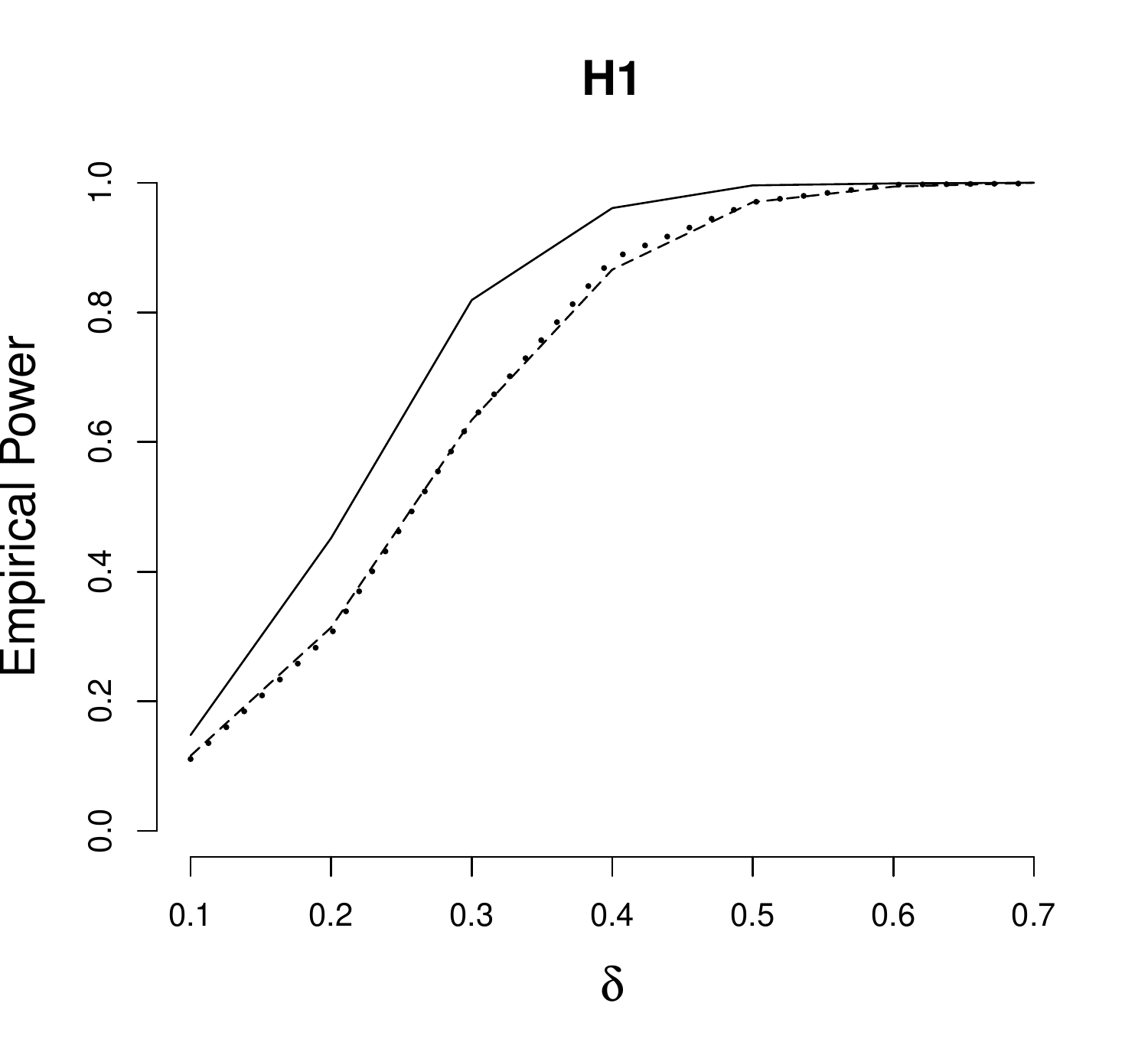}\quad
		\includegraphics[width=.49\textwidth]{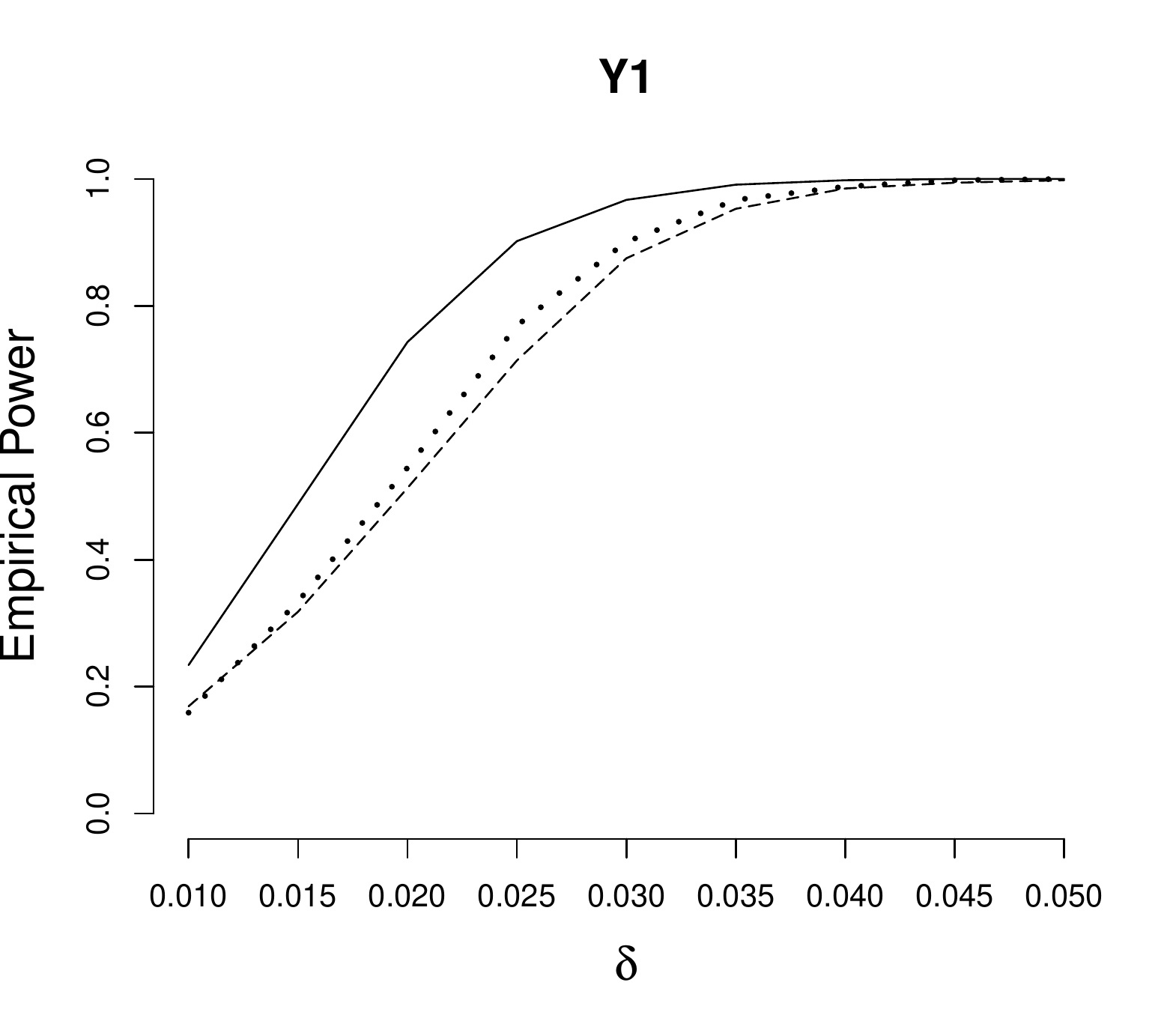}
		\renewcommand{\baselinestretch}{1.2}
		\caption{Empirical power of the competing GGF, MHR, and HR methods for 
			testing linear effect for the sparse sampling design. \textit{Solid lines} 
			indicate results for the MHR method, \textit{dashed lines} indicate results 
			for the GGF method, and \textit{dotted lines} indicate results for the HR method. 
			The significance level is $\alpha=0.05$. The number of Monte Carlo experiments is 
			1,000, and the sample size is $n=500$.}
		\label{plot_lin_spr_500}
	\end{figure}	
	% % % % % % % % % % % % % % % % % % % % % % % % % % % % % % % % % % % % % % % % % % % % % % % % % %
	% % % % % % % % % % % % % % % % % % % % % % % % % % % % % % % % % % % % % % % % % % % % % % % % % %
	
	% % % % % % % % % % % % % % % % % % % % % % % % % % % % % % % % % % % % % % % % % % % % % % % % % %
	% % % % % % % % % % % % % % % % % % % % % % % % % % % % % % % % % % % % % % % % % % % % % % % % % %
	% % % % % % % % % % % % % % % % % % % % % % % % % % % % % % % % % % % % % % % % % % % % % % % % % %
	% % % % % % % % % % % % % % % % % % % % % % % % % % % % % % % % % % % % % % % % % % % % % % % % % %
	% % % % % % % % % % % % % % % % % % % % % % % % % % % % % % % % % % % % % % % % % % % % % % % % % %
	
	\newpage
	\section{Simulation results for testing nullity}
	
	\subsection{\textbf{Power curves for dense sampling design}}
	
	\begin{figure}[htp]
		\centering
		\includegraphics[width=.48\textwidth]{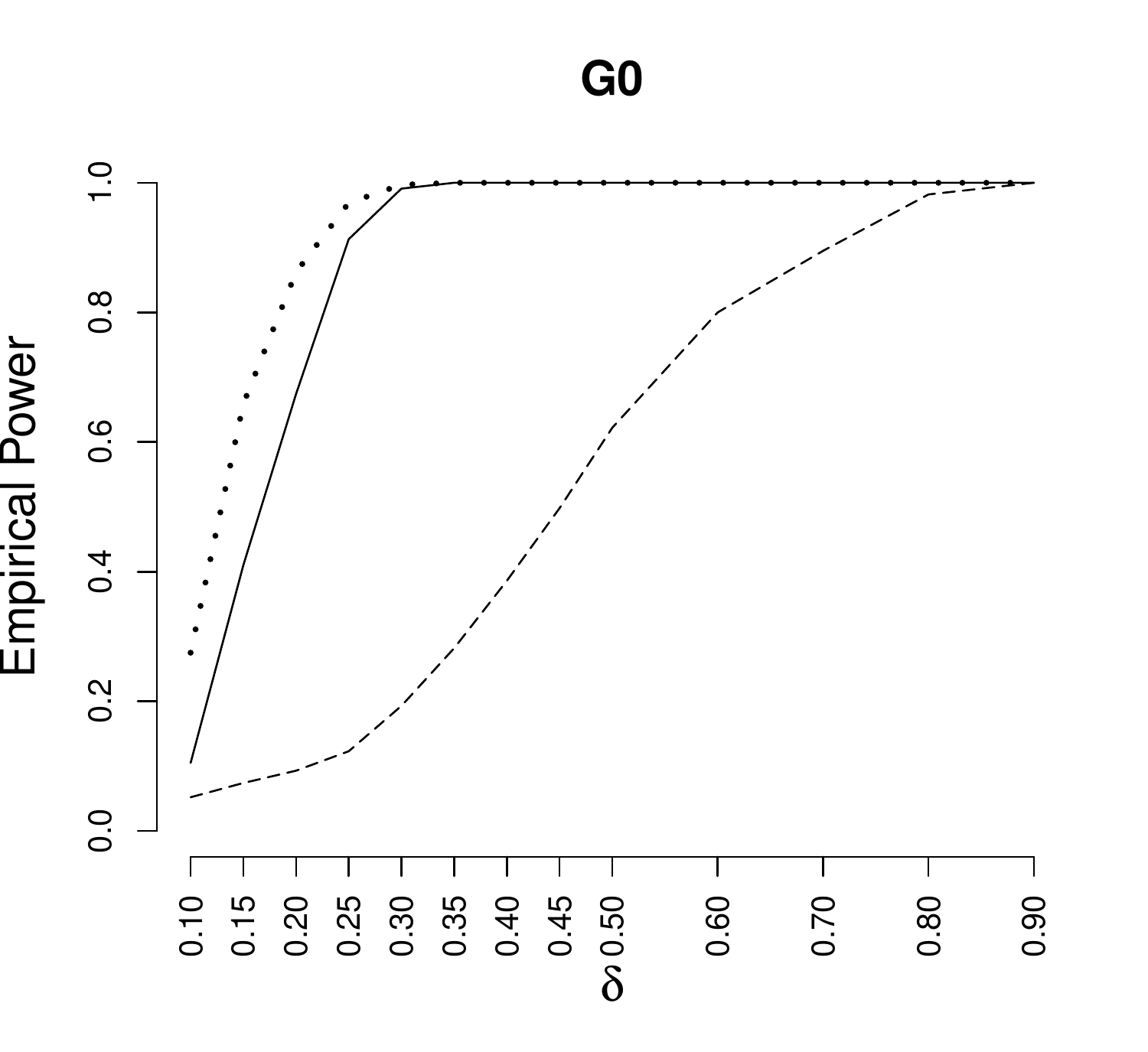}\quad
		\includegraphics[width=.49\textwidth]{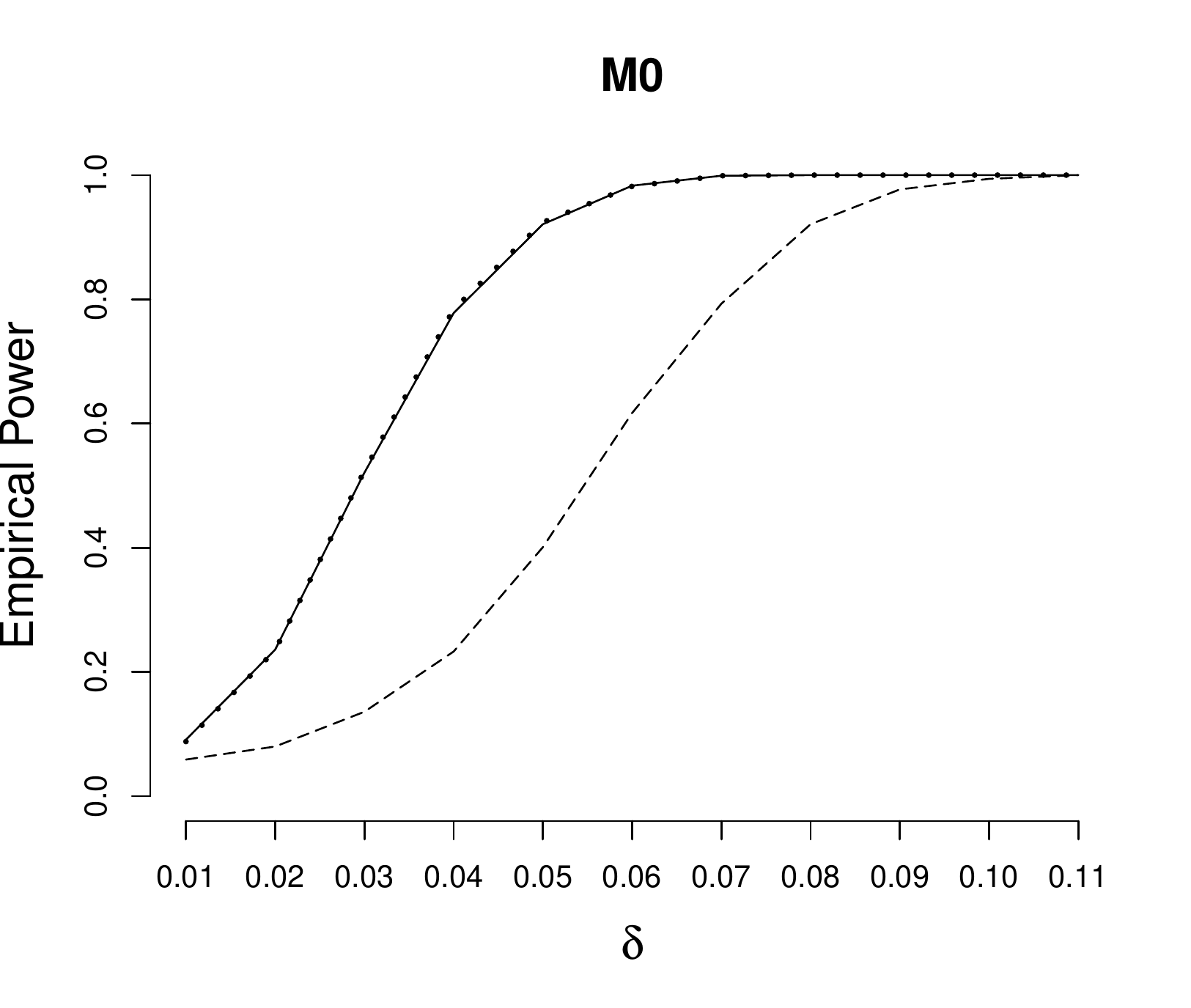}
		
		\medskip
		
		\includegraphics[width=.48\textwidth]{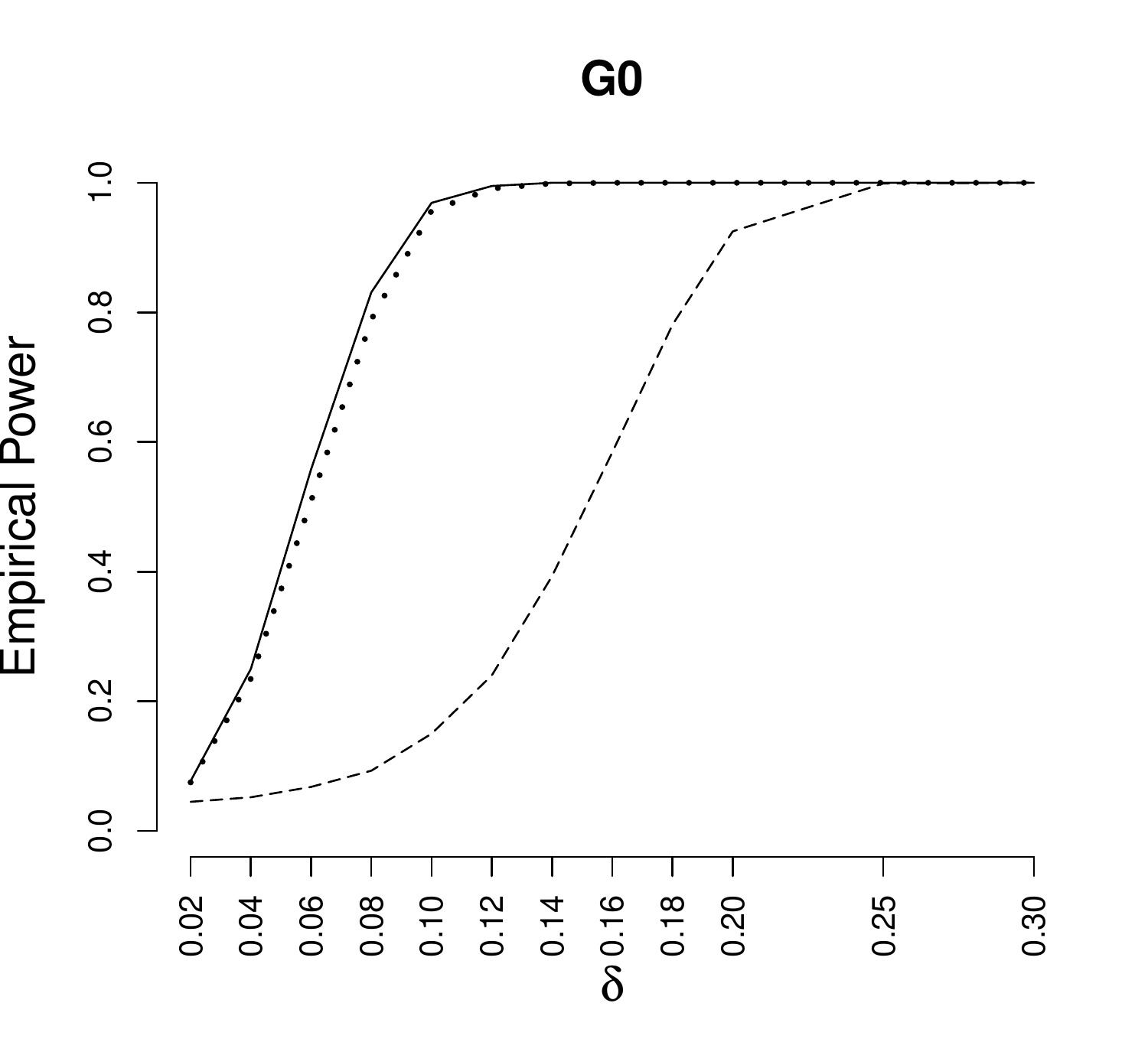}\quad
		\includegraphics[width=.49\textwidth]{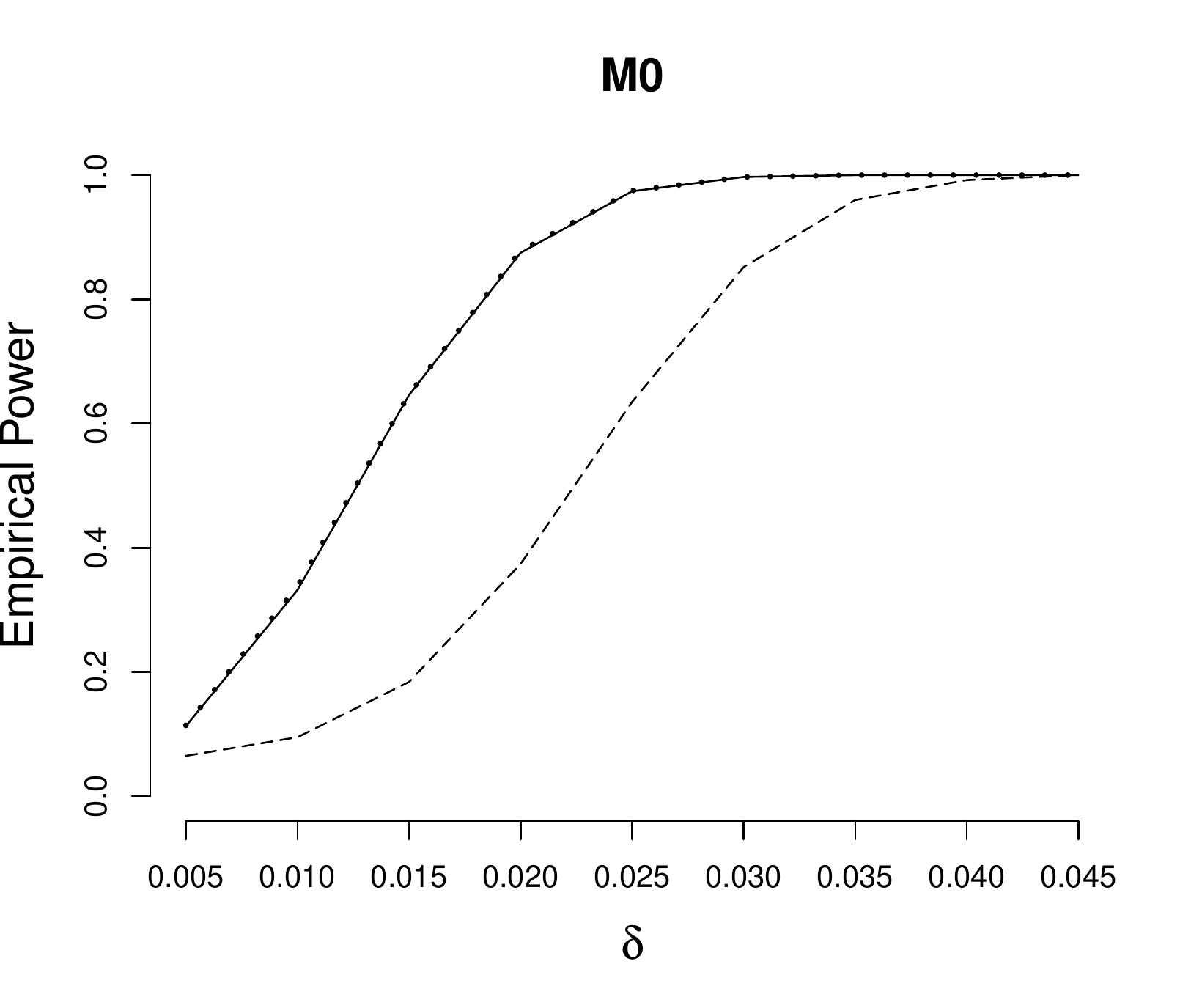}
		\renewcommand{\baselinestretch}{1.2}
		\caption{Empirical power of the competing GGF, MHR, and KSM methods for testing 
			no effect for the dense sampling design with sample sizes $n=100$ (1st row) 
			and $n=500$ (2nd row), respectively. \textit{Solid lines} indicate results for 
			the MHR method, \textit{dashed lines} indicate results for the GGF method, 
			and \textit{dotted lines} indicate results for the KSM method. The significance 
			level is $\alpha=0.05$. The number of Monte Carlo experiments is 1,000.}
		\label{plot_null_dns_100_500}
	\end{figure}
	%% % % % % % % % % % % % % % % % % % % % % % % % % % % % % % % % % % % % % % % % % % % % % % % % % %
	%% % % % % % % % % % % % % % % % % % % % % % % % % % % % % % % % % % % % % % % % % % % % % % % % % %
	\newpage
	
	\subsection{\textbf{Power curves for moderate sampling design}}
	
	\begin{figure}[htp]
		\centering
		\includegraphics[width=.48\textwidth]{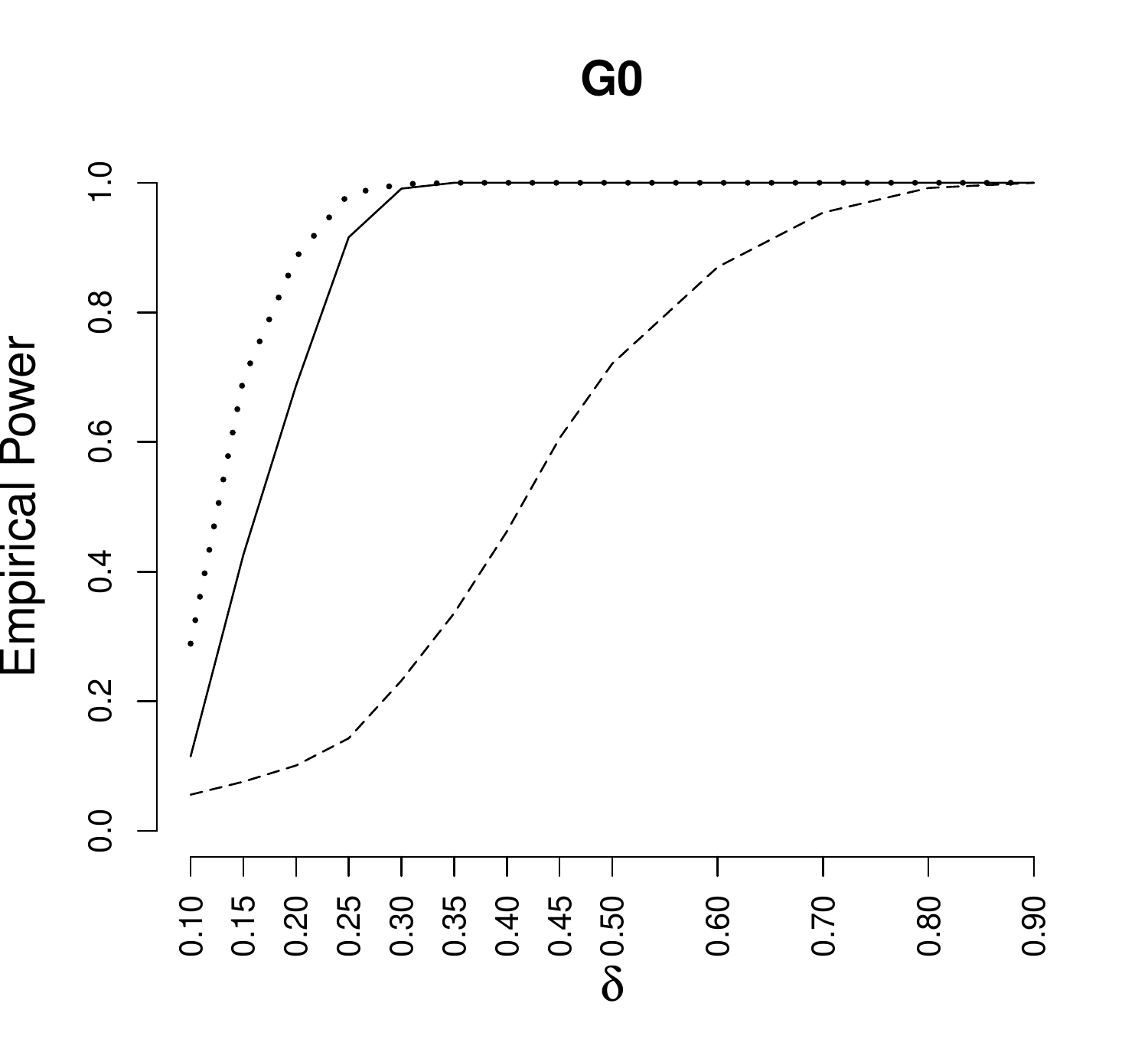}\quad
		\includegraphics[width=.48\textwidth]{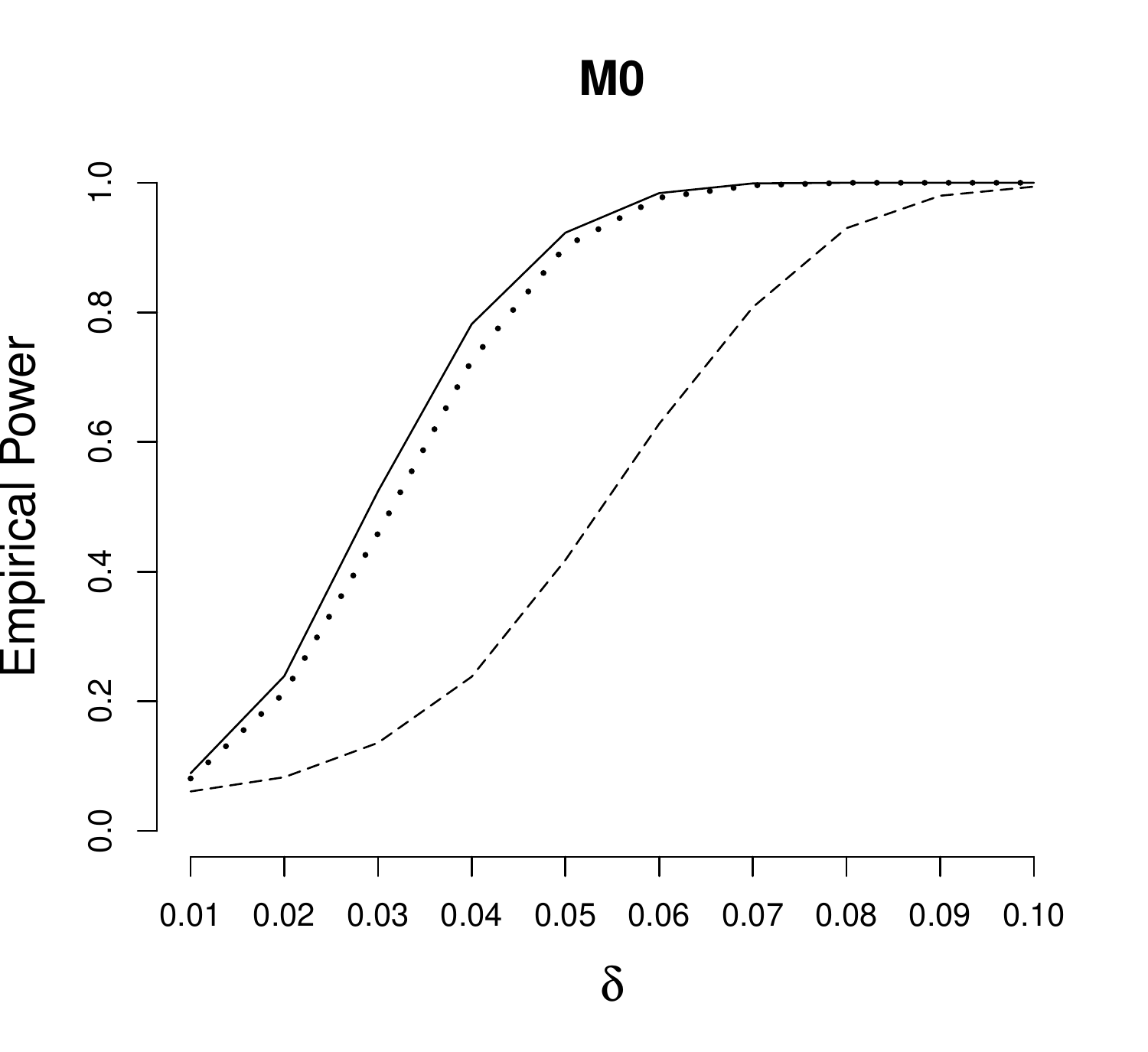}
		\renewcommand{\baselinestretch}{1.2}
		\caption{Empirical power of the competing GGF, MHR, and KSM methods for testing 
			no effect for the moderate sampling design. \textit{Solid lines} indicate results for 
			the MHR method, \textit{dashed lines} indicate results for the GGF method, 
			and \textit{dotted lines} indicate results for the KSM method. The significance 
			level is $\alpha=0.05$. The number of Monte Carlo experiments is 1,000, and 
			the sample size is $n=100$.}
		\label{plot_null_mod_100}
	\end{figure}
	%% % % % % % % % % % % % % % % % % % % % % % % % % % % % % % % % % % % % % % % % % % % % % % % % % %
	%% % % % % % % % % % % % % % % % % % % % % % % % % % % % % % % % % % % % % % % % % % % % % % % % % %
	\newpage
	
	\subsection{\textbf{Power curves for sparse sampling design}}
	
	\begin{figure}[htp]
		\centering
		\includegraphics[width=.48\textwidth]{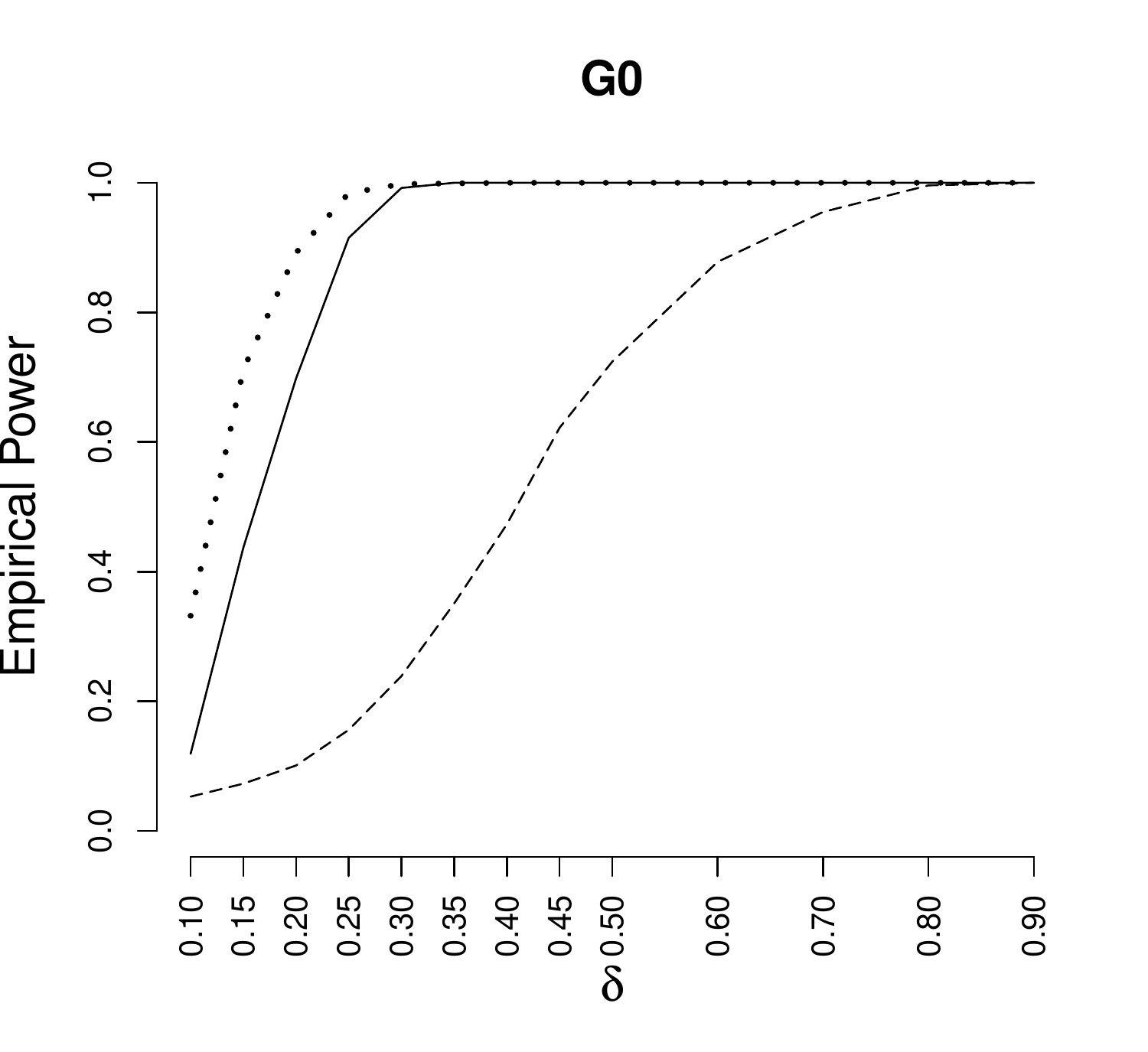}\quad
		\includegraphics[width=.48\textwidth]{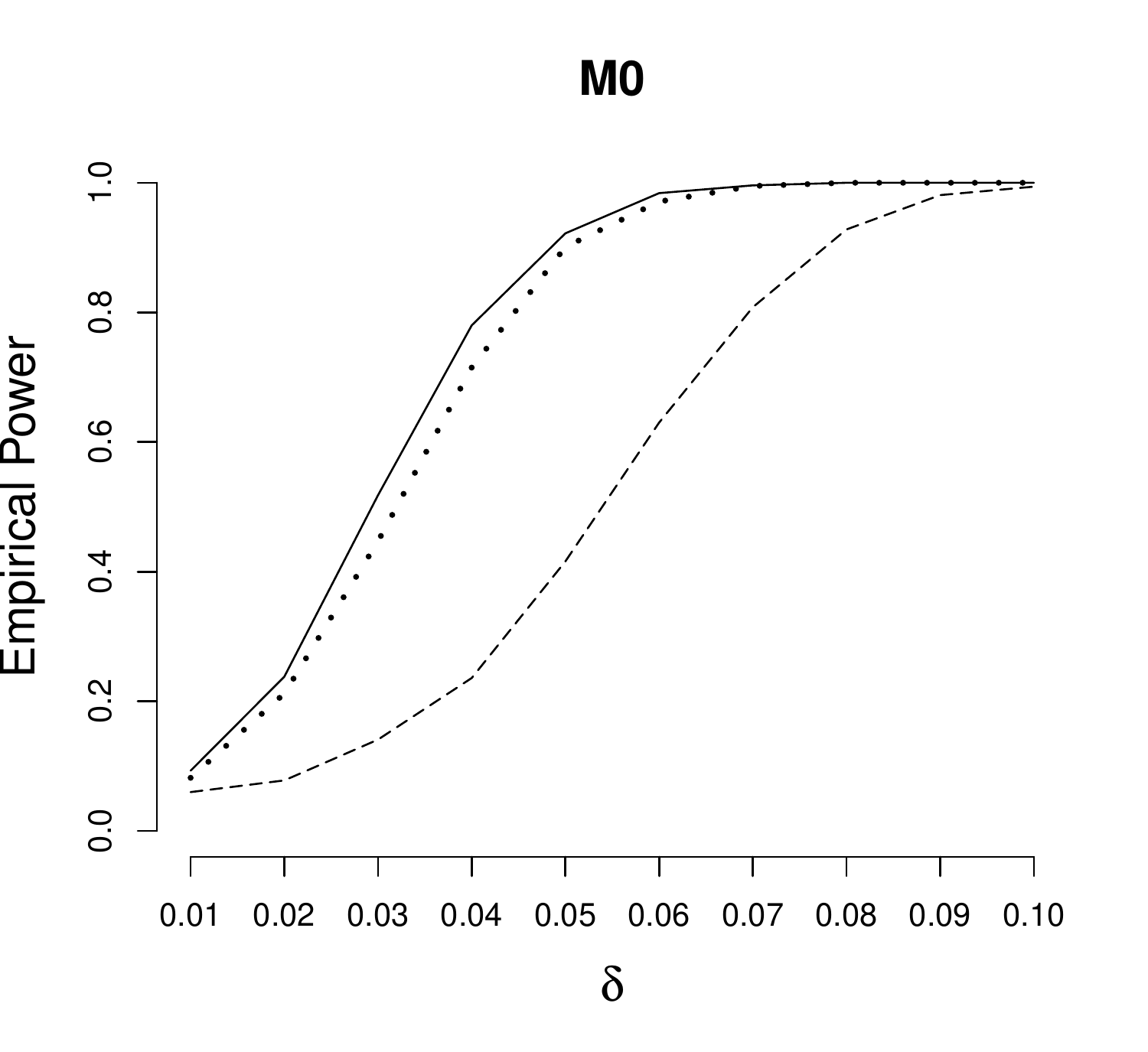}
		
		\medskip
		
		\includegraphics[width=.48\textwidth]{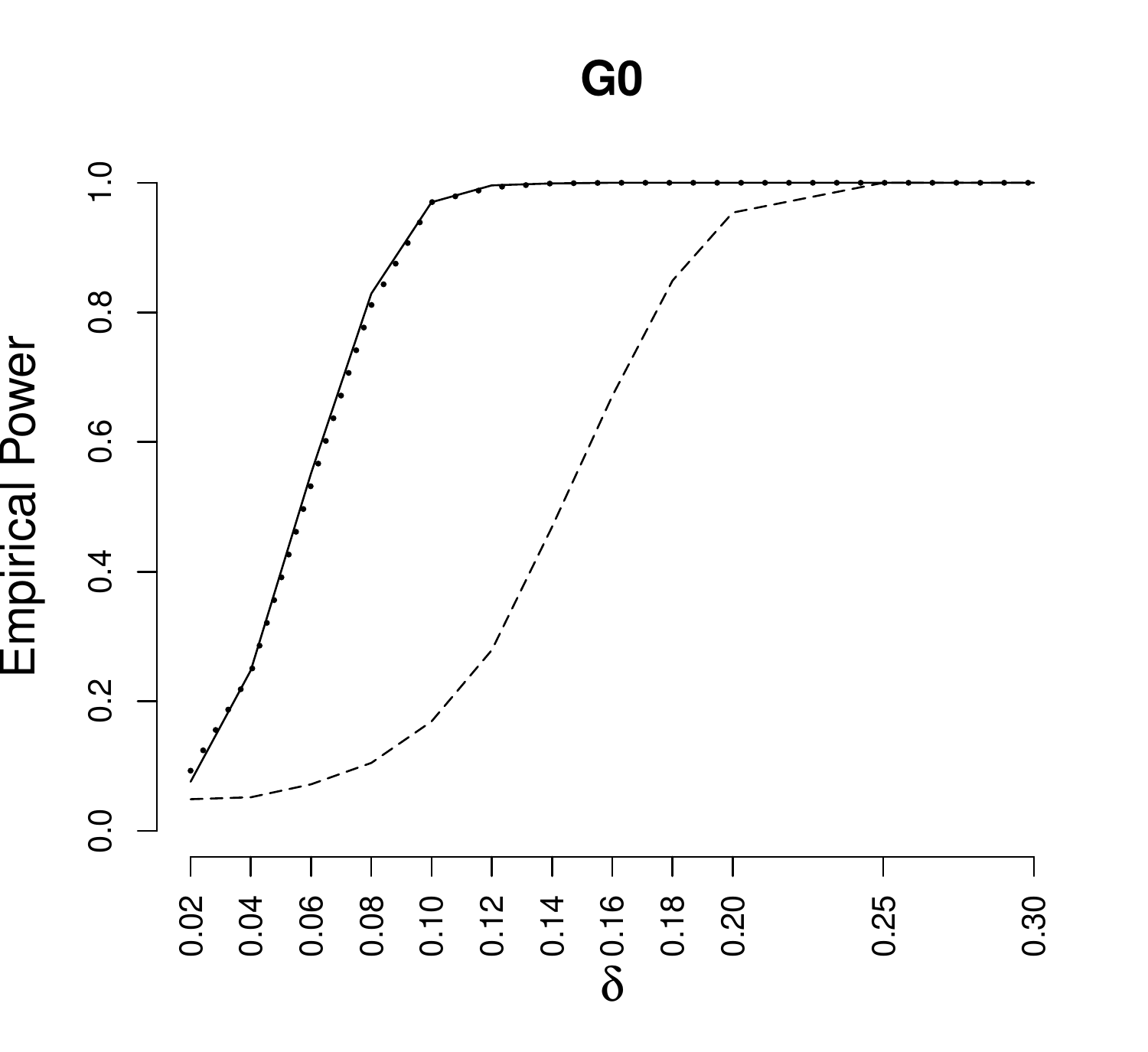}\quad
		\includegraphics[width=.49\textwidth]{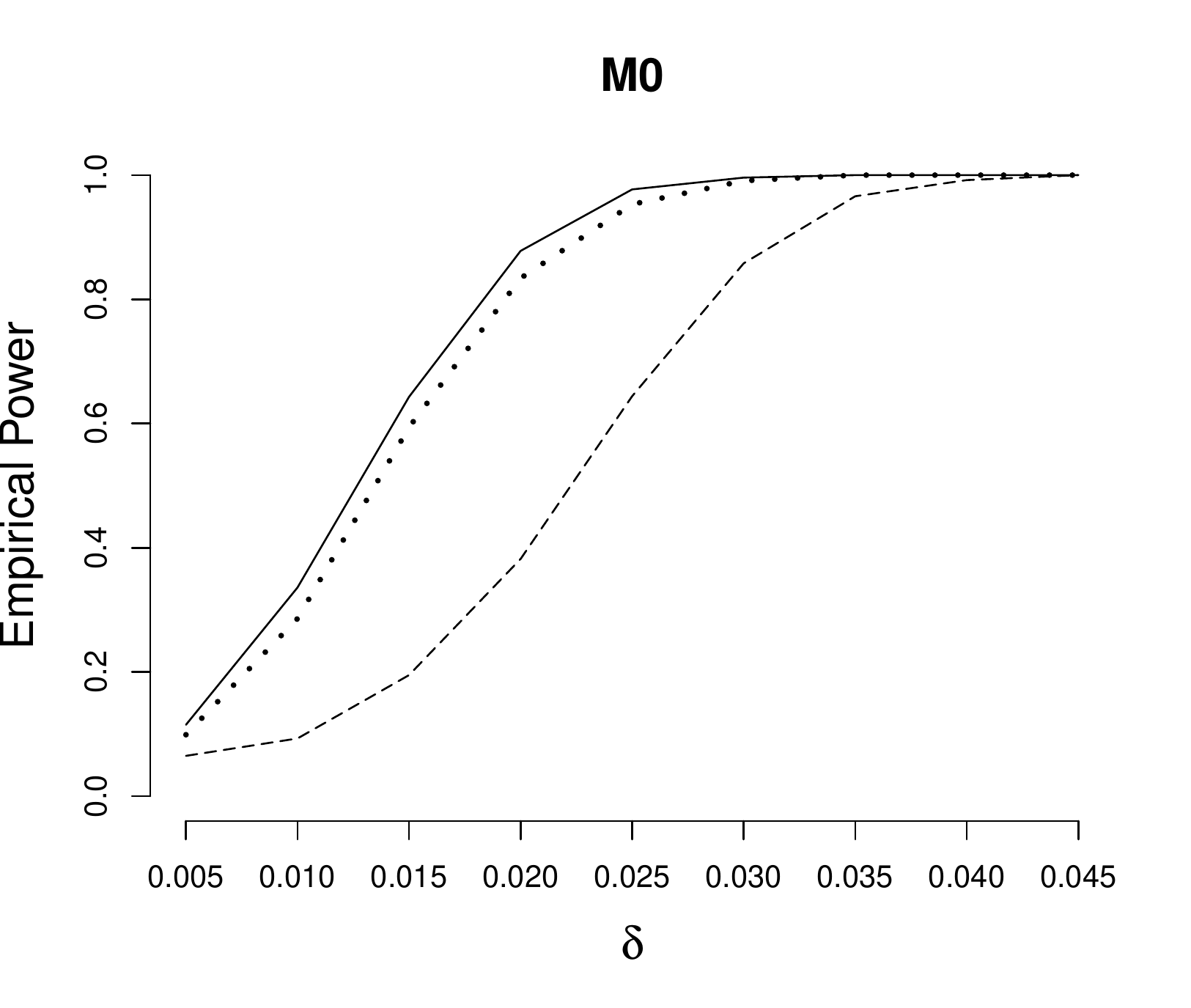}
		\renewcommand{\baselinestretch}{1.2}
		\caption{Empirical power of the competing GGF, MHR, and KSM methods for testing 
			no effect for the sparse sampling design with sample sizes $n=100$ (1st row) 
			and $n=500$ (2nd row), respectively. \textit{Solid lines} indicate results for 
			the MHR method, \textit{dashed lines} indicate results for the GGF method, 
			and \textit{dotted lines} indicate results for the KSM method. The significance 
			level is $\alpha=0.05$. The number of Monte Carlo experiments is 1,000.}
		\label{plot_null_spr_100_500}
	\end{figure}

\end{document}